\documentclass[fleqn,usenatbib]{mnras}
\usepackage[T1]{fontenc}
\usepackage[utf8]{inputenc}
\usepackage{graphicx}
\usepackage{subfig}
\usepackage{amsmath}
\usepackage{amssymb}
\usepackage{caption}
\usepackage{svg}
\usepackage{newtxtext,newtxmath}
\usepackage{xcolor}

\newcommand{\lrsb}[1]{\left[{#1}\right]}
\newcommand{\lara}[1]{\langle{#1}\rangle}

\def\mr{\mathrm}

\def\d{\mr{d}}

\def\Phiap{\Phi_{\rm ap}}
\def\PhiLT{\Phi_{\rm LT}}

\def\aorb{a_{\rm orb}}
\def\borb{b_{\rm orb}}

\def\rint{r_{\rm int}}

\usepackage[colorinlistoftodos]{todonotes}
\title[TDE stream intersection]{General Relativistic Stream Crossing in Tidal Disruption Events}
\author[Batra et al.]{Gauri Batra$^{1}$\thanks{gb377@cornell.edu}, Wenbin Lu$^{2, 3}$\thanks{wenbinlu@astro.princeton.edu}, Cl{\'e}ment Bonnerot$^{4, 3}$, E. Sterl Phinney$^3$ \\
$^1$Department of Physics, Cornell University, Ithaca, NY 14853, USA\\
$^2$Department of Astrophysical Sciences, Princeton University, Princeton, NJ 08544, USA\\
$^3$TAPIR, Walter Burke Institute for Theoretical Physics, Mail Code 350-17, Caltech, Pasadena, CA 91125, USA\\
$^4$Niels Bohr International Academy, Niels Bohr Institute, Blegdamsvej 17, DK-2100 Copenhagen Ø, Denmark
}
\date{December 2021}

\begin{document}

\maketitle

\begin{abstract}
    When a star is tidally disrupted by a supermassive black hole (BH) at the center of a galaxy, the gas debris is stretched into an elongated cold stream. The longitudinal motion of the stream accurately follows geodesics in the Kerr spacetime and the stream evolution in the transverse dimensions is decoupled from the longitudinal motion.
    Using an approximate tidal equation, we calculate the evolution of the stream thickness along the geodesic, during which we treat the effect of nozzle shocks as a perfect bounce. Self-intersection occurs when the closest approach separation is smaller than the stream thickness.
    This algorithm allows us to explore a wide parameter space of orbital angular momenta, inclinations, and BH spins to obtain the properties of stream intersection. We identify two collision modes: in about half of the cases, the collision occurs near the pericenter at an angle close to $0^{\rm o}$ (the ``rear-end'' mode) and the other half has collisions far from the pericenter with collision angles close to $180^{\rm o}$ (the ``head-on'' mode). The intersection typically occurs between consecutive half-orbits with a delay time that spans a wide range (from months up to a decade). The intersection radius generally increases with the orbital angular momentum and depends less strongly on the inclination and BH spin. The thickness ratio of the two colliding ends is of order unity and the transverse separation is a small fraction of the sum of the two thicknesses, so a large fraction of the stream is shock heated in an offset collision. Many of the numerical results can be analytically understood in a post-Newtonian picture, where the orientation of an elliptical orbit undergoes apsidal and Lense-Thirring precession. Instead of thickness inflation due to energy dissipation at nozzle shocks as invoked in earlier works, we find the physical reason for stream collision to be a geometric one.
    After the collision, we expect the gas to undergo secondary shocks and form an accretion disk, generating bright electromagnetic emission.
\end{abstract}

\begin{keywords}
transients: tidal disruption events --- black hole physics --- methods: numerical --- galaxies: nuclei 
\end{keywords}

\section{Introduction}
An inevitable consequence of the N-body dynamics near a supermassive black hole (BH) is that some stars will be scattered into highly eccentric orbits and get disrupted by the BH's enormous tidal forces \citep{hills75_tdes_agn, frank76_tde_rate, magorrian99_tde_rate, alexander05_stellar_dynamics, stone20_tde_rate_review}. Such tidal disruption events (TDEs) feed a transient episode of accretion onto the otherwise dormant BH and generate a bright flare across the electromagnetic spectrum \citep{lacy82_tdes_Galactic_Center, rees88_tdes, phinney89_tde_fallback_rate,cannizzo90_tde_accretion, ulmer99_tde_flare, strubbe09_tde_flare}. They provide our best chance of probing the demographics of supermassive BHs at the centers of quiescent galaxies at cosmological distances \citep{kormendy13_SMBH_demographics}.

The hydrodynamics of the tidal disruption process has now been reasonably well understood, thanks to a large volume of analytical and numerical works \citep{carter83_tidal_compression, evans89_tde_simulation, ayal00_tde_simulation, lodato09_tde_simulation, guillochon13_tde_simulation, stone13_tde_fb_rate, cheng14_relativistic_tdes, tejeda17_relativistic_tdes, steinberg19_deep_tdes, ryu20_tde_in_GR, rossi20_disruption_process}. The consequence is that the star is tidally stretched into a long, thin stream of debris, the evolution of which is controlled by the tidal gravity, self-gravity, as well as internal pressure forces \citep{kochanek94_stream_evolution,coughlin16_stream_evolution, Bonnerot21_stream_width_evolution}. In the longitudinal direction of the stream, the evolution is almost entirely controlled by tidal gravity, meaning that each segment of the stream accurately follows its own geodesic like a test particle. Each time a fluid element passes near the pericenter, it undergoes general relativistic (GR) precession, which will eventually lead to the self-intersection of the stream \citep{rees88_tdes, dai13_gr_precession, dai15_pn1_precession, guillochon15_dark_year, Bonnerot17_stream_crossings, liptai19_GRSPH_simulation, lu20_self-intersection}. If this intersection is sufficiently violent, it strongly broadens the angular momentum and orbital energy distributions of the bound debris and leads to numerous secondary shocks and then the formation of a coherently rotating accretion flow \citep{shiokawa15_GR_precession_intersection, bonnerot16_precession_schwarzschild, hayasaki16_precession_kerr, sadowski16_relativistic_tde, jiang16_stream_collision, lu20_self-intersection, bonnerot20_disk_formation, andalman20_disk-formation, curd21_global_TDE, bonnerot21_accretion_flow_review}. The processes of disk formation and accretion generate bright electromagnetic emission that is observable from cosmological distances \citep{strubbe09_tde_flare, coughlin14_ZEBRA, piran15_stream_collision, metzger16_bright_year, dai18_unified_model, lu20_self-intersection, bonnerot21_first_light}.

On the observational side, of the order $10^2$ TDE candidates have been identified by various transient searching strategies in different bands, including X-ray surveys \citep{bade96_rosat_tde, esquej08_xmm_newton_tdes, saxton12_x-ray_tde, saxton17_xmm_tde, lin15_xmm_tde, brightman21_X-ray_TDE, sazonov21_erosita_TDEs}, UV-optical surveys \citep{gezari06_GALEX_tde, gezari12_ps-10jh, chornock14_ps1-11af, holoien16_as14li, blagorodnova17_iptf16fnl, hung17_iPTF16axa, hinkle21_AS19dj, vanvelzen21_ztf_tdes}, and infrared surveys \citep{wang18_infrared_tdes, jiang21_MIRONG}.
To extract the physical information from the increasing number and diversity of TDEs, a better understanding of the overall hydrodynamics is needed. A crucial ingredient is the self-intersection of the fallback stream, as it initiates the formation of the accretion disk and all subsequent electromagnetic emission. 

In this work, we compute the geodesic motion of the fallback stream in the Kerr spacetime for a broad range of BH spins, orbital angular momenta and inclinations. An important feature of an inclined Kerr geodesic is that Lense-Thirring (LT) precession may cause missed prompt collision right after the first pericenter passage \citep{dai13_gr_precession, guillochon15_dark_year, hayasaki16_precession_kerr}, unlike in the Schwarzschild case. While the motion of the stream center is described by a Kerr geodesic, we model the stream thickness by integrating an approximated tidal equation. Then, the position of the stream collision is where the separation between the stream centers is less than the sum of the thicknesses of the two approaching ends. One of the key differences between our model and the work of \citet{guillochon15_dark_year} is the treatment of stream thickness evolution. They assume that the stream thickness increases linearly with orbital winding number (their Eq. 2), which they attribute to some unspecified energy dissipation by the nozzle shock near the pericenter. Consequently, the stream collision is mainly the result of the continuing growth of the stream thickness in their study. However, recent numerical simulations by \citet{Bonnerot21_stream_width_evolution} show that the stream thickness does not dramatically increase after it passes through the nozzle shock. In this work, we treat the effect of the nozzle shock as a perfect bounce, and the evolution of the stream thickness is determined by the relativistic tidal forces.
It will be shown that the physical reason for the stream collision is a geometrical one that can be analytically described in a post-Newtonian picture for mildly relativistic orbits.

This paper is organized as follows. In \S \ref{sec:methods}, we describe our algorithm, which takes as input the BH spin and the stream’s orbital parameters and then computes the position of the stream intersection. A parameter space exploration is presented in \S \ref{sec:parameter_exploration}, where we show how the properties of the stream intersection depend on the BH spin and the stream's orbital angular momentum and inclination. In \S\ref{sec:physical_reason}, we provide a post-Newtonian description that unveils the physical reason for the stream collision and analytically predicts the delay time before the collision. The limitations of our model are discussed in \S \ref{sec:discussion}. A summary of the main results is provided in \S \ref{sec:summary}. In this work, we only consider a representative TDE case of a $M=10^6M_\odot$ BH and a $1M_\odot$ star. The exploration of a wider parameter space is left for a future work. Our calculations are fully general relativistic, so we use geometrical units with $G=M=c=1$ throughout the paper, unless otherwise specified.

\section{Methods}\label{sec:methods}

\subsection{Geodesic motion of stream center}
\label{sec:orbit}

The geodesic of a representative debris particle at the center of the fallback stream is specified by the following parameters:
\begin{enumerate}
    \item $\tilde E$, the total specific energy (including rest mass),
    \item $L$, the total specific angular momentum,
    \item $I$, the inclination of the orbit (where $I=0$ is for the equatorial case and $I=\pi/2$ is for an initially polar orbit), and
    \item $a$, the BH spin (where $a=0$ is the Schwarzschild case and $a<0$ is for retrograde orbits),
\end{enumerate}
The inclination angle is defined by $\cos I = L_z/L$ ($L_z$ being the specific angular momentum component along the spin axis) and the Carter constant is $Q=L^2(\sin I)^2$.

The geodesic equations of motion for a Kerr spacetime are numerically integrated to machine precision using a program based on \cite{1994ApJ...421...46R}. We work in Boyer-Lindquist coordinates ($t$, $r$, $\theta$, $\phi$) with the metric given by
\begin{eqnarray}
\begin{split}
    d s^2 = \,& -\left(1-{2r\over \Sigma}\right) dt^2 - {4ar\sin^2\theta \over \Sigma} dt d\phi + {\Sigma \over \Delta} dr^2 + \Sigma d \theta^2 \\
    &+ \left(r^2 + a^2 + {2a^2 r\sin^2\theta \over \Sigma}\right)\sin^2\theta\, d \phi^2,
\end{split}
\end{eqnarray}
where
\begin{eqnarray}
    \Sigma=r^2+a^2\mu^2,
    \label{eq:sigma}
\end{eqnarray}
\begin{eqnarray}
    \Delta=r^2-2r+a^2.
    \label{eq:delta}
\end{eqnarray}
The orbit is initialized at the apocenter at $\tau=\phi=0$ and $\theta=\pi/2$, where $\tau$ is the proper time. Note that starting from $\theta=\pi/2$ (in the BH's equatorial plane) loses the generality of the initial conditions, but as we show in \S\ref{sec:physical_reason}, this choice is representative for most TDEs. The main effect of other choices of initial $\theta$ is to change the delay time before the stream collision, and the dependence can be analytically predicted under the post-Newtonian approximation (see \S\ref{sec:physical_reason}).

\begin{figure}
    \centering
    \includegraphics[width=0.48\textwidth]{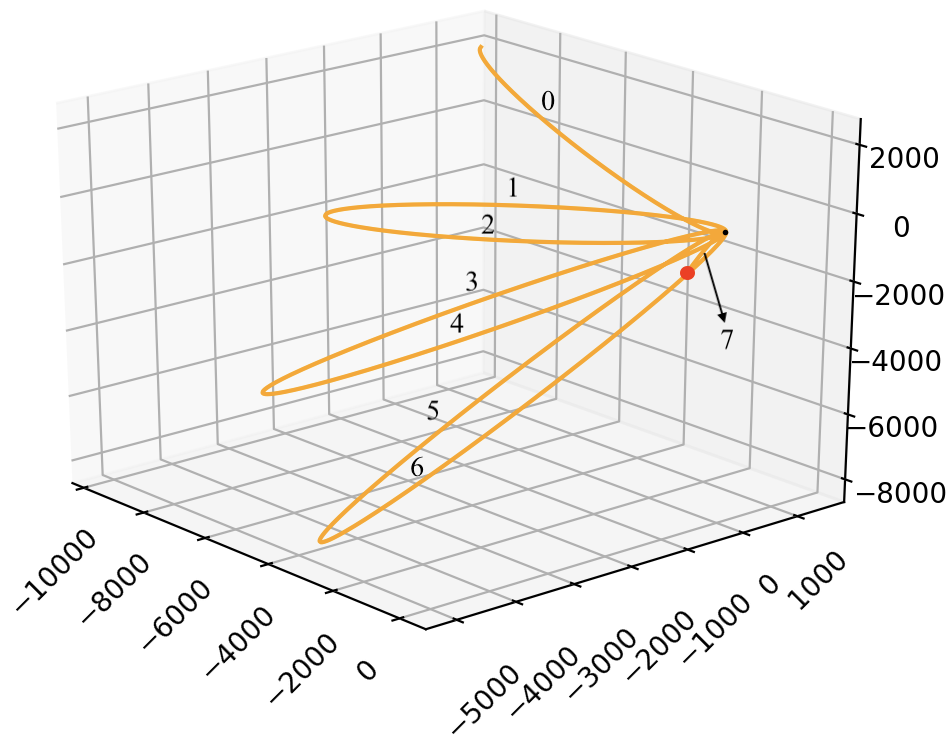}
    \caption{The orbit around a Kerr BH for a test particle at the center of the fallback stream. The parameters for this case are: total energy $\tilde E=0.9999$, total angular momentum $L=6.5$, inclination angle $\cos I=0.5$, BH spin $a=0.9$. The position of the BH, the origin of the Cartesian coordinate system, is marked by a black dot. The evolution is terminated when the stream crossing occurs, as marked by a red dot (the parameters for the stream thickness evolution are described in \S \ref{sec:stream_thickness}). All axes are in units of the gravitational radius $GM/c^2$. 
    } 
    \label{fig:fullorbit}
\end{figure}

For $u=1/r$ and $\mu=\cos\theta$, points on the orbit are sampled in intervals of $du$, which is related to proper-time intervals $d\tau$ by \citep{bardeen72_Kerr_geodesic}
\begin{eqnarray}
    \label{eq:dudtau}
    \frac{du}{d\tau}=u^2\frac{\sqrt{V_r}}{\Sigma},
\end{eqnarray}
where
\begin{eqnarray}
    \label{eq:Vr}
    V_r=T^2-\Delta \left(r^2+K\right),
\end{eqnarray}
\begin{eqnarray}
   K = (L_z - a\tilde{E})^2 + Q,
\end{eqnarray}
\begin{eqnarray}\label{eq:T}
   T=\tilde E \left(r^2+a^2\right)-aL_z.
\end{eqnarray}
An example of the geodesic motion for total energy $\tilde E=0.9999$, total angular momentum $L=6.5$, inclination angle $\cos I=0.5$ and BH spin $a=0.9$ is shown in Fig. \ref{fig:fullorbit}.

\subsection{Stream thickness evolution}\label{sec:stream_thickness}

\begin{figure}
    \centering
    \includegraphics[width=0.48\textwidth]{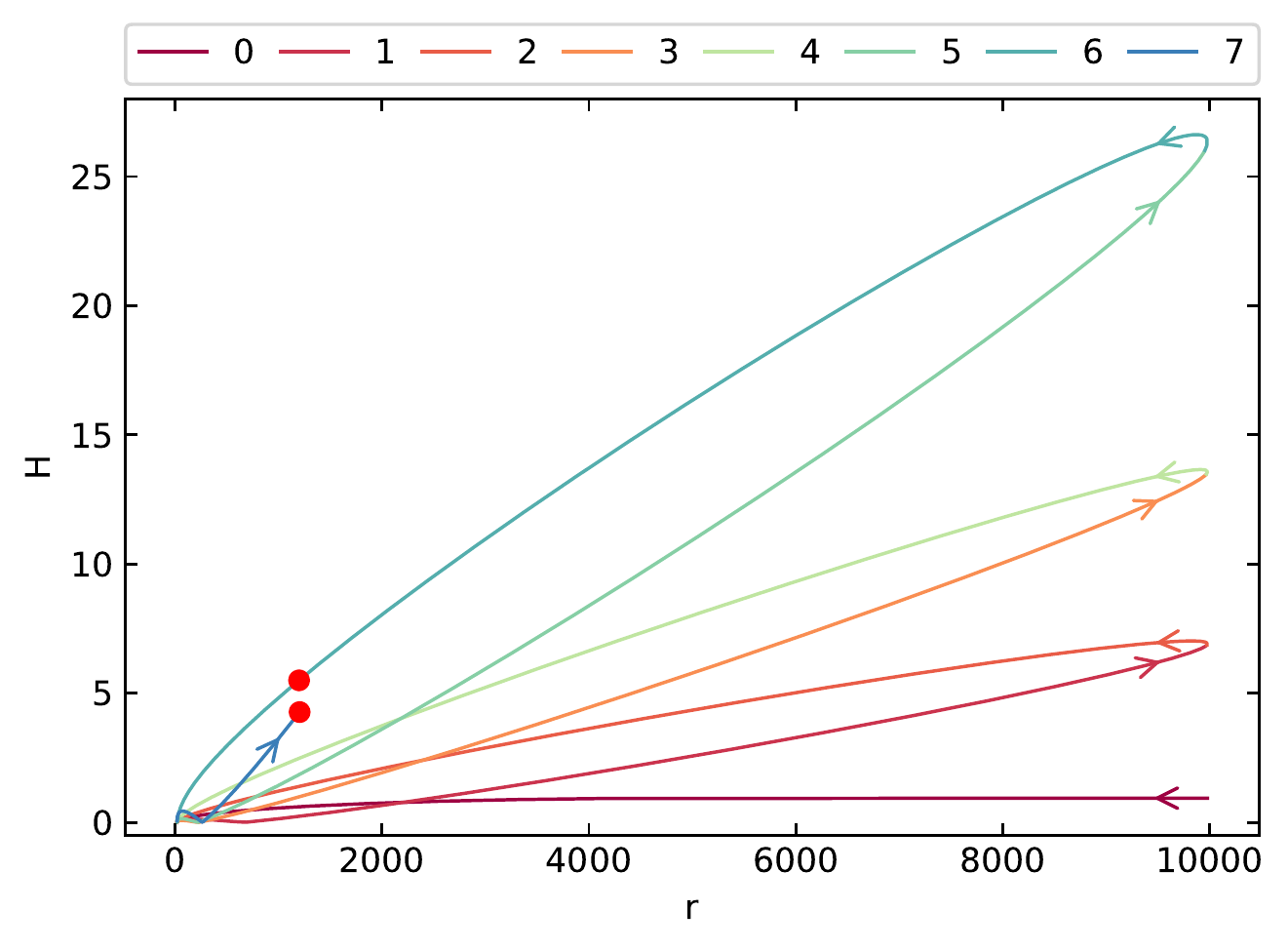}
    \includegraphics[width=0.48\textwidth]{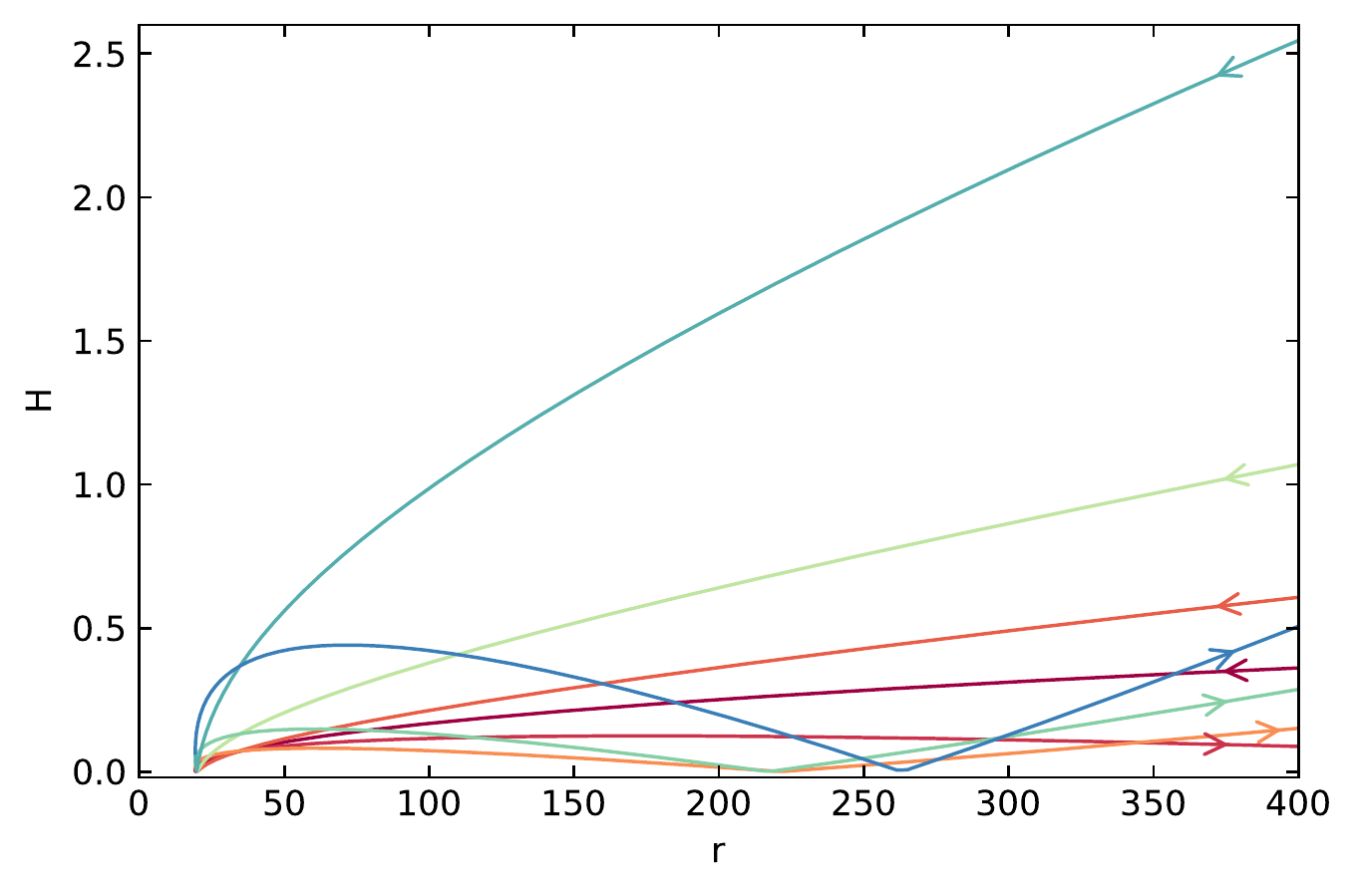}
    \includegraphics[width=0.48\textwidth]{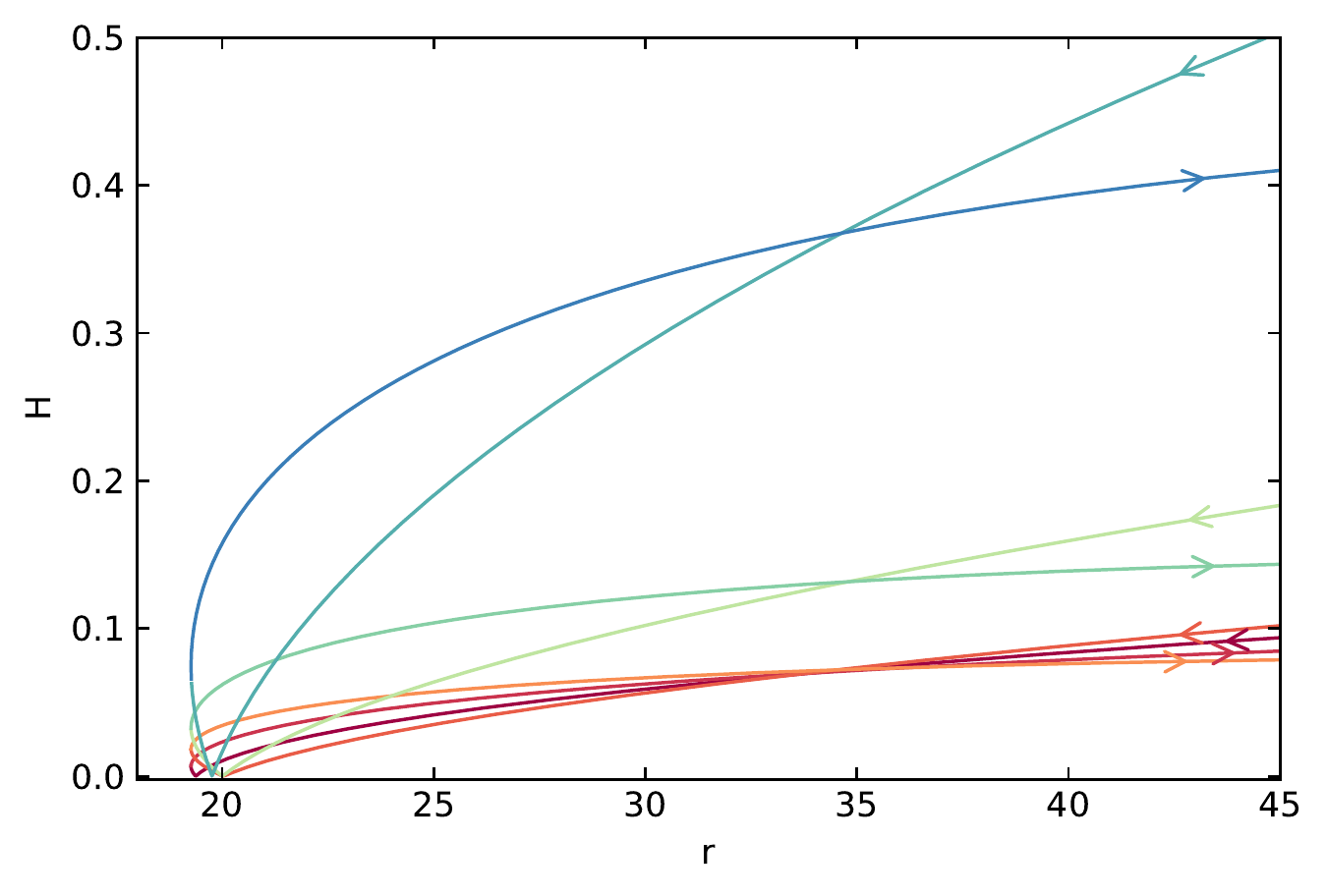}
    \caption{The evolution of stream thickness $H$ as a function of $r$, magnified further to show details. To read the plot, start at the apocenter, and follow the plot continuously to see how the thickness would evolve in time and with $r$. On average, the stream becomes thicker with each successive winding, and the thickness ratio between the outward and inward segments of the same winding decreases with time. The geodesic is the same as shown in Fig. \ref{fig:fullorbit}, with $\tilde E=0.9999$, $L=6.5$, $\cos I=0.5$, $a=0.9$. Here we set $\sigma=1$ (see \S \ref{sec:stream_thickness} for its definition). The evolution is terminated when the stream crossing occurs, as marked by the red dots. Both axes are in units of the gravitational radius $GM/c^2$. 
    }
    \label{fig:H_evolve}
\end{figure}

After tidal disruption, the star is stretched into a long, thin stream. In the previous subsection, we compute the orbit of a test particle at the center of mass of the stream. In the plane transverse to the velocity vector, the stream has finite thickness, the evolution of which must be calculated to obtain the position of stream crossing. The long-term evolution of the stream thickness over multiple pericenter passages is currently still an open question, due to the prohibitive computational costs of global GR (magneto-) hydrodynamic simulations \citep[see][and references therein]{bonnerot21_accretion_flow_review}. Right after the tidal disruption, self-gravity plays an important role confining the stream in the transverse direction, but it becomes unimportant soon after the first apocenter passage. We therefore neglect the influence of pressure forces in the subsequent evolution except at the nozzle shocks caused by short episodes of strong compression in the transverse direction.

These considerations motivate us to model the evolution of the stream thickness $H$ according to the following approximate tidal force equation based on the $C_{22}$ component of the tidal tensor given by \citep[][their Eq. (56)]{marck83_tidal_tensor}
\begin{eqnarray}\label{eq:tfe}
\begin{split}
    \frac{d^2H}{d\tau^2} &\approx -H {r\over \Sigma^3}(r^2 - 3a^2\mu^2)
    \left(1 + 3{r^2 R^2 - a^2\mu^2 S^2 \over K \Sigma^2}\right),\\
    R &= K - a^2\mu^2,\ S = r^2 + K,
\end{split}
\end{eqnarray}
where the stream thickness $H$ is considered to be in the direction perpendicular to the local orbital plane (as defined far from the BH where GR is unimportant). Eq. (\ref{eq:tfe}) is exact in the Schwarzschild limit ($a\approx 0$), which is a good approximation because most of the stream-thickness evolution occurs far from the BH where the spin effects are minor. For the majority of the parameter space considered in this work, the LT precession angle is small ($\ll 1\rm\, rad$) and stream crossing typically occurs between adjacent orbital windings, so the thickness perpendicular to the local orbital plane is the main physical quantity determining the occurrence of intersection. We do not consider the evolution of stream thickness within the local orbital plane, which is affected by the angular momentum spread across the stream\footnote{The angular momentum spread across the stream \citep[as can be seen from the simulations by][]{cheng14_relativistic_tdes} may cause the stream to look like a fan, with in-plane width typically larger than the vertical width. } as well as more complicated tidal and non-inertial forces than in Eq. (\ref{eq:tfe}). A full treatment of the 2D stream cross-sectional structure in Kerr spacetime is much more complicated and computationally expensive, and this is left for future works.



We consider the part of the stream that is initially self-gravitating right after the disruption.
The initial conditions for the thickness evolution are taken to roughly correspond to the physical situation that the stream's self-gravity becomes sub-dominant compared to tidal forces shortly after the first apocenter passage. Motivated by the hydrodynamic simulations of \citet[][their Fig. 5]{Bonnerot21_stream_width_evolution}, we set the initial stream thickness $H_0=5R_\odot$ and the initial vertical expansion velocity $\dot H_0=0$ when the stream center reaches a distance $r=r_{\rm a}/2$ ($r_{\rm a}$ being the apocenter radius) from the BH after the first apocenter passage. Our results are insensitive to the exact position of maximum thickness (where $\dot H=0$), because the thickness only depends weakly on the radius for this part of the evolution.
For the same reason, we also take the stream thickness between $r=r_{\rm a}/2$ and $r=r_{\rm a}$ (the apocenter) to be constant at $H=H_0$. As we will discuss later, the line tangent to the orbit at the point of maximum stream thickness (where $\dot H = 0$) determines the positions of the nozzle shocks (where $H=0$). Since we choose the initial point for the thickness evolution to be on the semi-\textit{minor} axis, this tangent line is parallel to the semi-major axis. For other choices of the initial point somewhere between $r=r_{\rm a}/2$ and $r_{\rm a}$ (but not extremely close to the apocenter), the tangent line will also be approximately parallel to the major axis, because the orbit is nearly radial due to its high eccentricity. In the subsequent evolution, the positions of the nozzle shocks are strongly affected by apsidal precession and are insensitive to our choice of the initial point.

For a given geodesic with known $r(\tau)$ and $\theta(\tau)$, we numerically integrate Eq. (\ref{eq:tfe}) to find the stream thickness evolution $H(\tau)$. The integration is performed using the fourth-order Runge-Kutta method with an adaptive step size of $d\tau=10^{-4}r^{3/2}$ (and accuracy is checked through a numerical convergence study).
We then map $d\tau$ to $du$ using Eq. (\ref{eq:dudtau}).

Another aspect of the stream thickness evolution is the occurrence of nozzle shocks when the thickness reaches $H=0$. Near the point of maximum compression, hydrodynamic shocks convert the bulk kinetic energy associated with the motion perpendicular to the local orbital plane into heat. The pressure of the shocked gas then leads to outward expansion. As long as the timescale for the expansion of the shocked gas is much shorter than the orbital time near the pericenter, the effect of the nozzle shocks may be approximately taken into account by instantaneously flipping the sign of $\dot H$. Our model assumes the post-bounce vertical velocity to be $-\sigma \dot{H}$, where $\sigma$ is an adjustable parameter describing the strength of the bounce. Motivated by the Newtonian hydrodynamic simulations of \citet{Bonnerot21_stream_width_evolution}, we set $\sigma=1$ in this paper. If a small fraction of the orbital kinetic energy is dissipated by nozzle shocks \citep[as speculated by][]{leloudas16_15lh_TDE}, then it is possible to have $\sigma >1$, which would make the stream thicker (and self-intersection would occur earlier).



An example of the stream thickness evolution is shown in Fig. \ref{fig:H_evolve}, for the geodesic considered in Fig. \ref{fig:fullorbit}. An interesting result of our simulations is that the positions of nozzle shocks are not at the pericenter or apocenter. This is mainly due to GR apsidal precession \citep[][see their Figs. 4 and 5]{luminet85_nozzle_shock}. The center of mass of the stream and its upper edge move as test particles on two different orbital planes. The two planes intersect at a line\footnote{This will later on be referred to as ``Intersection Line 1'' so as to differentiate it with the other ``Intersection Line 2'' between the two (LT-precessed) orbital planes of the stream centers. Both intersection lines are important in our geometric understanding of the physical reason for stream collision. } which goes through the BH. The stream thickness $H$ is dictated by the projected distance of the stream position $\vec{r}$ to this intersection line, and hence $H=0$ when the stream crosses this line. Due to apsidal precession, it is not possible for this intersection line to go through both the pericenter and apocenter. We find that each orbital winding has two nozzle shocks, one of which is near the pericenter and the other one is far from the pericenter and the apocenter, and that LT precession further causes variations in the nozzle shock positions in each successive winding.



It is important to note a limitation of this model: the effect of nozzle shocks on the stream thickness is treated as an instantaneous rebound when the stream undergoes maximum compression at $H=0$. In reality, pressure forces may continue to be important long after the nozzle shocks, and future works may incorporate the results of pressure forces from hydrodynamic simulations \citep[see][]{Bonnerot21_stream_width_evolution} into the evolution of stream thickness.



\subsection{Computing stream crossing}\label{sec:intersect}

\begin{table}
    \centering
    \begin{tabular}{c c}
    \hline
    $i$ & 7\\
    $j$ & 6\\
    $r$ & 1200.8\\
    $\theta\rm\,[rad]$ & 2.60\\
    $\phi\rm\,[rad]$ & 4.57\\
    $\Delta s$ & 8.18\\
    $H_1+H_2$ & 9.76\\
    $\Delta s/(H_1+H_2)$ & 0.84\\
    $H_2/H_1$ & 0.77\\
    $H_1$ & 5.51\\
    $\cos \gamma$ & -0.96\\
    $v_{\rm rel}$ & 0.08\\
    \hline
    \end{tabular}
    \caption{Properties of the intersection region, including the spatial coordinates ($r,\, \theta,\, \phi$), minimum separation between the two colliding segments $\Delta s$, the stream width $H_1$ and $H_2$, the collision angle $\gamma$, and the relative speed of collision $v_{\rm rel}$, for the parameters $\tilde E=0.9999$, $L=6.5$, $\cos I=0.5$, $a=0.9$ and $\sigma=1$. All results are in units where $G=M=c=1$.
    }
    \label{tab:finalresult}
\end{table}

To find the point of stream collision, we first project the 3D orbit onto the $r$-$\phi$ plane and find the points where the trajectory intersects itself in that 2D projection. These points are labelled as intersection candidates, which are found using the following procedure:

\begin{enumerate}
    \item Split the entire orbit into many half-orbit arrays, going from pericenter to apocenter or from apocenter to pericenter.
    \item Compare the half-orbit array labelled $i=0, 1, 2, \ldots$, to each half-orbit $j(<i)$ before it.
    Consider two consecutive points sampled on each orbit in the order of increasing $r$: $A=(r_i^1, \phi_i^1)$ and $B=(r_i^2, \phi_i^2)$ on $i$, and $C=(r_j^1, \phi_j^1)$ and $D=(r_j^2, \phi_j^2)$ on $j$. For $\phi\in(0,2\pi)$, let $\Delta \phi_1=\phi_j^1-\phi_i^1$ and $\Delta \phi_2=\phi_j^2-\phi_i^2$. Intersection in 2D occurs when $\Delta \phi_1\Delta \phi_2<0$. The intersection candidate lies between $A$ and $B$ on $i$, and between $C$ and $D$ on $j$.
\end{enumerate}

The resolution in $r$ ($dr$) and in $\phi$ ($d\phi$) are given by the step size $d\tau=10^{-4}r^{3/2}$ using the geodesic equations of motion, $dr/d\tau=\sqrt{V_r}/\Sigma$ and $d\phi/d\tau=(-a\tilde E + L_z + aT/\Delta)/\Sigma$ \citep{bardeen72_Kerr_geodesic}, where the quantities $\Sigma$, $\Delta$, $V_r$, and $T$ are defined in Eqs. (\ref{eq:sigma}) -- (\ref{eq:T}). This resolution means that the fractional error in the position of stream collision is less than $10^{-4}$, which is negligible compared to $H/r$ (which sets the physical ambiguity of the collision position).


Once we have the intersection candidates, each given by four points as above, the following procedure is used to determine whether collision has physically taken place:

\begin{enumerate}
    \item Compute the shortest distance between the two line segments $AB$ and $CD$, and call this $\Delta s$, which is computed in the local frame near the intersection candidate.
    \item Let $H_1$ be the average of the stream thicknesses at $C$ and $D$. Let $H_2$ be the average of the stream thicknesses at $A$ and $B$. Collision occurs if $\Delta s<H_1+H_2$.
\end{enumerate}


\begin{figure}
    \centering
    \includegraphics[width=0.4\textwidth]{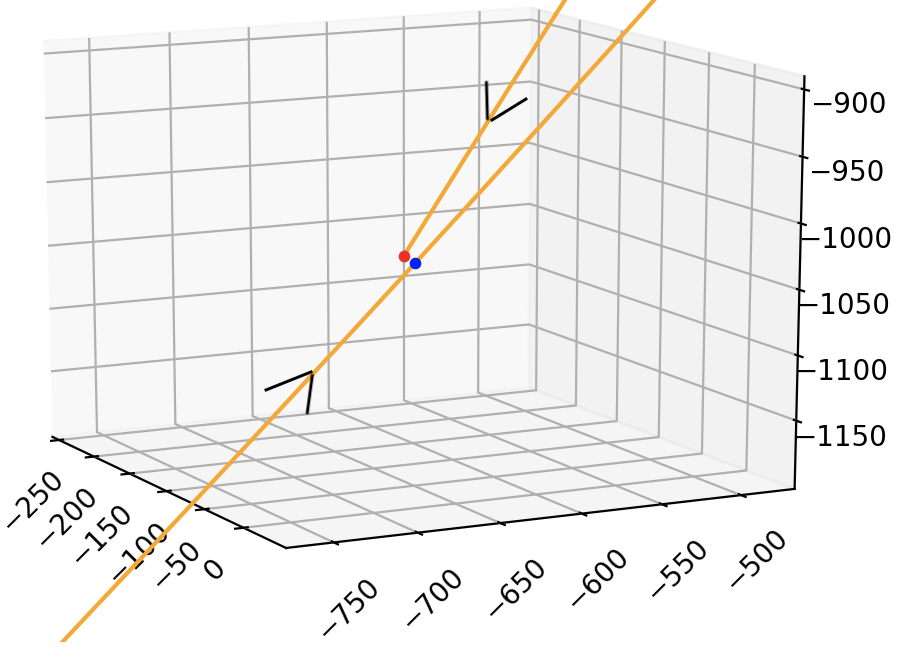}
    \includegraphics[width=0.4\textwidth]{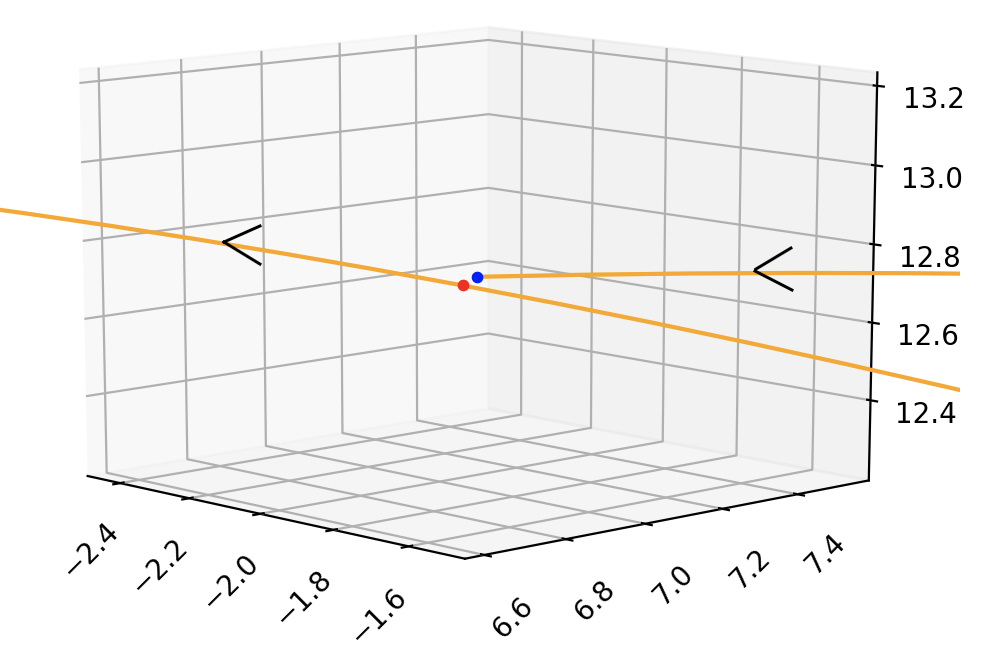}
    \caption{Two different modes of intersection for $a=0.9$. The top panel shows a case of the ``head-on collision'' mode with $\cos\gamma$ close to $-1$ ($\gamma$ being the intersection angle), for inclination angle $\cos I=0.5$ and total angular momentum $L=6.5$. The bottom panel shows a ``rear-end collision'' case with $\cos\gamma$ close to 1, for the same inclination $\cos (I)=0.5$ but slightly smaller angular momentum $L=5.7$. In both panels, the points corresponding to the nearest separation between the two stream segments are marked by red and blue dots.
    }
    \label{fig:modes}
\end{figure}

\begin{figure*}
    \centering
    \includegraphics[width=0.9\textwidth]{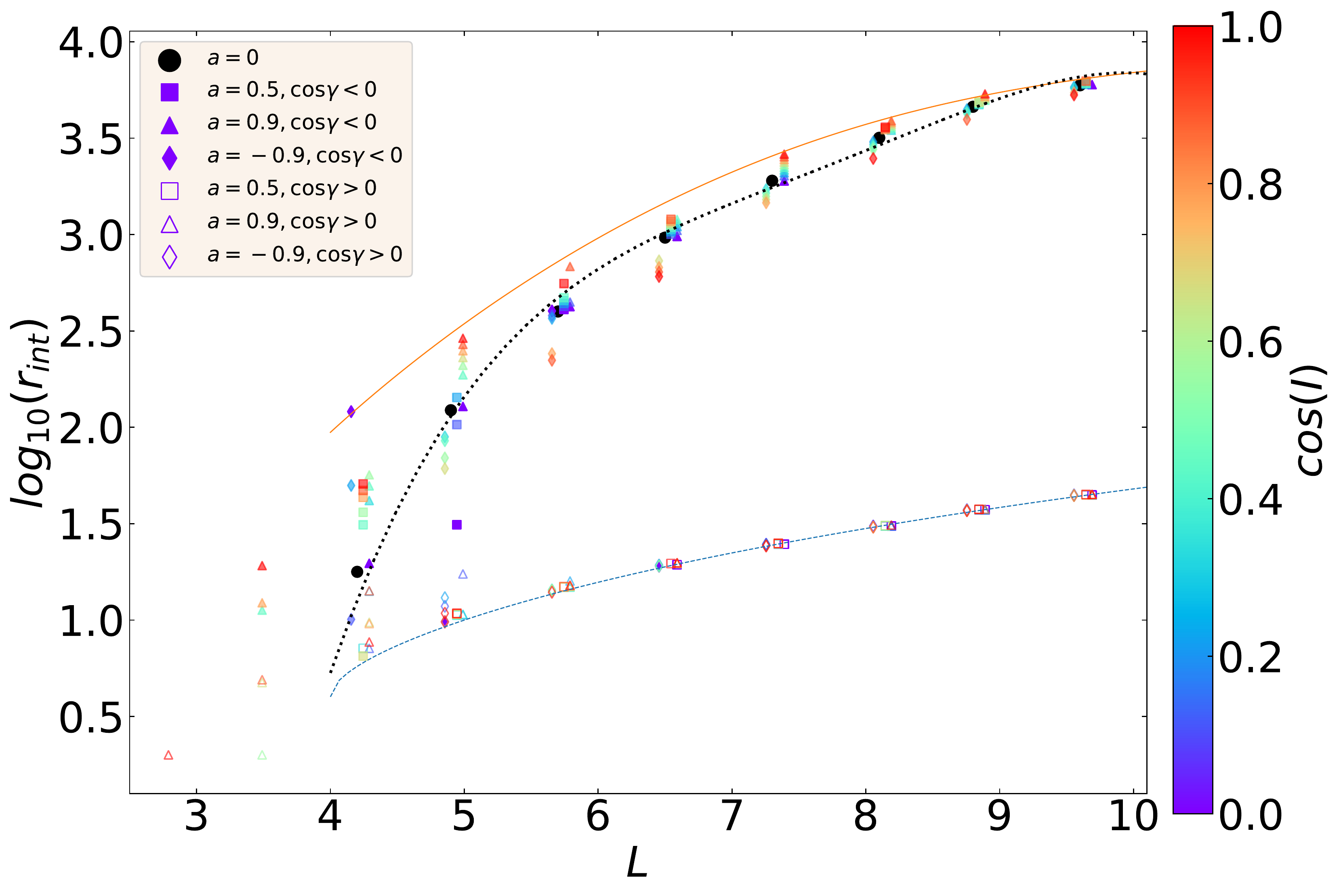}
    \caption{The intersection radius $r_{\rm int}$ as a function of total angular momentum $L$. Different BH spins are represented by different symbol shapes and the colors indicate the inclination angles $I$. For clarity, we have added small horizontal offsets for each case of BH spin: $\Delta L=+0.1$ ($a=0.9$, triangles), $+0.05$ ($a=0.5$, squares), $0$ ($a=0$, black circles), $-0.05$ ($a=-0.9$, diamonds). The thin blue dashed line shows the pericenter radius as a function of angular momentum for the $a=0$ case. The black dotted line is a polynomial fit (Eq. (\ref{eq:fit})) to the $\mathrm{log}\,r_{\rm int}$ values for the $a=0$ case. The intersection radius predicted by  \citet{dai15_pn1_precession} is shown by a thin solid orange line, which is only accurate for large $L$. 
    }
    \label{fig:rint}
\end{figure*}

\subsection{A case study}\label{sec:case_study}
Let us illustrate the procedure outlined above in a concrete example. The parameters input into the routine are $\tilde E=0.9999$, $L=6.5$, $\cos I=0.5$, $a=0.9$, and $\sigma=1$. The geodesic followed by the center of the stream is shown in Fig. \ref{fig:fullorbit}. 
We first project the 3D orbit into the $r$-$\phi$ plane and find all intersection \textit{candidates} between different parts of the orbit. Each intersection candidate is given by the four points around it, two on each half-orbit.

Afterwards, we find the minimum physical distance $\Delta s$ between $AB$ and $CD$, where $A/B$ are points on half-orbit $i$ and $C/D$ are points on half-orbit $j$.


Next, using the stream thickness evolution (as determined by Eq. (\ref{eq:tfe}) and shown in Fig. \ref{fig:H_evolve}), we find the true collision point where $\Delta s<H_1+H_2$. The details of this collision are given in Table \ref{tab:finalresult}. For instance, the two segments of the fallback stream collide almost head-on at an angle of $\gamma\simeq 164$ degrees. Note that the collision occurs between the 6th and 7th half-orbits, and the delay time is slightly more than 2 orbital periods after the first pericenter passage of the fallback stream (or 3 orbital periods after the stellar disruption). As shown later by our full parameter space exploration, such a delay time is quite typical for TDEs. We also note that the stream is very thin and elongated at the time of collision with $H_1\simeq H_2 \simeq 10^{-2}r$, which means that, after the collision, the shock-heated gas will expand nearly adiabatically and hence the energy/angular momentum distribution will be strongly broadened \citep[see][]{lu20_self-intersection}.

\section{Results}\label{sec:parameter_exploration}

The algorithm above can be used to find various quantities associated with collision for any set of parameters. In this section, results from a detailed parameter space exploration are presented.

For each of four BH spins ($a=-0.9$, $0$, $0.5$, $0.9$), we consider a two-dimensional parameter space --- cases with varying total angular momentum $L$ and inclination $\cos (I)$ are explored. The specific energy is fixed to be $\tilde E = 0.9999$ for all cases. We use a $10\times10$ grid for $L$ and $\cos I$: $L\in\{2.71 , 3.41, 4.19, 4.96, 5.74, 6.52 , 7.29, 8.07, 8.84, 9.62\}$, and $\cos I\in\{0.05 , 0.15, 0.25, 0.35, 0.45, 0.55, 0.65, 0.75, 0.85, 0.95\}$. The minimum angular momentum is given by that of a marginally bound equatorial orbit for $a=0.9$ (the most deeply penetrating case without plunging): $L_{\rm min} \approx 2.71$ in geometrized units, which is obtained by numerically solving $V_r=d_rV_r=0$ (where $V_r$ is given by Eq. (\ref{eq:Vr}))\footnote{The minimum angular momentum as a function of $a$ and $\cos I$ is described in Appendix \ref{sec:Lmin}.}. Note that, for some of the lowest angular momentum points, the geodesics are plunging and the star is completely swallowed by the BH. The maximum angular momentum is set by the requirement of tidal disruption, $L_{\rm max} = \sqrt{2 r_{\rm T}}\simeq 9.7$, where $r_{\rm T}=R_*(M/M_*)^{1/3}\simeq 46.7$ in geometrized units is the (Newtonian) tidal disruption radius for a $10^6M_\odot$ BH and $1M_\odot$ star. 

Interesting systematic properties of TDE stream intersections are summarized in the following subsections.






\begin{figure*}
\centering
\begin{tabular}{ccc}
\subfloat[$a=0$]{\includegraphics[width = 0.32\textwidth]{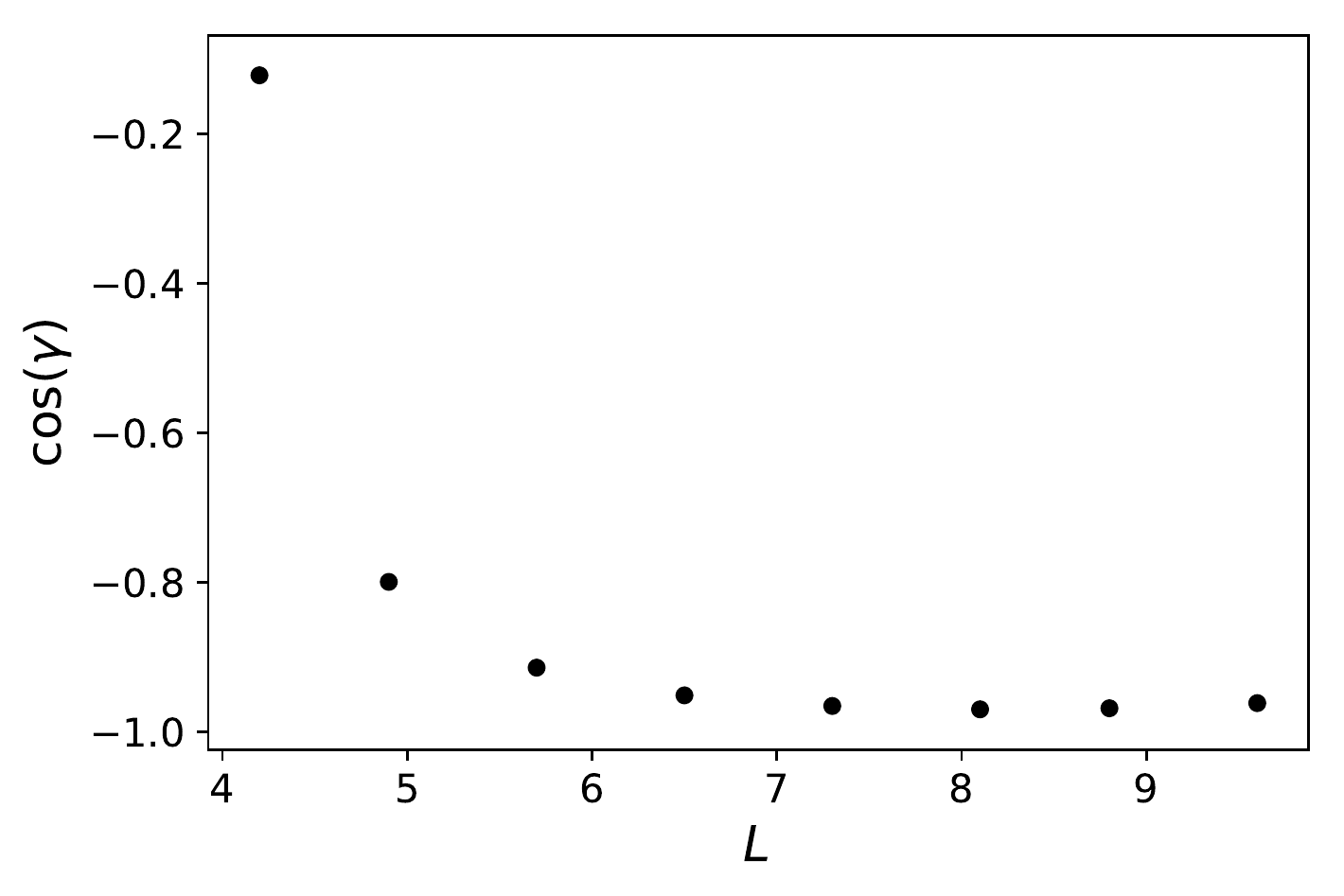}} &
\subfloat[$a=0$]{\includegraphics[width = 0.32\textwidth]{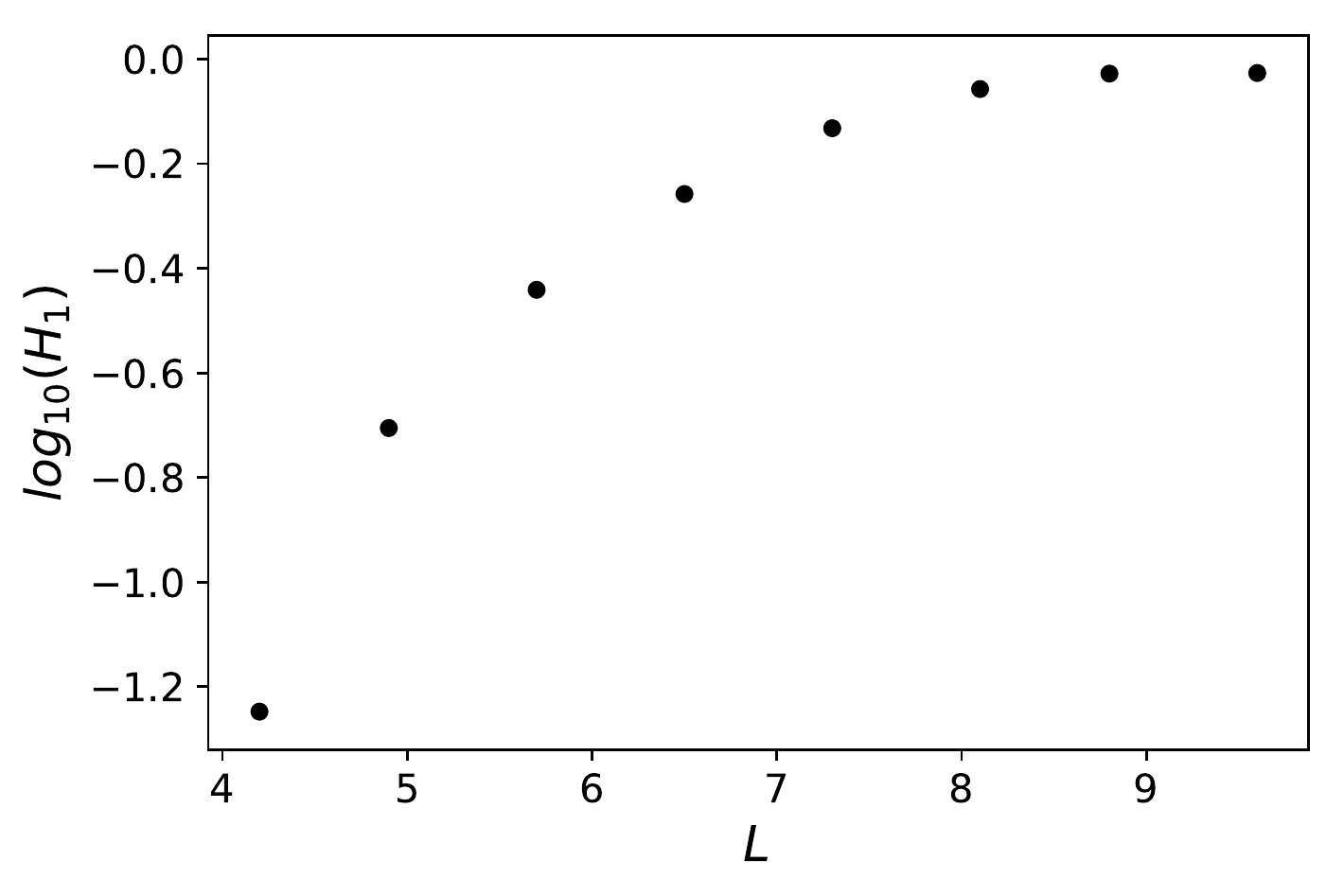}} &
\subfloat[$a=0$]{\includegraphics[width = 0.32\textwidth]{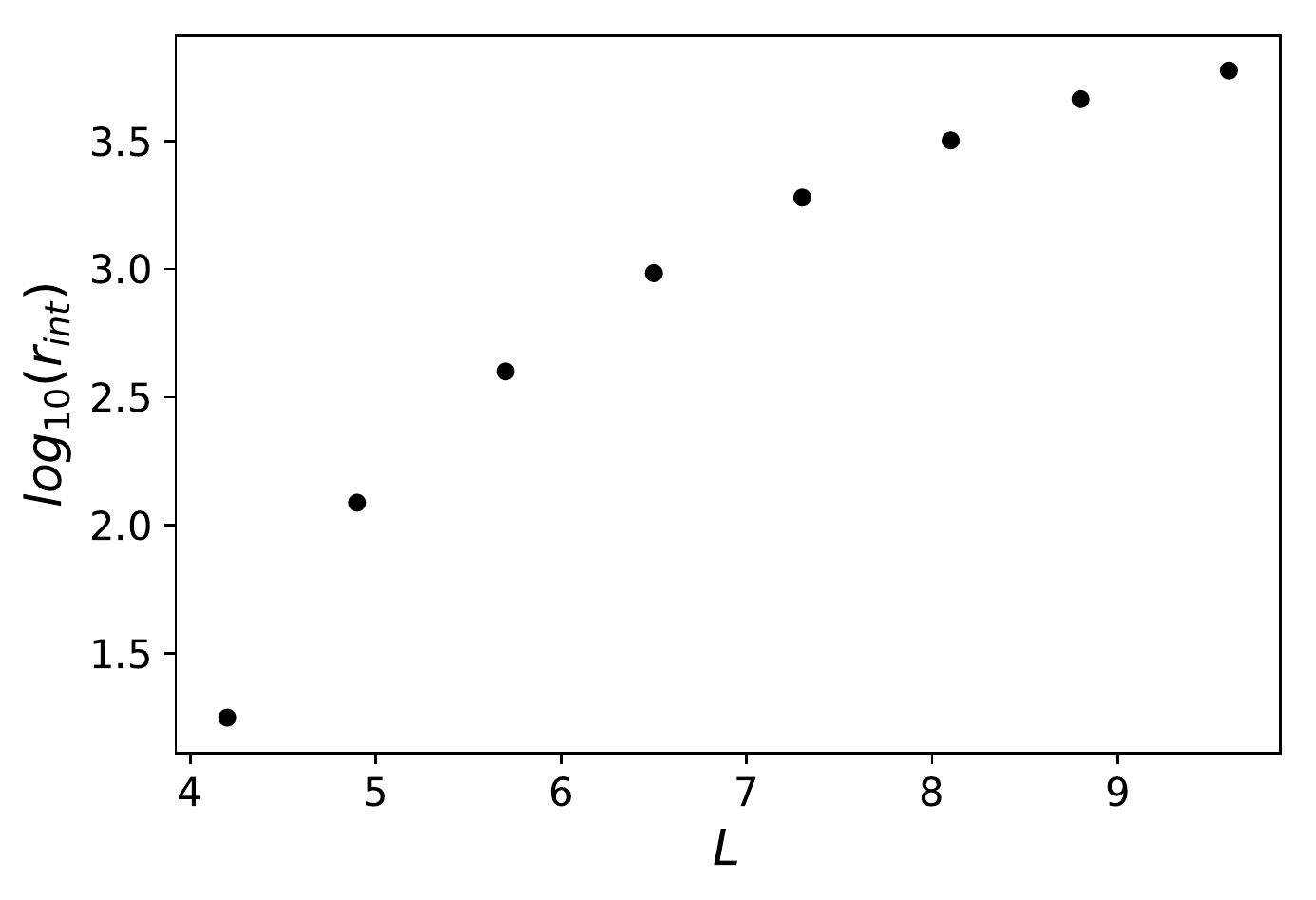}}\\
\subfloat[$a=-0.9$]{\includegraphics[width = 0.32\textwidth]{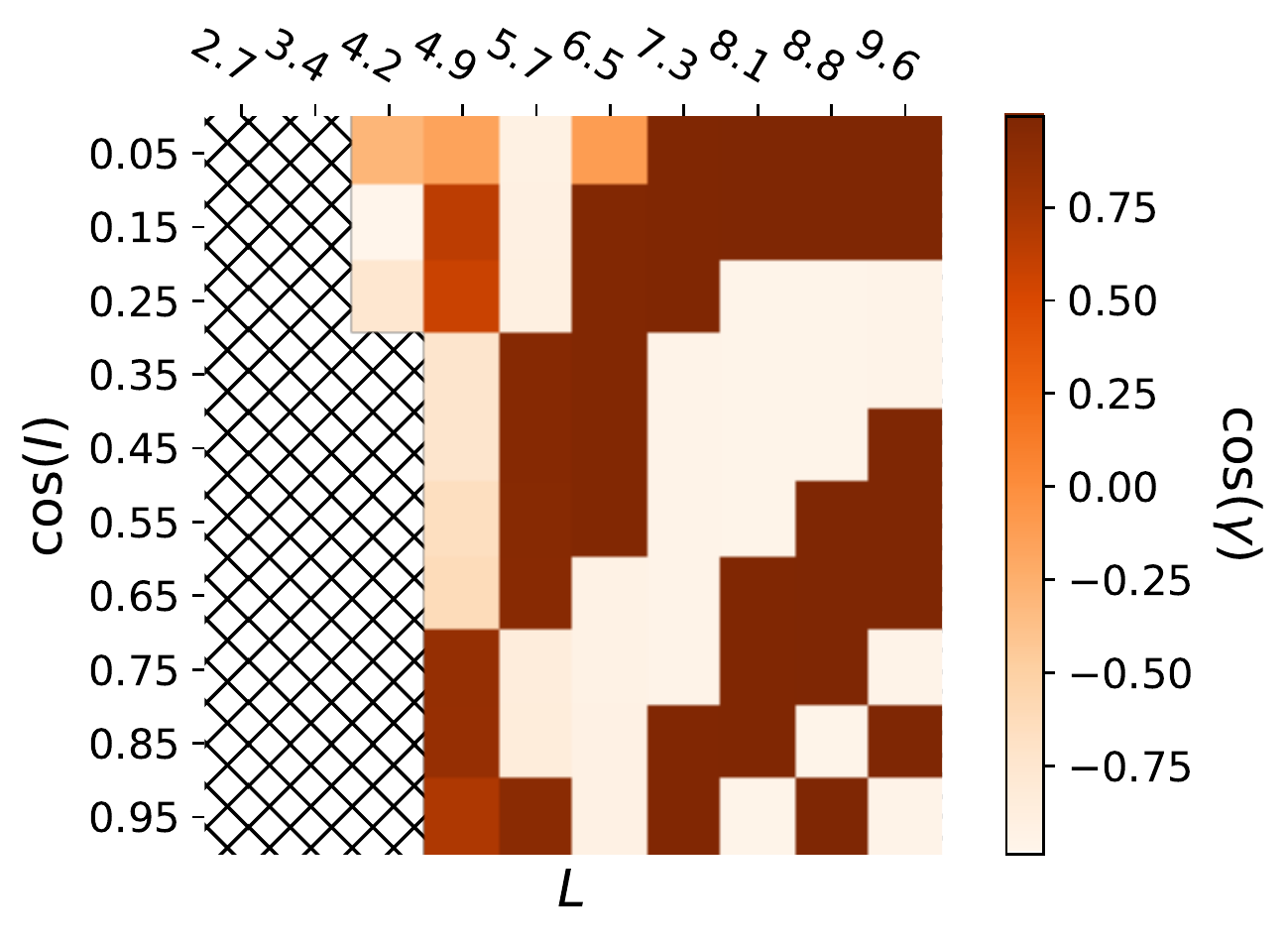}} &
\subfloat[$a=-0.9$]{\includegraphics[width = 0.32\textwidth]{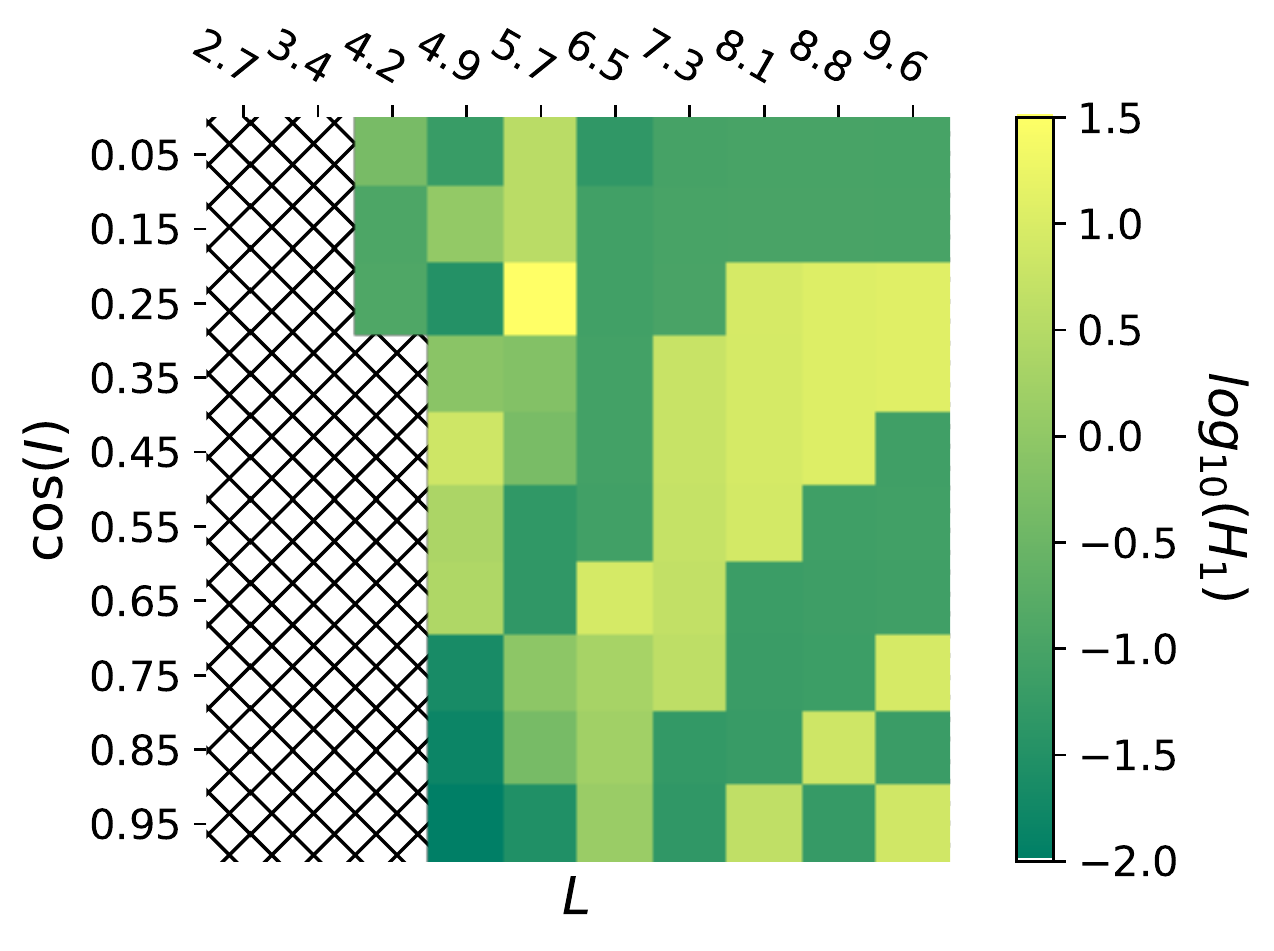}} &
\subfloat[$a=-0.9$]{\includegraphics[width = 0.32\textwidth]{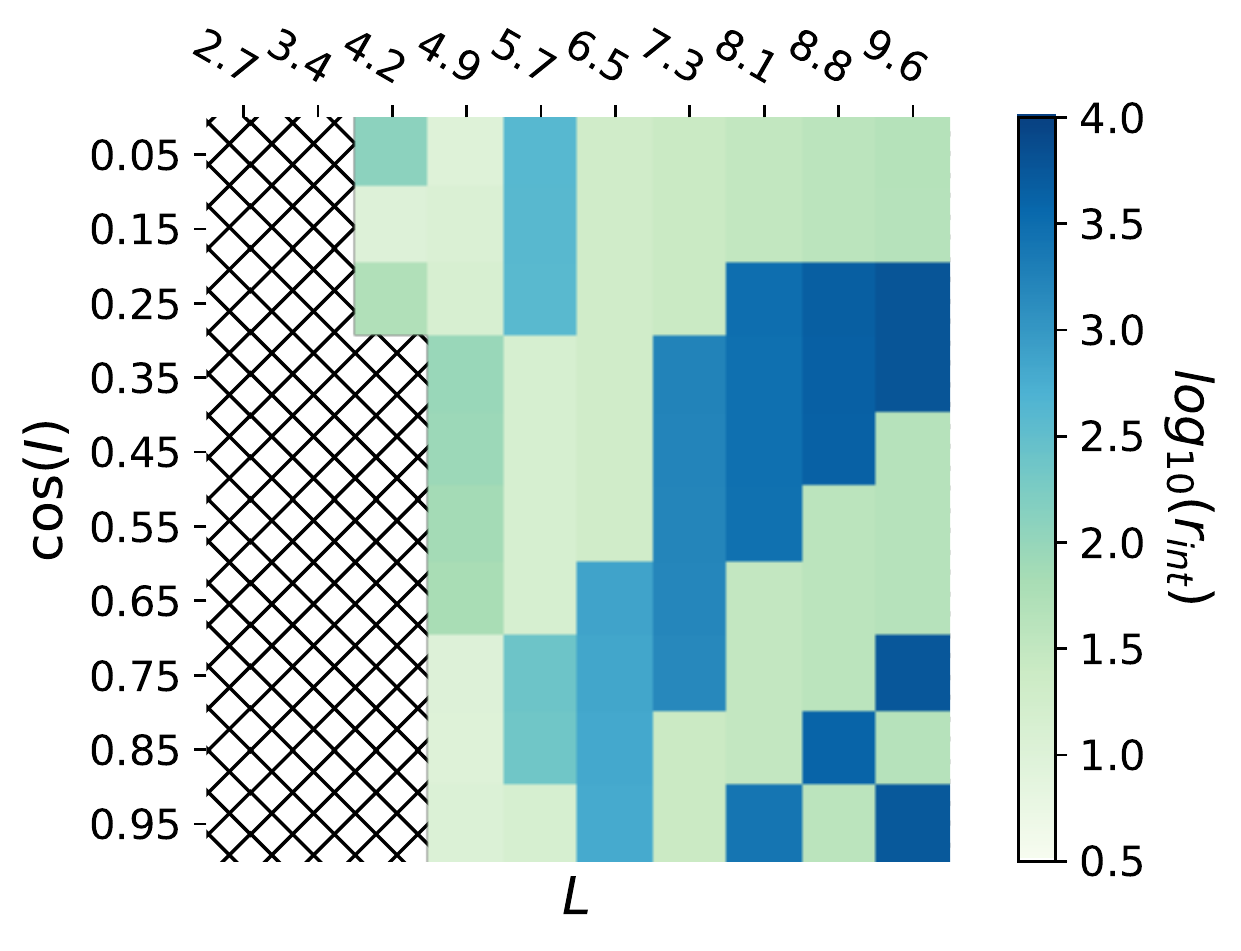}} \\
\subfloat[$a=0.5$]{\includegraphics[width = 0.32\textwidth]{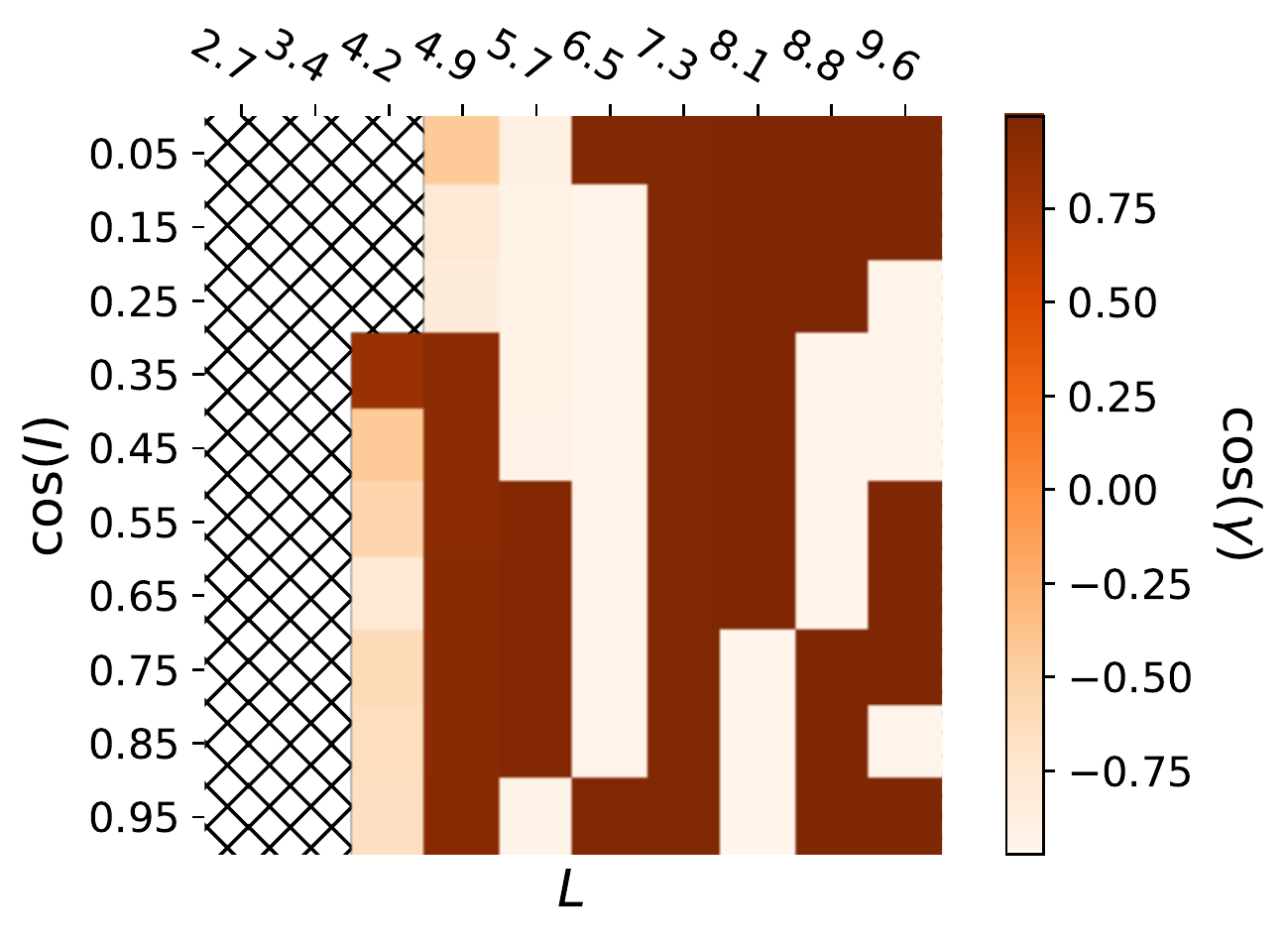}} &
\subfloat[$a=0.5$]{\includegraphics[width = 0.32\textwidth]{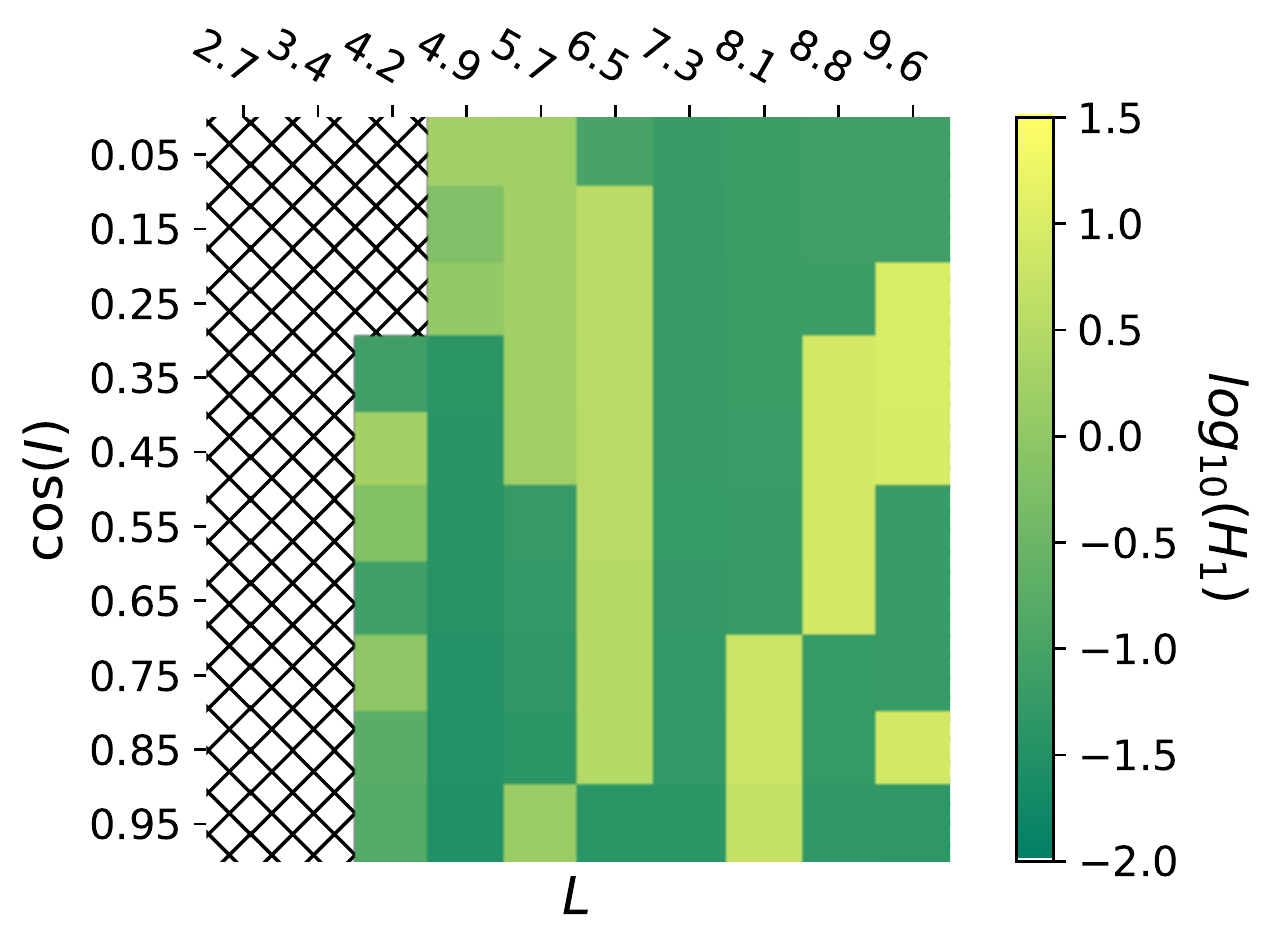}}&
\subfloat[$a=0.5$]{\includegraphics[width = 0.32\textwidth]{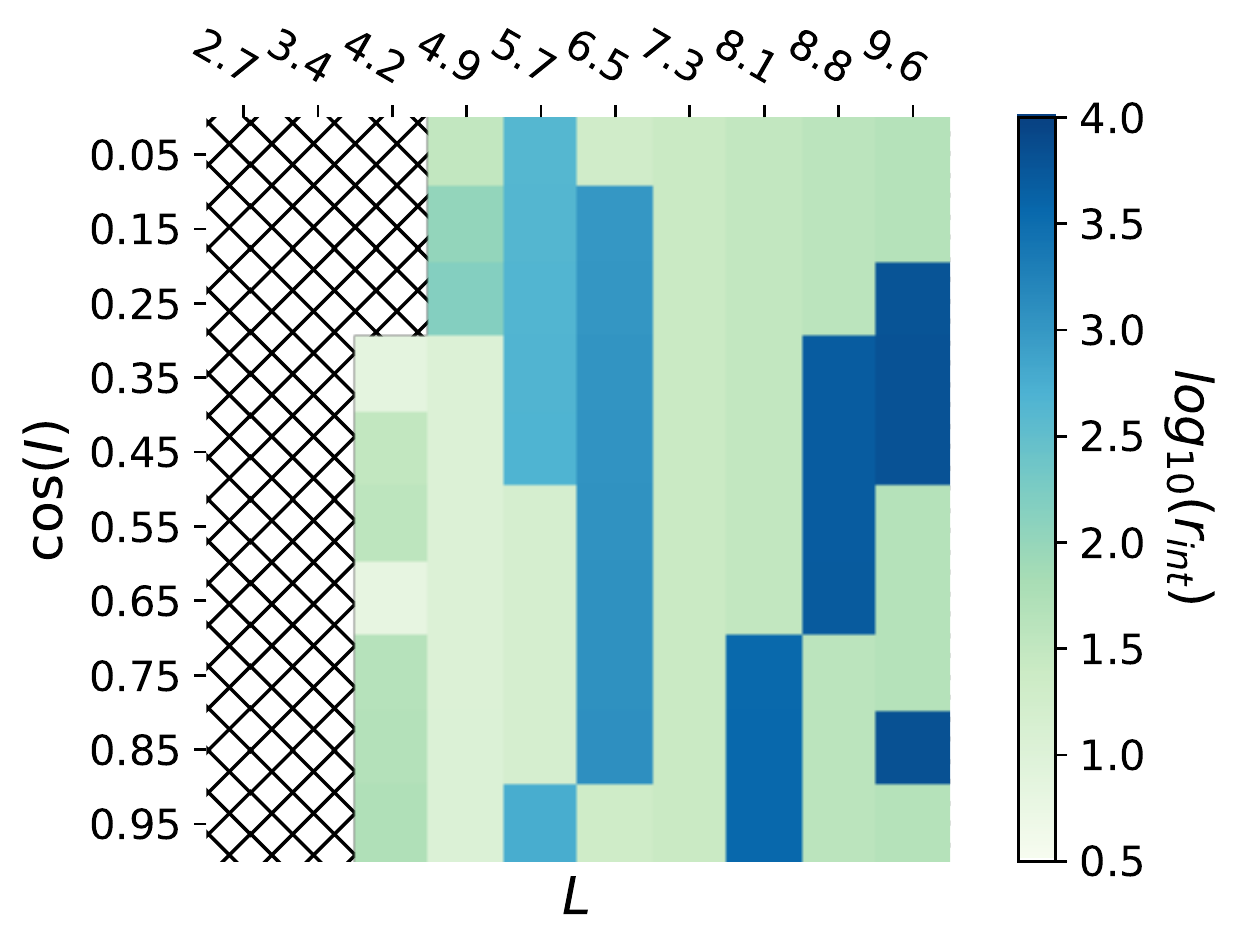}} \\
\subfloat[$a=0.9$]{\includegraphics[width = 0.32\textwidth]{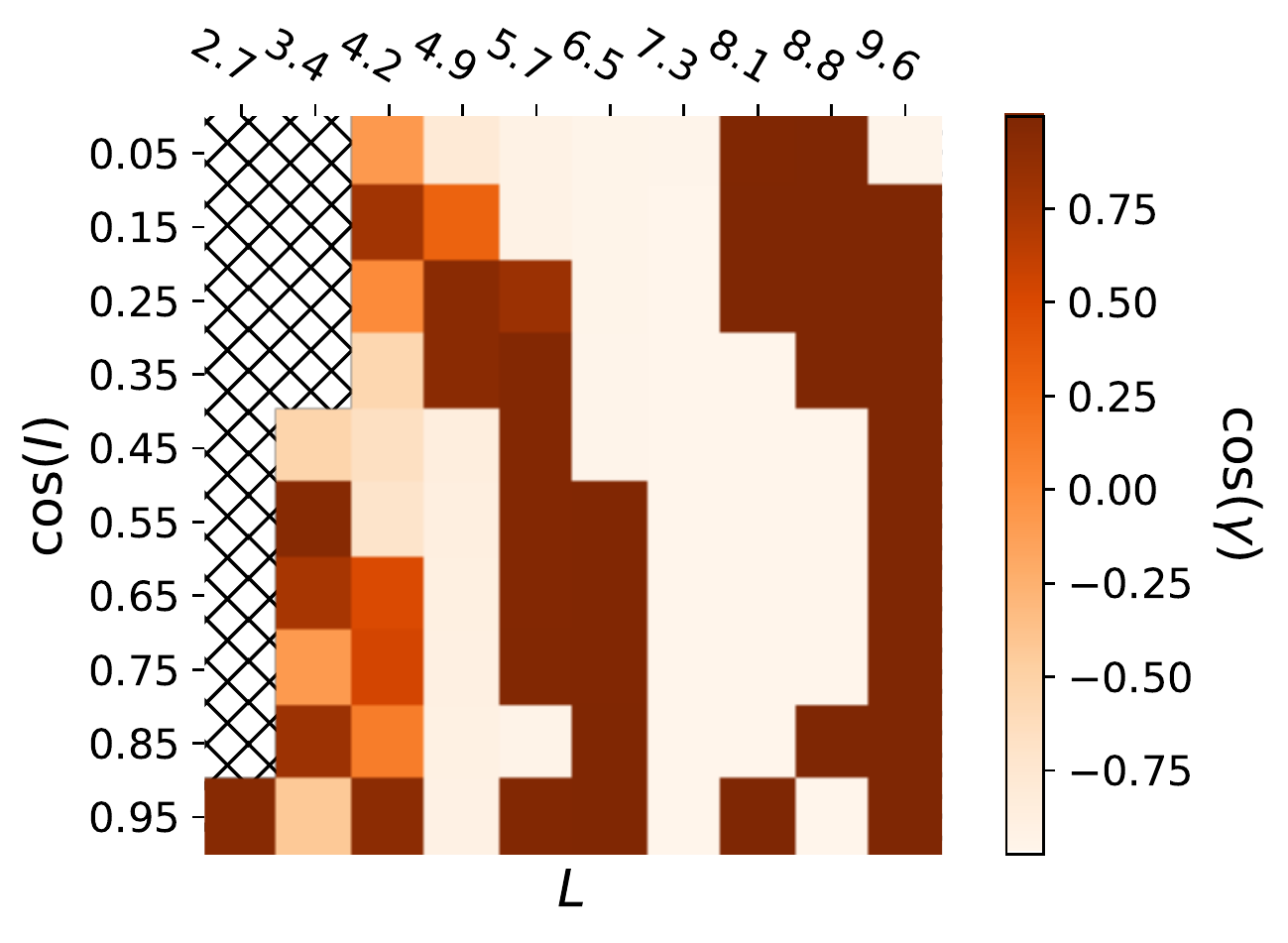}} &
\subfloat[$a=0.9$]{\includegraphics[width = 0.32\textwidth]{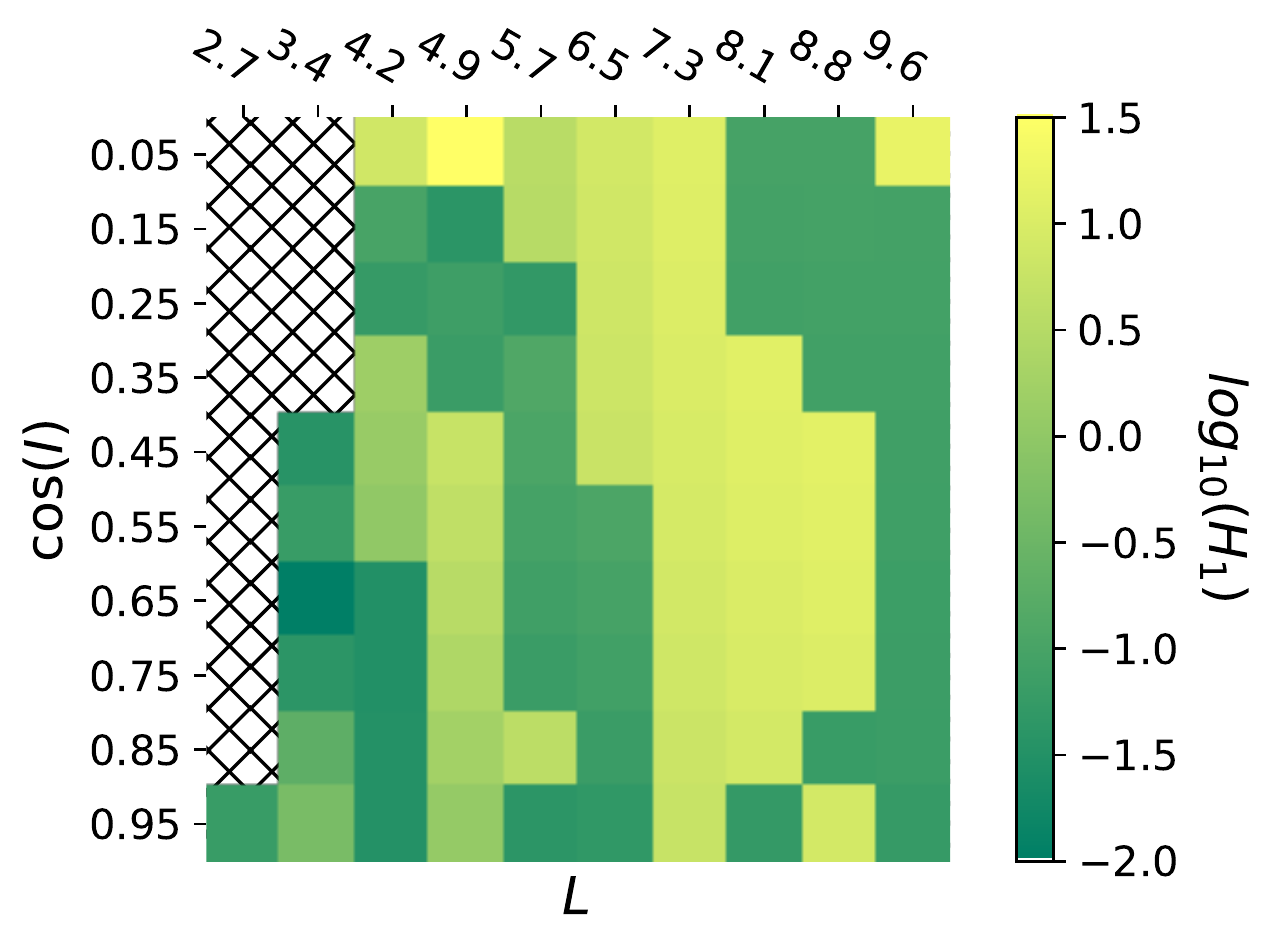}} &
\subfloat[$a=0.9$]{\includegraphics[width = 0.32\textwidth]{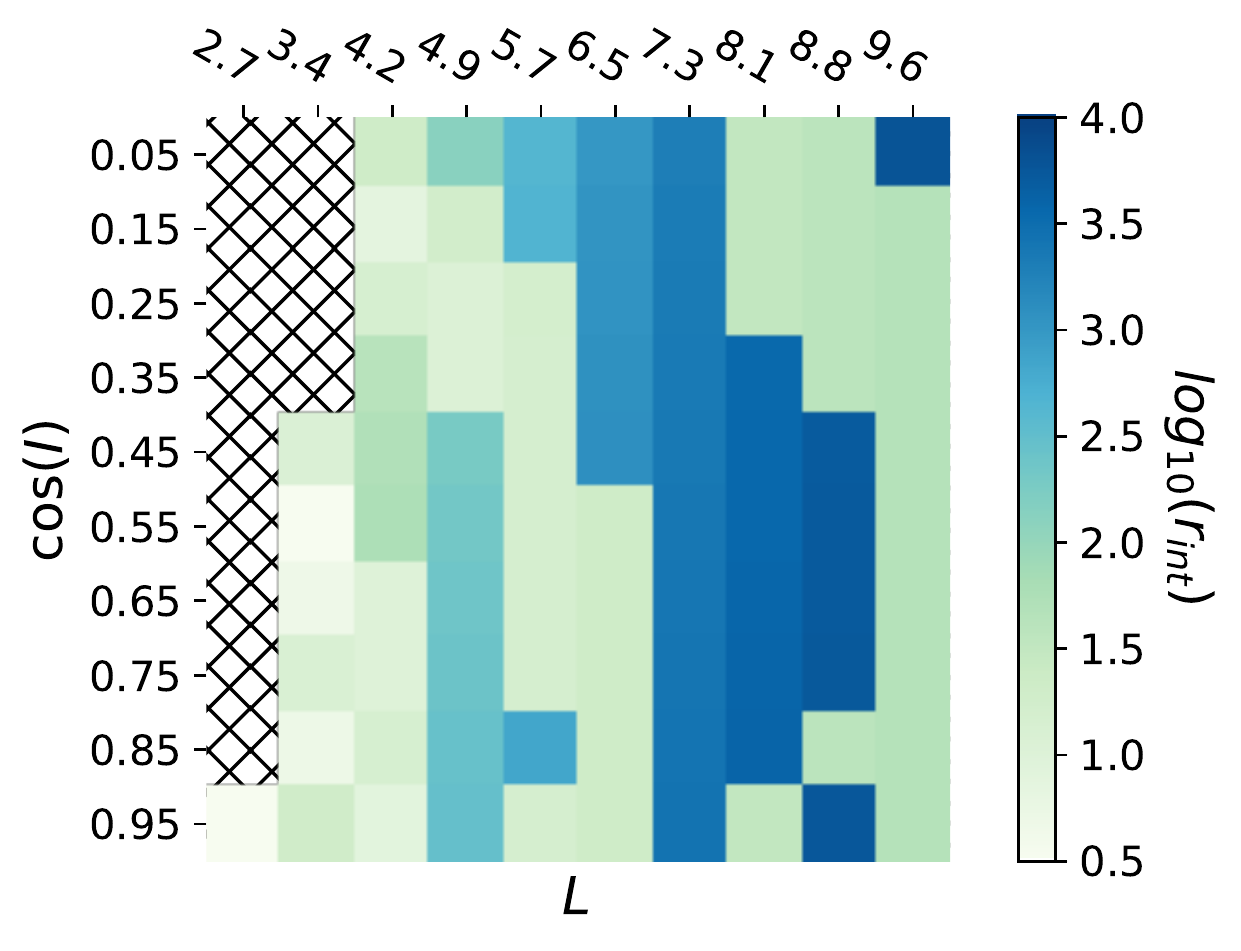}}\\
\end{tabular}
\caption{The intersection angle $\cos(\gamma)$ (first column), thickness of one of the colliding streams $H_1$ (middle column), and intersection radius $\log_{10}(r_{\rm int})$ (right column), as a function of angular momentum $L$ and inclination angle $\cos(I)$. The BH spins in each panel are indicated at the bottom of each plot. }
\label{fig:corplots}
\end{figure*}

\subsection{Two modes of intersection}

We find two modes of intersection: about half of the TDEs have rear-end collisions ($\cos\gamma>0$) near the pericenter and the other half have head-on collisions ($\cos\gamma<0$) typically further from the pericenter. This is illustrated in Fig. \ref{fig:modes}, where we show two representative cases of these two modes.

There is a bimodal distribution of intersection angles (see Fig. \ref{fig:corplots}), which are either close to 180$^{\rm o}$ (``head-on'' mode) or 0$^{\rm o}$ (``rear-end'' mode), with very few cases near 90$^{\rm o}$. This is expected as we consider a post-Newtonian picture of two elliptical orbits with different orbital planes (due to LT precession) but the same focal point --- the orbits must have two closest approaches and in the limit of high eccentricity, one of the closest approaches is near the pericenter and the other one is far away.
When the collision occurs near the pericenter, the two colliding ends have the same sense of rotation and this leads to $\cos\gamma\simeq 1$. When the intersection point is far away from the pericenter and the apocenter\footnote{The maximum angular momentum (as required by the tidal disruption of a $1M_\odot$ star by a $10^6M_\odot$ BH) gives a minimum apsidal precession angle of about $12^{\rm o}$, and this means that the intersection point cannot be very close to the apocenter \citep{dai15_pn1_precession}.}, the nearly radial orbits directly lead to $\cos\gamma\simeq -1$. Later on in \S \ref{sec:vrel}, we provide analytical expressions for the radial coordinate of the collision point and the collision angle in the post-Newtonian limit.


We also note that the work by \citet{guillochon15_dark_year} did not find any ``rear-end'' collisions near the pericenter, and that is because they consider each of the orbital windings to be a series of unconnected ellipses without explicitly resolving the (physically connected) near-pericenter segments between neighboring ellipses --- their setup prevents them from finding the ``rear-end'' collisions that we have identified. As we demonstrate later, the self-crossing shock in the rear-end mode is sufficiently strong to thicken the stream significantly and redistribute the angular momentum of the stream, so we expect rapid circularization of the bound gas due to subsequent shocks.

\subsection{Intersection radius}

In about half of the cases we studied, the intersection occurs near the pericenter (the rear-end mode), so the intersection radius is trivial to obtain. For the other half of TDEs in the head-on mode, we find the intersection radius depends mainly on the total angular momentum but less strongly on other parameters. This is illustrated in Fig. \ref{fig:rint} (it can also be clearly seen on the rightmost panels of Fig. \ref{fig:corplots}), where we show that the intersection radius $r_{\rm int}$ generally increases with the total angular momentum $L$, for all BH spins $a$ (different symbol styles) and inclination angles $I$ (different colors). This is caused by weaker apsidal precession at larger angular momentum (or larger pericenter radius). For the mildly relativistic cases where the angular momentum is much larger than the minimum $L_{\rm min}$, the BH spin and orbital inclination play minor roles in determining the intersection radius. This is because intersection typically occurs between adjacent orbital windings (as detailed in \S \ref{sec:delay_time}) and the out-of-plane precession angle\footnote{Note that $\Phi_{\rm LT}$ is defined as the angle between the angular momentum vectors between adjacent windings, which is different from the nodal shift $\Delta \Omega_{\rm nod} \approx 4\pi a/L^3$.} $\Phi_{\rm LT}$ (due to the LT effect) is much smaller than the angle of in-plane (apsidal) precession $\Phi_{\rm ap}$. In the post-Newtonian approximation, these two angles are given by \citep[][including the lowest order term due to spin]{merritt13_precession_angles}
\begin{eqnarray}\label{eq:LT_precession}
    \Phi_{\rm LT} \approx {4\pi a\sin I\over L^3},
\end{eqnarray}
\begin{eqnarray}\label{eq:apsidal_precession}
    \Phi_{\rm ap} \approx {6\pi \over L^2} - {8\pi a\cos I \over L^3}.
\end{eqnarray}
We note that these two are the angles between the orbital angular momentum vectors ($\Phi_{\rm LT}$) and between the orbital periapses $(\Phi_{\rm ap})$ in a fixed reference frame, and it can be seen that the in-plane precession is larger than the out-of-plane precession by a factor of $1.5L$ or more.

A polynomial fit to the $\log(r_{\rm int})$ values for the $a=0$ cases is given by the following:
\begin{multline}
    \log(r_{\rm int})=-0.0081L^4+0.253L^3-2.96L^2+15.6L-28.5.
    \label{eq:fit}
\end{multline}

However, for the highly relativistic cases with angular momenta only slightly above $L_{\rm min}$, the intersection radius increases with spin such that $a=-0.9$ (or $0.9$) gives the smallest (largest) $r_{\rm int}$. This is mainly because the 1.5-order post-Newtonian contribution to apsidal precession, $-8\pi L^{-3} a \cos I$ (see Eq. (\ref{eq:apsidal_precession})), is negative for prograde orbits, and the reduction in apsidal precession angle increases the intersection radius \citep[see also][]{stone19_TDE_in_GR}. The dependence of $r_{\rm int}$ on the inclination angle can largely be explained by this effect as well.




\begin{figure}
\centering
\begin{tabular}{c}
\subfloat[$a=-0.9$]{\includegraphics[width = 0.4\textwidth]{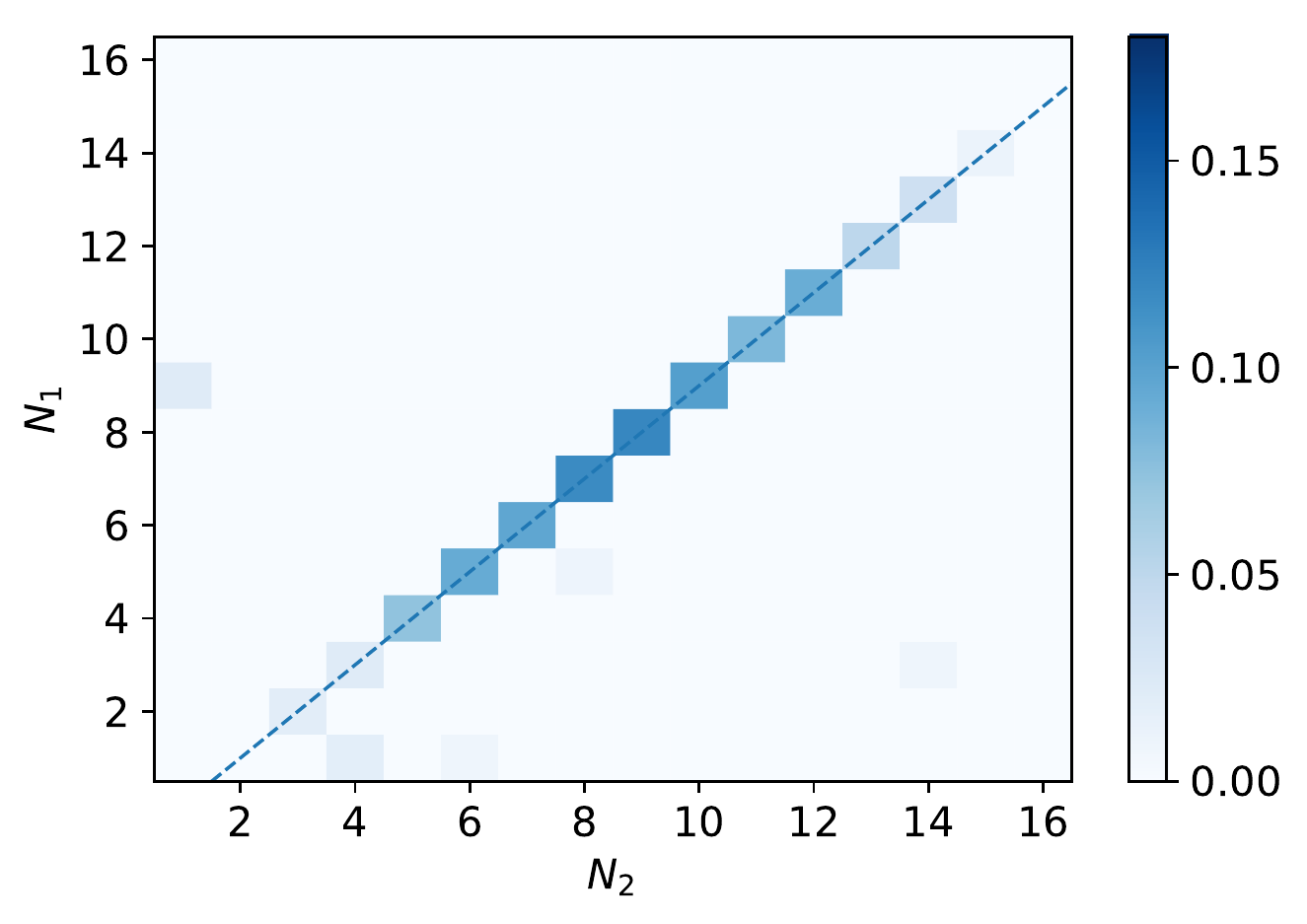}}\\
\subfloat[$a=0.5$]{\includegraphics[width = 0.4\textwidth]{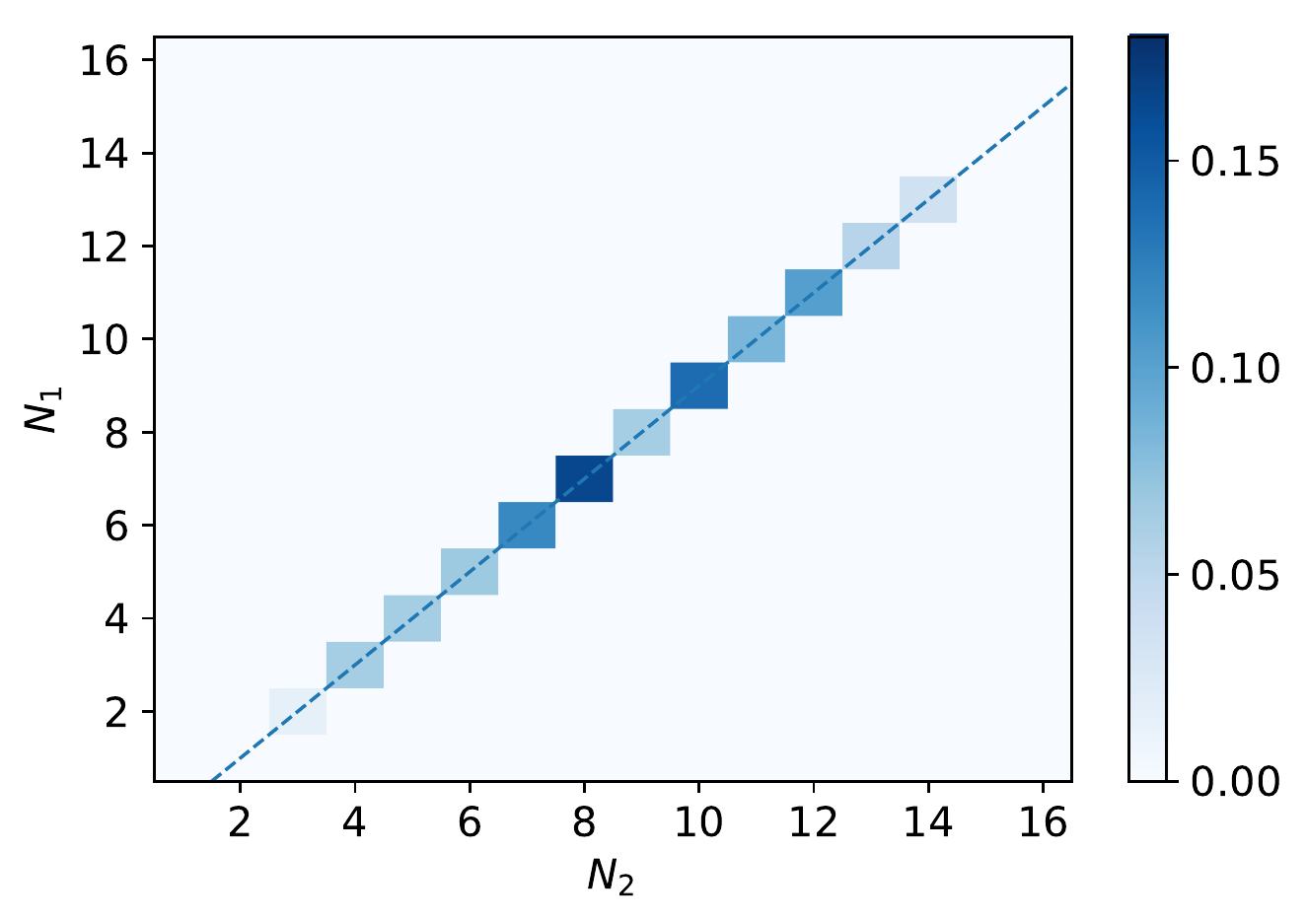}}\\
\subfloat[$a=0.9$]{\includegraphics[width = 0.4\textwidth]{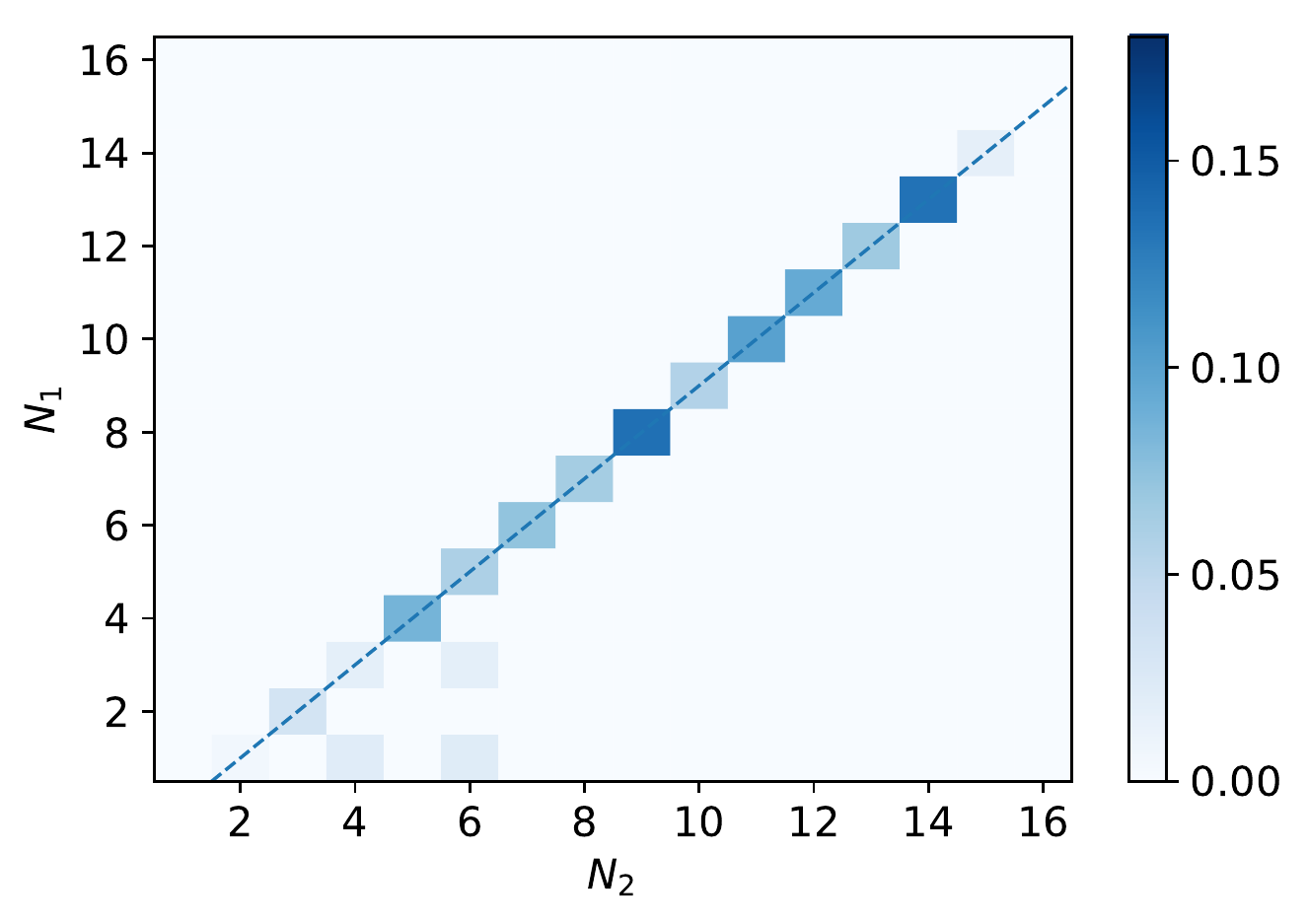}}\\
\end{tabular}
\caption{The distribution of orbital winding numbers $N_2$ vs. $N_1$ for three different BH spins, under the full ``loss cone'' assumption. We do not find any prompt collisions (for which $N_1=0$ and $N_2 = 1$) on our grid points for all BH spins.
For this case of $\tilde E=0.9999$, each half-orbit revolution takes 2 months.
}
\label{fig:N12}
\end{figure}

\subsection{Delay time}
\label{sec:delay_time}
The delay between tidal disruption and stream intersection causes a ``dark phase'' before bright emission is generated by shocks and accretion \citep{dai13_gr_precession, guillochon15_dark_year}. In this subsection, we aim to estimate, in a statistical way, the typical delay time for TDEs of a $1M_\odot$ star by a $10^6M_\odot$ BH.

We denote the half-orbit numbers of the two colliding ends as $N_1=j$ and $N_2=i$, with $N_2>N_1$ by definition. For each BH spin $a$, we simulate about 100 cases (ignoring the plunging ones) of different \{$L$, $\cos I$\}, for each of which we obtain \{$N_1$, $N_2$\}. We assigned a statistical weight $w=L$ to each case, which is based on the assumption of the ``full loss cone'' angular momentum distribution and an isotropic distribution of incoming stars \citep{stone20_tde_rate_review}. The resulting normalized distributions of \{$N_1$, $N_2$\} for three different BH spins are shown in Fig. \ref{fig:N12}. 

In our simulations, the collision typically occurs between consecutive half-orbits, i.e. $N_2=N_1+1$, which is in agreement with \citet{guillochon15_dark_year}. This can be easily understood as follows. Since the LT precession angle is usually small ($\ll 1\rm\, rad$), the orbital planes of consecutive windings are spatially the closest, which means a collision eventually occurs when the angular width of the stream gradually grows above the angular separation between two adjacent orbital planes (cf. Fig. \ref{fig:H_evolve}).

\begin{figure}
    \centering
    \includegraphics[width=3.3in]{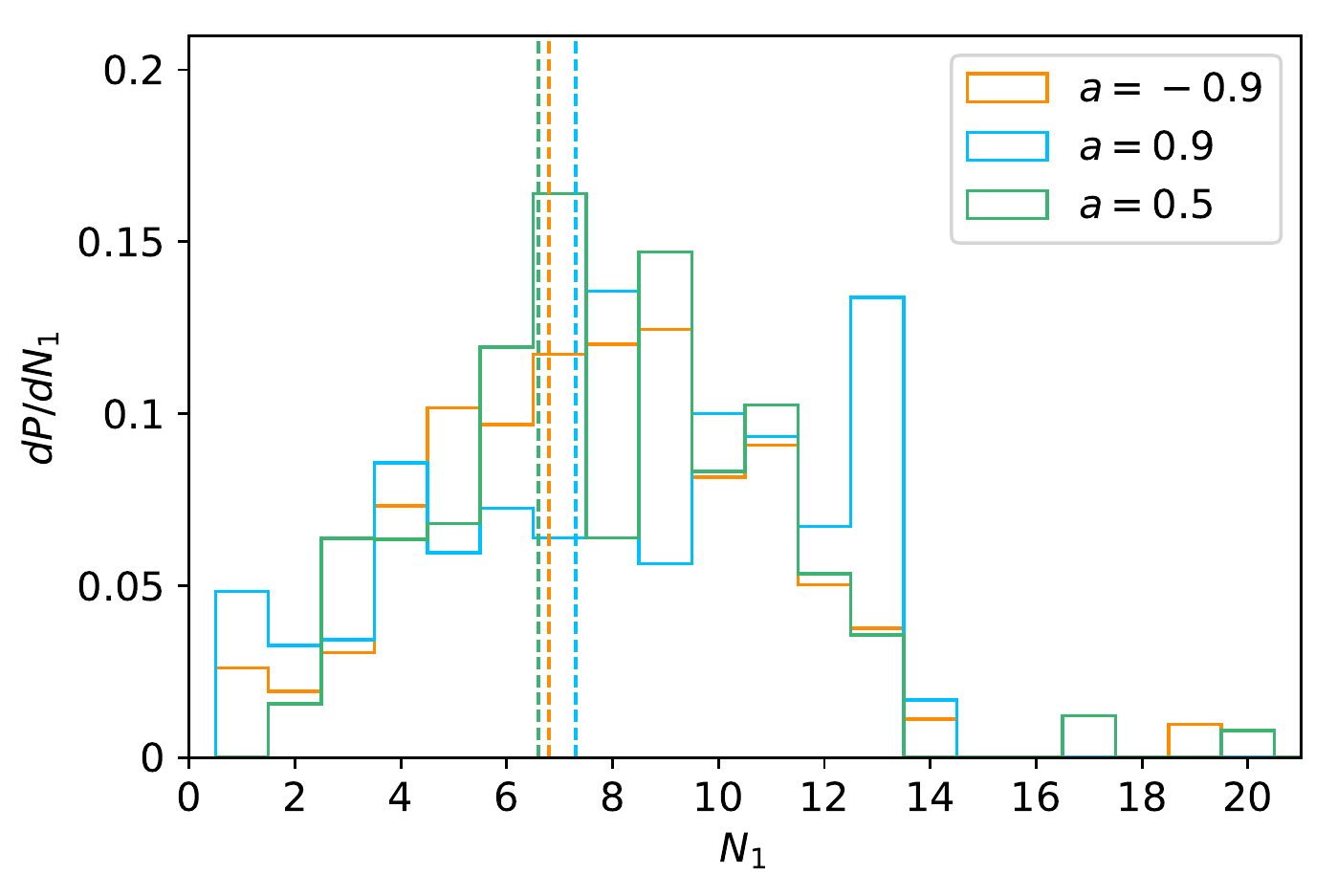}
    \caption{The probability distribution of the smaller winding number of the earlier colliding end, $dP/dN_1$, for three BH spins, under the full ``loss cone'' assumption. The vertical dashed lines show the median for each distribution. The largest angular momentum considered in our grid is $L=9.62$, for which the apsidal precession angle is about 12 degrees. This means that the orbit has precessed by $90^{\rm o}$ to be roughly perpendicular to the initial semimajor axis of the incoming star when $N_1\sim 13$. We show in \S \ref{sec:physical_reason} that, under the post-Newtonian approximation, stream collisions occur before this.}
    \label{fig:dpdn1}
\end{figure}

We note that one can tell which mode of intersection the collision follows by looking at $N_1$ and $N_2$: for odd $N_1$ and even $N_2$, the collision occurs in the head-on mode; and for even $N_1$ and odd $N_2$, the collision occurs in the rear-end mode. Fig. \ref{fig:dpdn1} shows $dP/dN_1$ for three different BH spins, where $P$ is the probability of occurrence of a TDE with $j=N_1$. For our choice of initial stream width $H_0=5R_\odot$, we do not find any case of prompt collision for which $N_1 = 0$ on our grid points. This means that less than a few percent of TDEs (those with $\cos I > 0.95$) may have prompt collisions for any of the three different BH spins ($-0.9, 0.5, 0.9$).

We also find that the delay between tidal disruption and stream intersection is typically between 2 and 7 orbital periods, with longest median delay in the highest spin $a=0.9$ case. We took the orbital energy to be $0.9999$, which corresponds to a semimajor axis of $5000$ gravitational radii and Keplerian period of about 4 months (a half-orbit takes 2 months). This means that the initially self-gravitating part of the fallback stream with initial thickness $H_0=5R_\odot$ only joins the accretion flow about one to three years after the tidal disruption. A caveat is that the delay time strongly depends on our choice of initial stream width $H_0$ as well as the initial polar angle $\theta$, but the delay times shown in Fig. \ref{fig:dpdn1} are representative for most TDEs. This will be discussed in \S \ref{sec:physical_reason}.

\begin{figure}
    \centering
    \includegraphics[width = 0.4\textwidth]{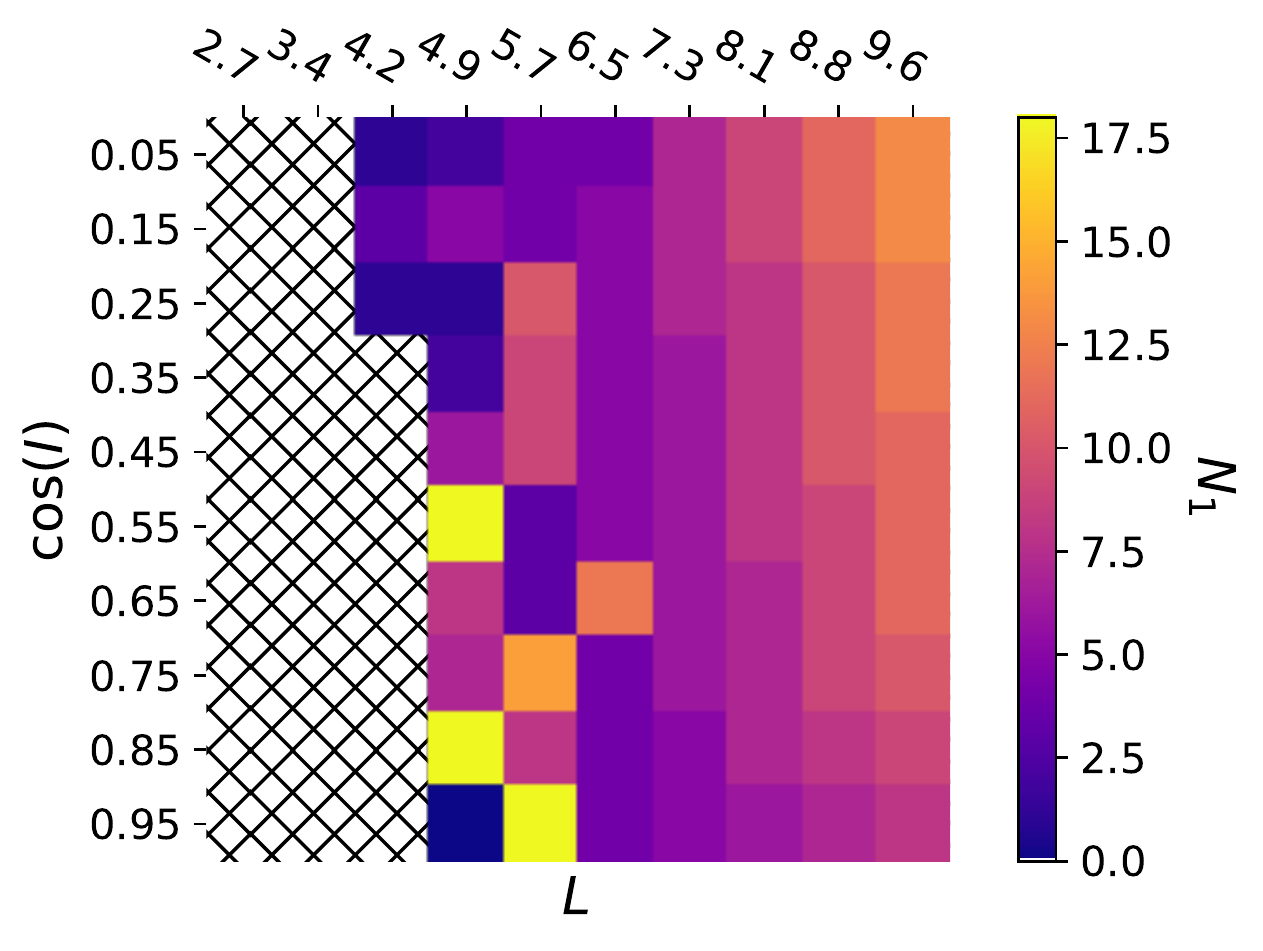}
    \includegraphics[width = 0.4\textwidth]{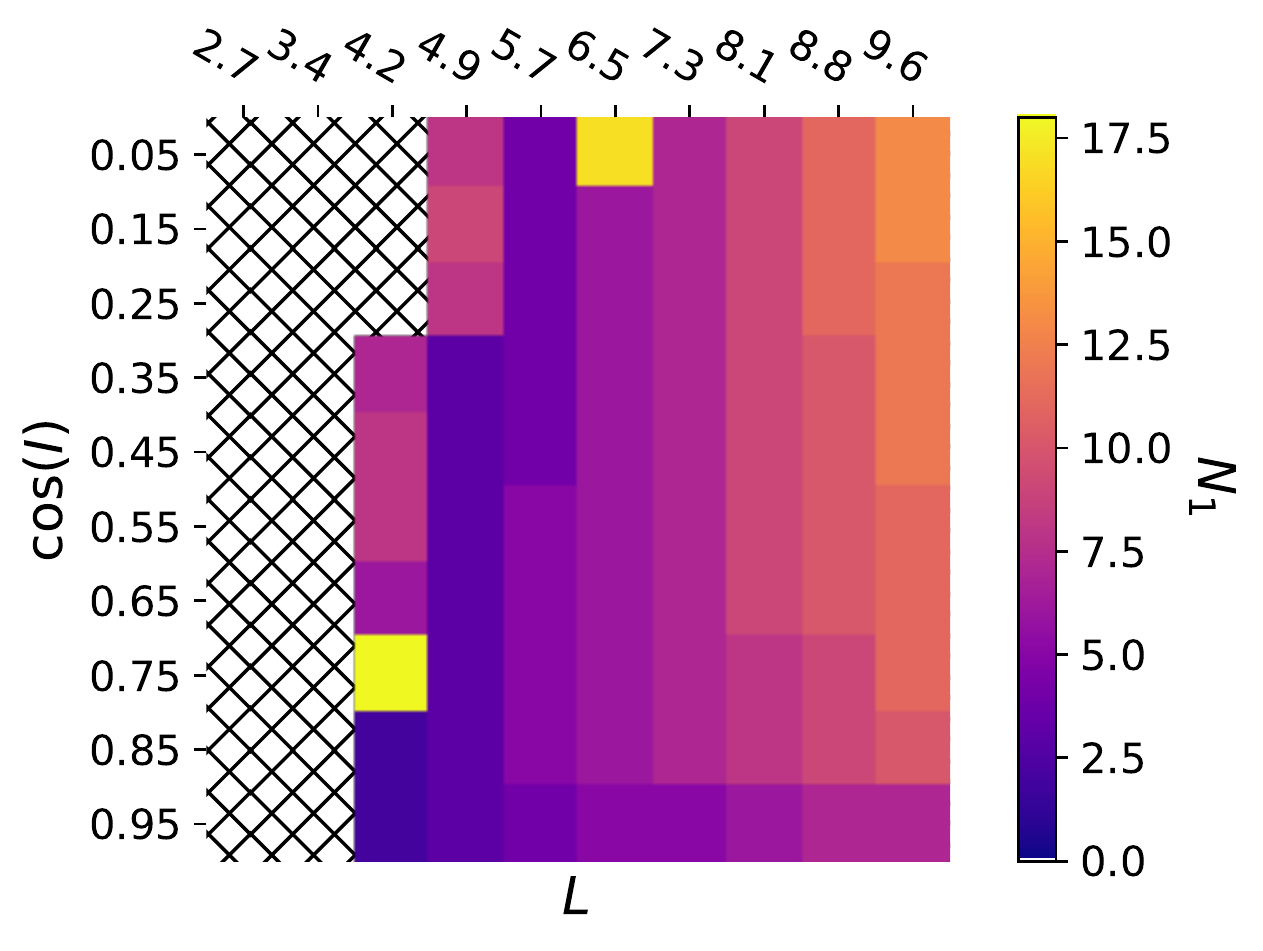}
    \includegraphics[width = 0.4\textwidth]{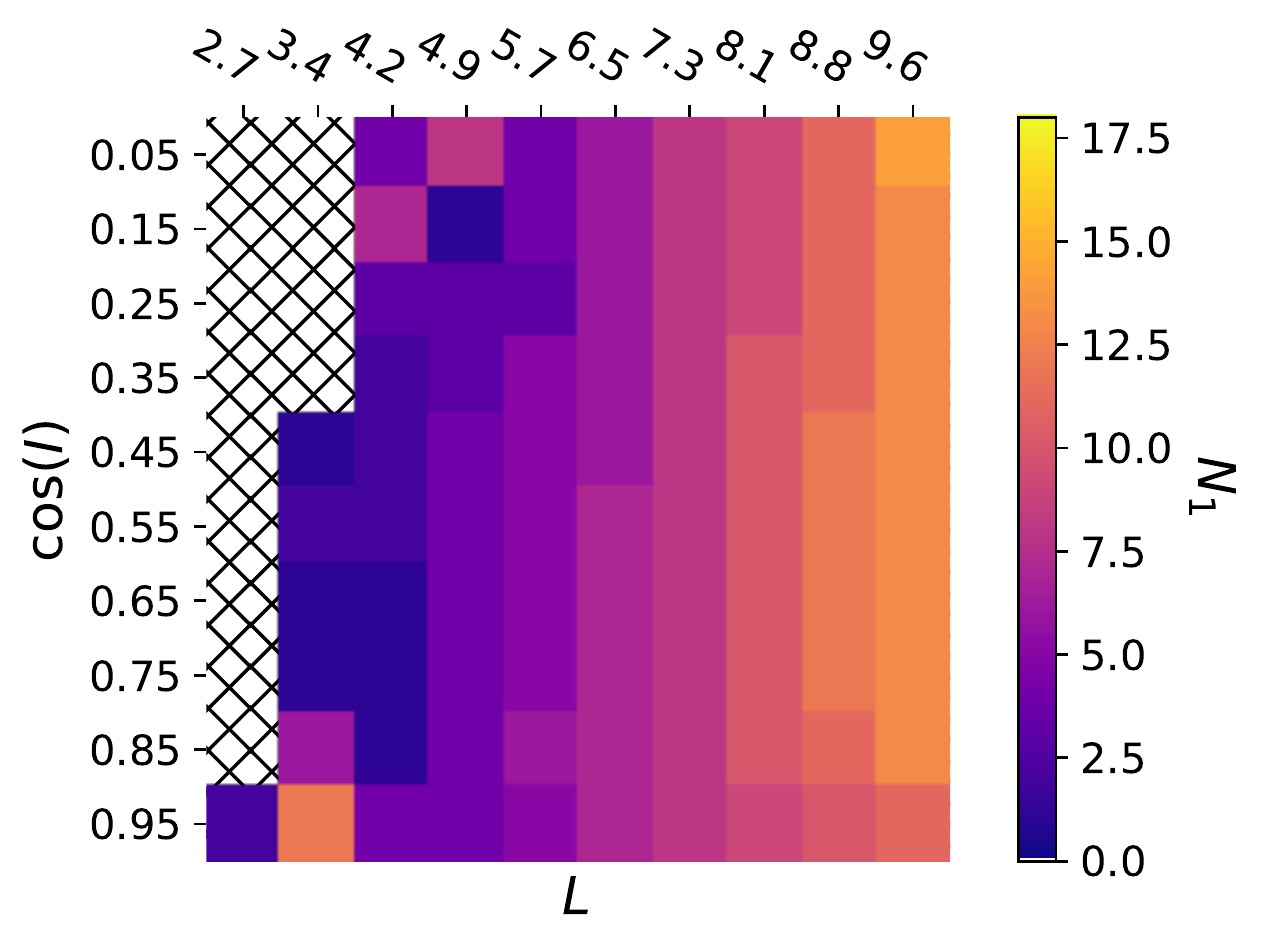}
    \caption{The distribution of the orbital winding number $N_1$ for different BH spins: $a=-0.9$ (top panel), $a=0.5$ (middle panel) and $a=0.9$ (bottom panel).}
    \label{fig:n1}
\end{figure}

The full distribution of $N_1$ as a function of the angular momentum $L$ and inclination $\cos I$ is shown in Fig. \ref{fig:n1}, where we find $N_1$ rapidly increases with $L$.
This can be qualitatively understood as follows.
For a larger $L$, the apsidal precession angle is much smaller and hence the stream thickness increases slower with winding number --- the stream thickness is maximized when the apsidal precession brings the major axis to be nearly perpendicular to \textit{Intersection Line 1}, which is coincident with the \textit{initial} major axis according to our initial conditions (as it is defined as the intersection line between the orbital plane of the stream center and that of the upper edge of the stream in the \textit{initial} orbit). This explains the strong dependence of $N_1$ on $L$. The dependence on the inclination angle $I$ is mainly affected by the LT precession angle $\PhiLT$ --- the vertical separation $\Delta s$ is linearly proportional to $|\Phi_{\rm LT}|\propto |a| \sin I$ --- a larger inclination angle tends to delay the stream collision. The weak dependence of the apsidal precession angle $\Phiap$ on $I$ also plays a minor role, causing $N_1(I)$ to be non-monotonic for large positive spins $a\sim 1$ and large inclination angles $I\gtrsim 45^{\rm o}$ (see the bottom panel of Fig. 8). A quantitative model for $N_1(L, \cos I)$ will be described in \S \ref{sec:physical_reason}.

\subsection{Thickness ratio and transverse offset}
The hydrodynamics of stream collision strongly depends on the thickness ratio $H_2/H_1$ as well as the dimensionless transverse offset $\Delta s/(H_1+H_2)$ \citep[e.g.,][]{jiang16_stream_collision}, where $\Delta s$ is the closest approach distance between the centers of the two colliding geodesics. The ratio $\Delta s/(H_1+H_2)$ is an indicator of the fraction of kinetic energy dissipated in the collision -- a higher ratio means less energy dissipation. Subsequently, the results of the stream collision affect the formation of the accretion disk \citep{bonnerot21_accretion_flow_review}. 

\begin{figure}
    \centering
    \includegraphics[width = 0.4\textwidth]{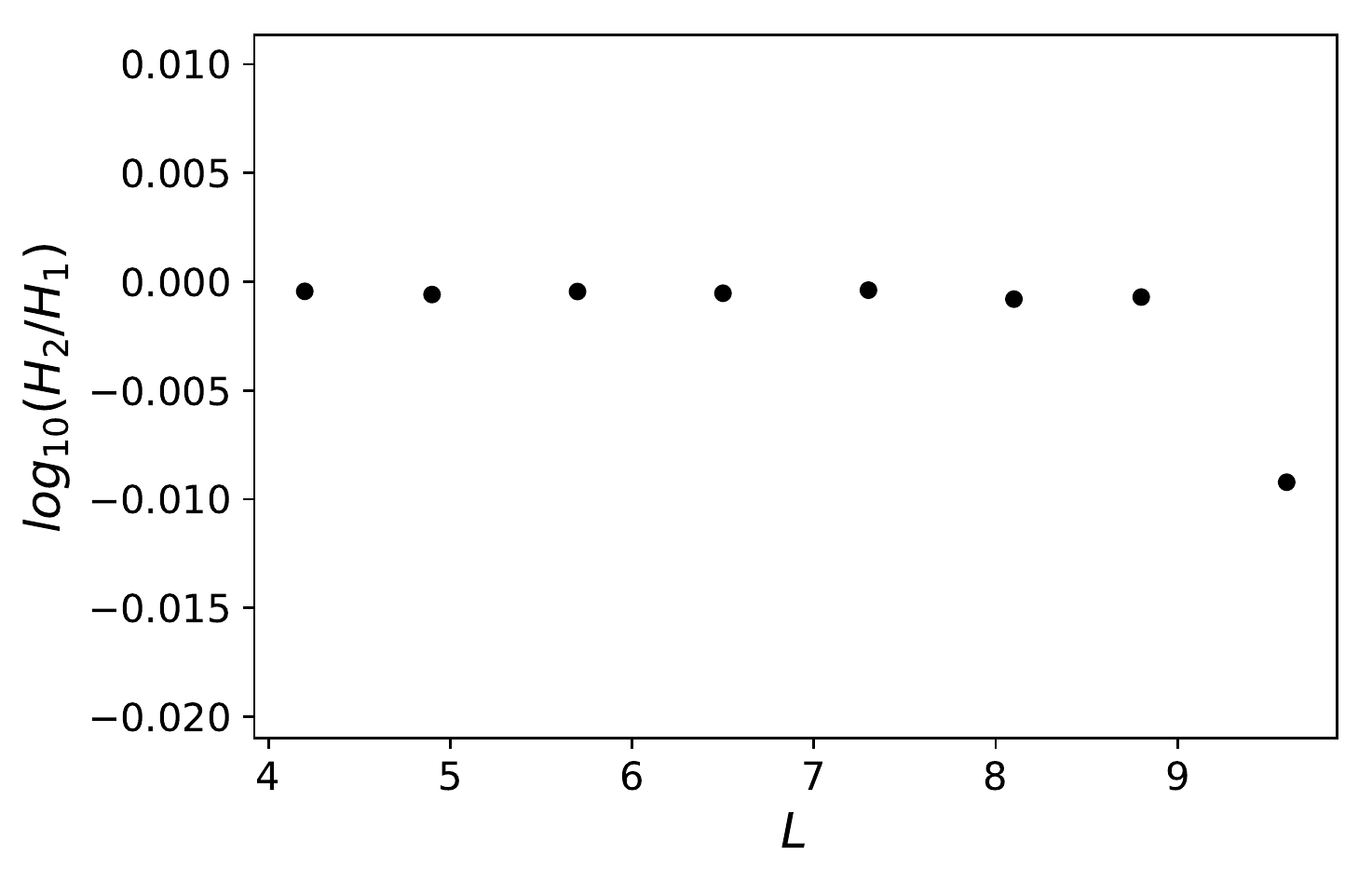}
    \includegraphics[width = 0.4\textwidth]{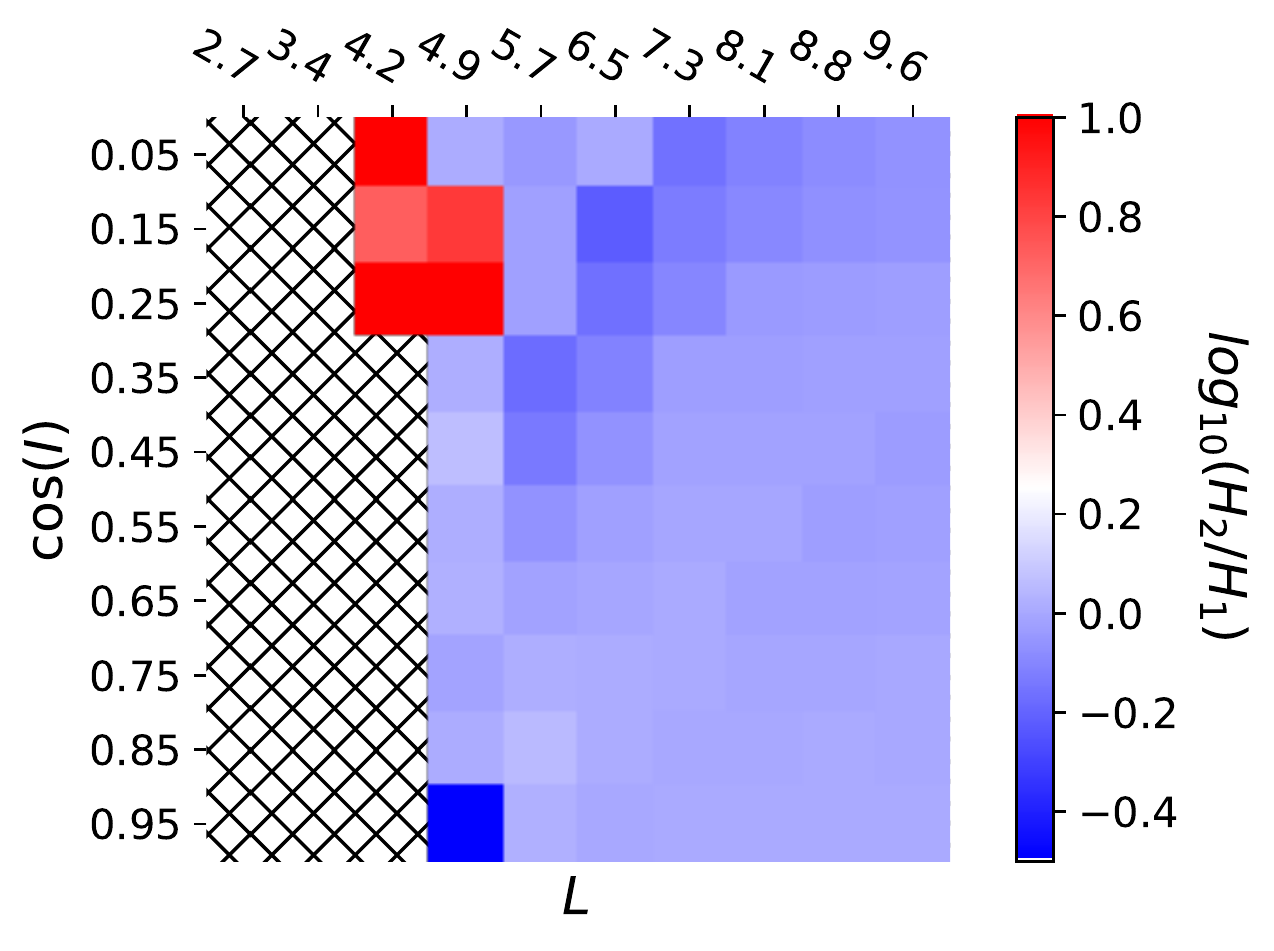}
    \includegraphics[width = 0.4\textwidth]{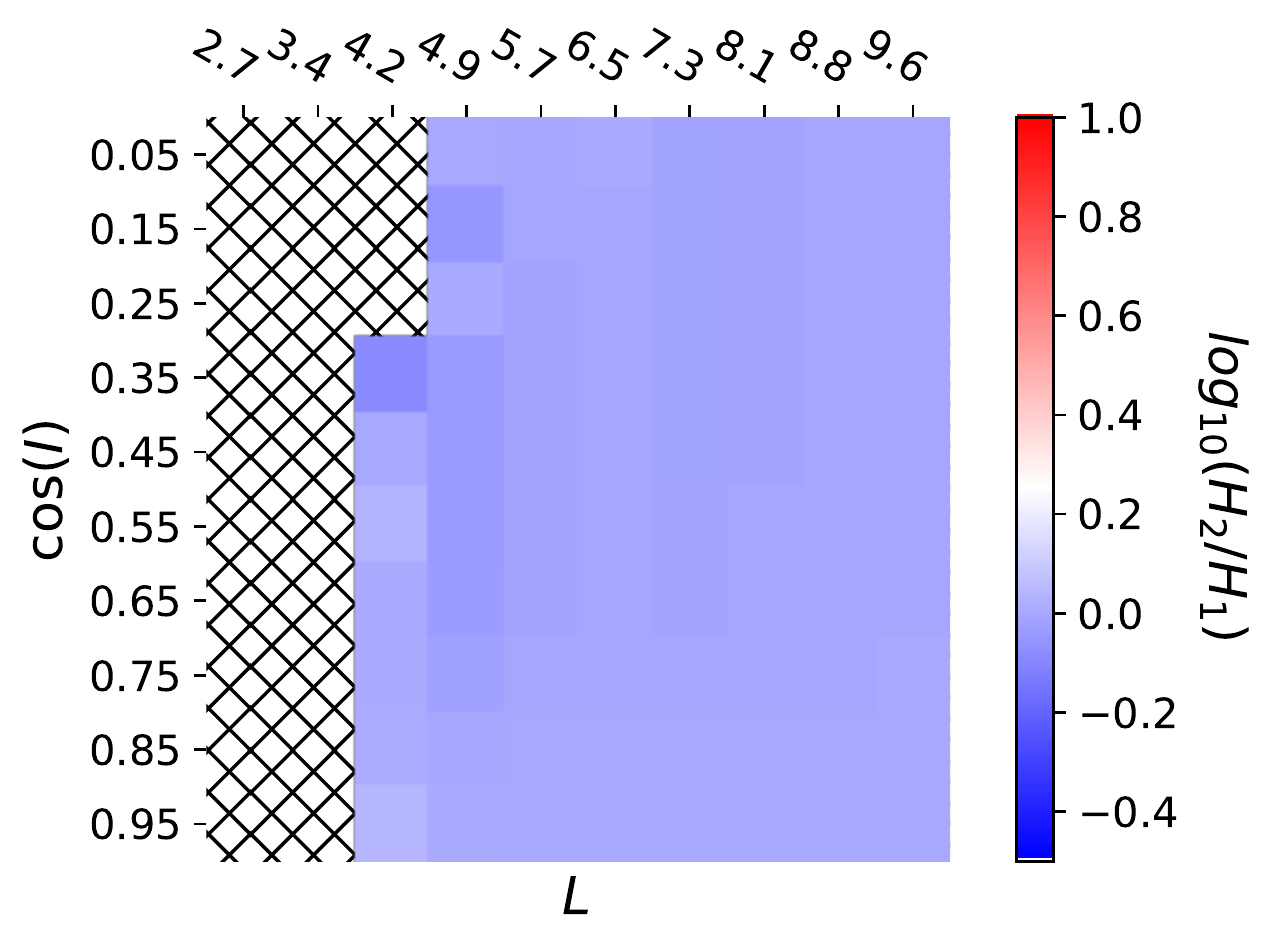}
    \includegraphics[width = 0.4\textwidth]{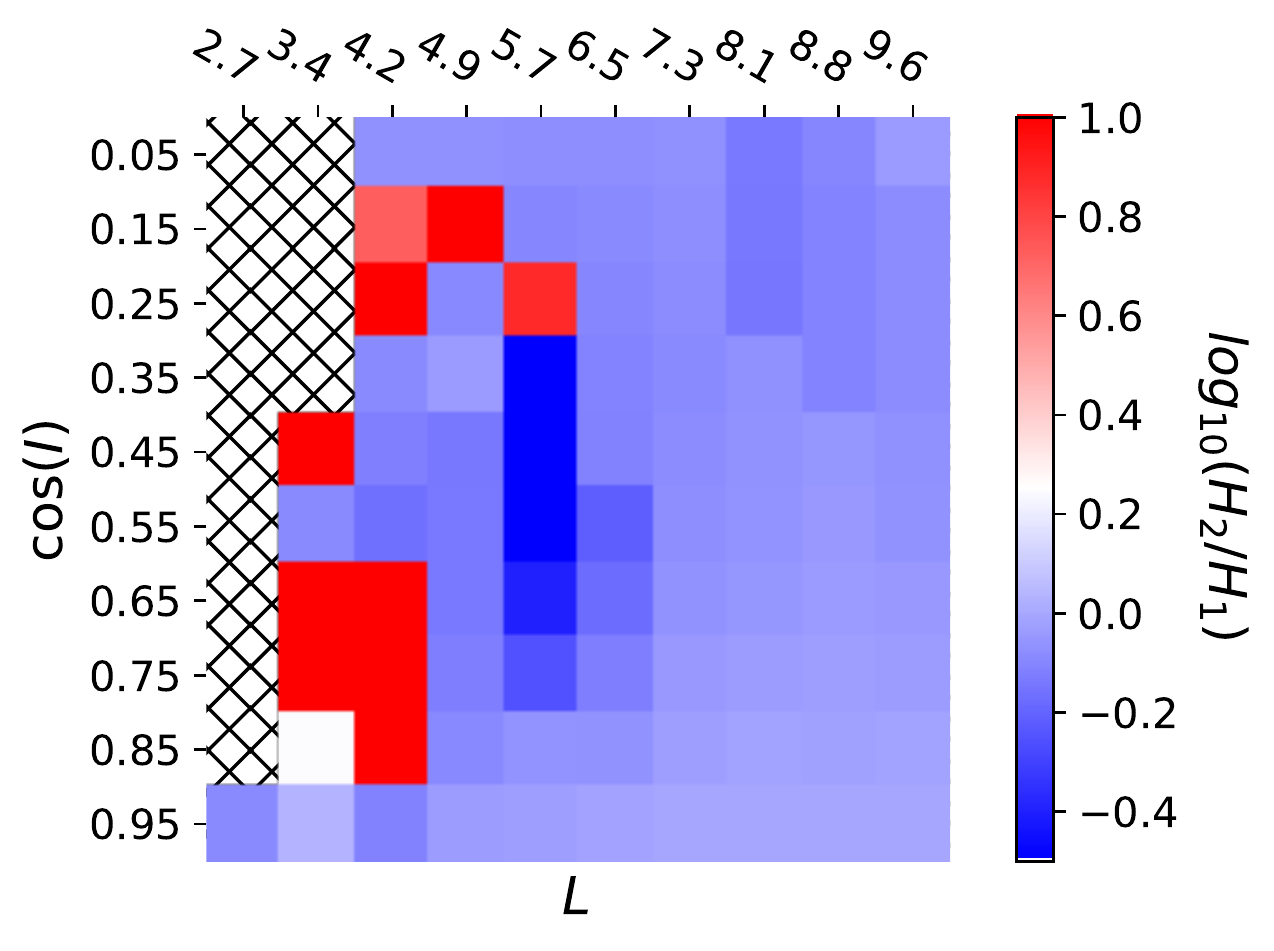}
    \caption{The distribution of thickness ratios $H_2/H_1$ for different BH spins: $a=0$ (top panel), $a=-0.9$ (second panel), $a=0.5$ (third panel) and $a=0.9$ (bottom panel).}
    \label{fig:H21}
\end{figure}

\begin{figure}
    \centering
    \includegraphics[width = 0.4\textwidth]{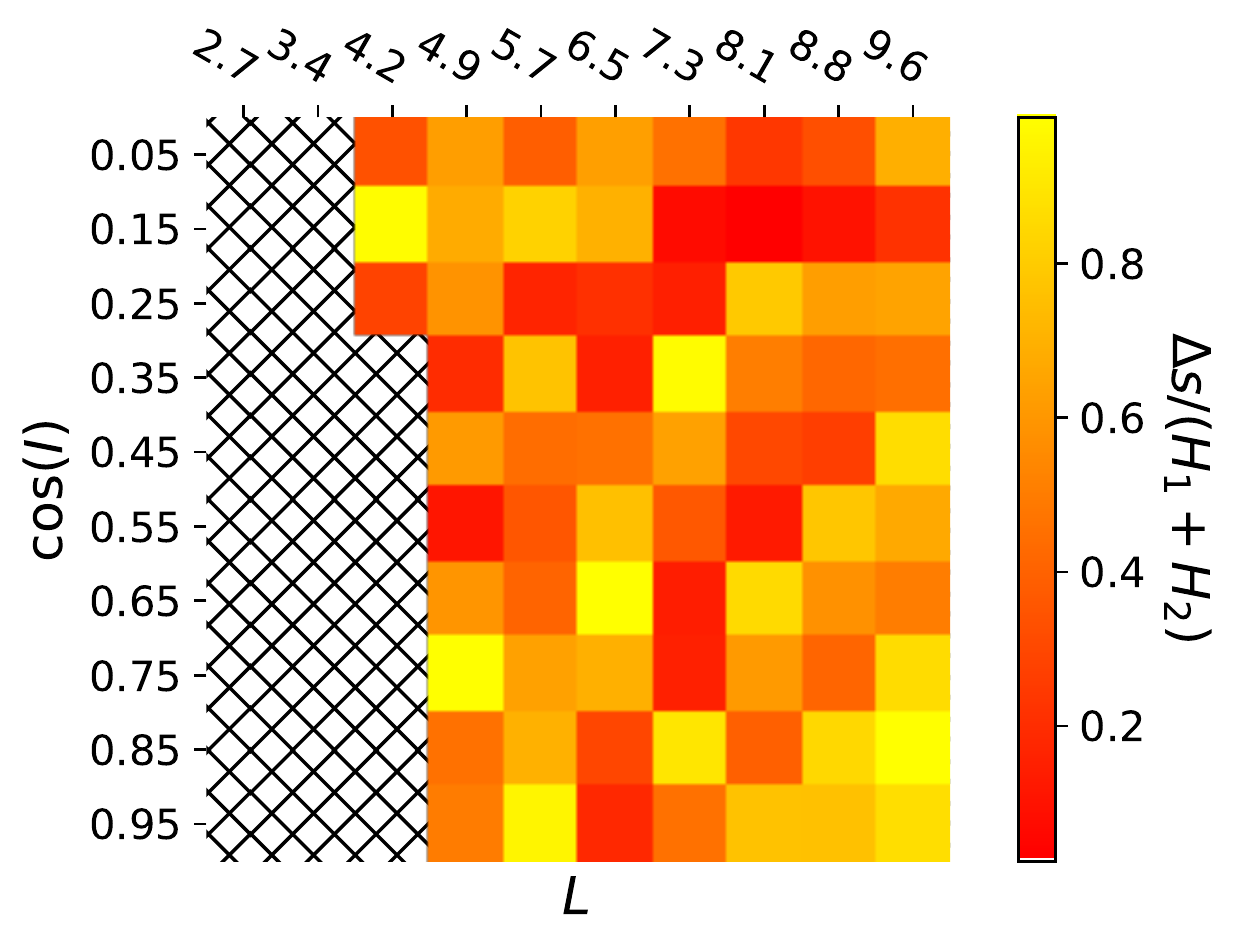}
    \includegraphics[width = 0.4\textwidth]{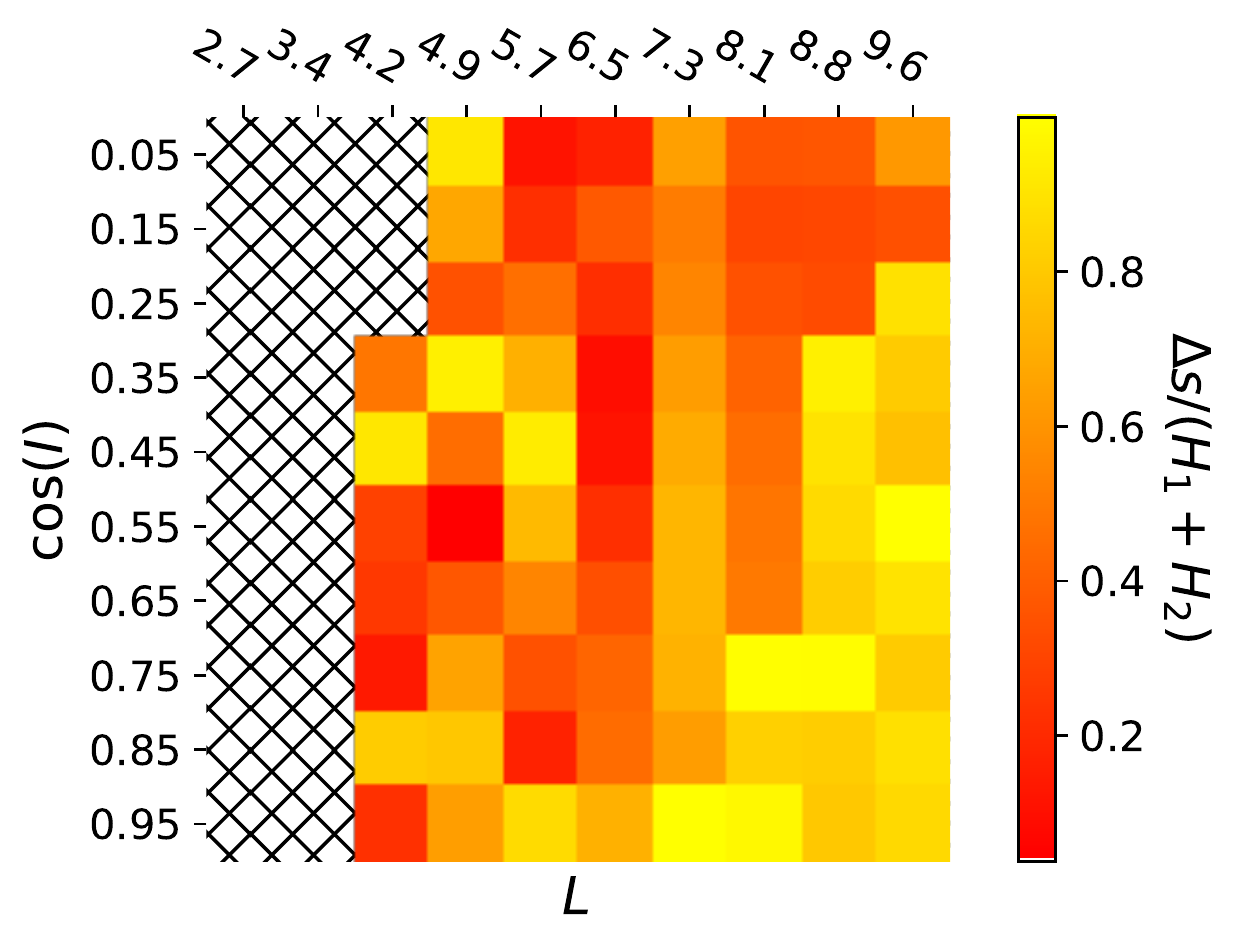}
    \includegraphics[width = 0.4\textwidth]{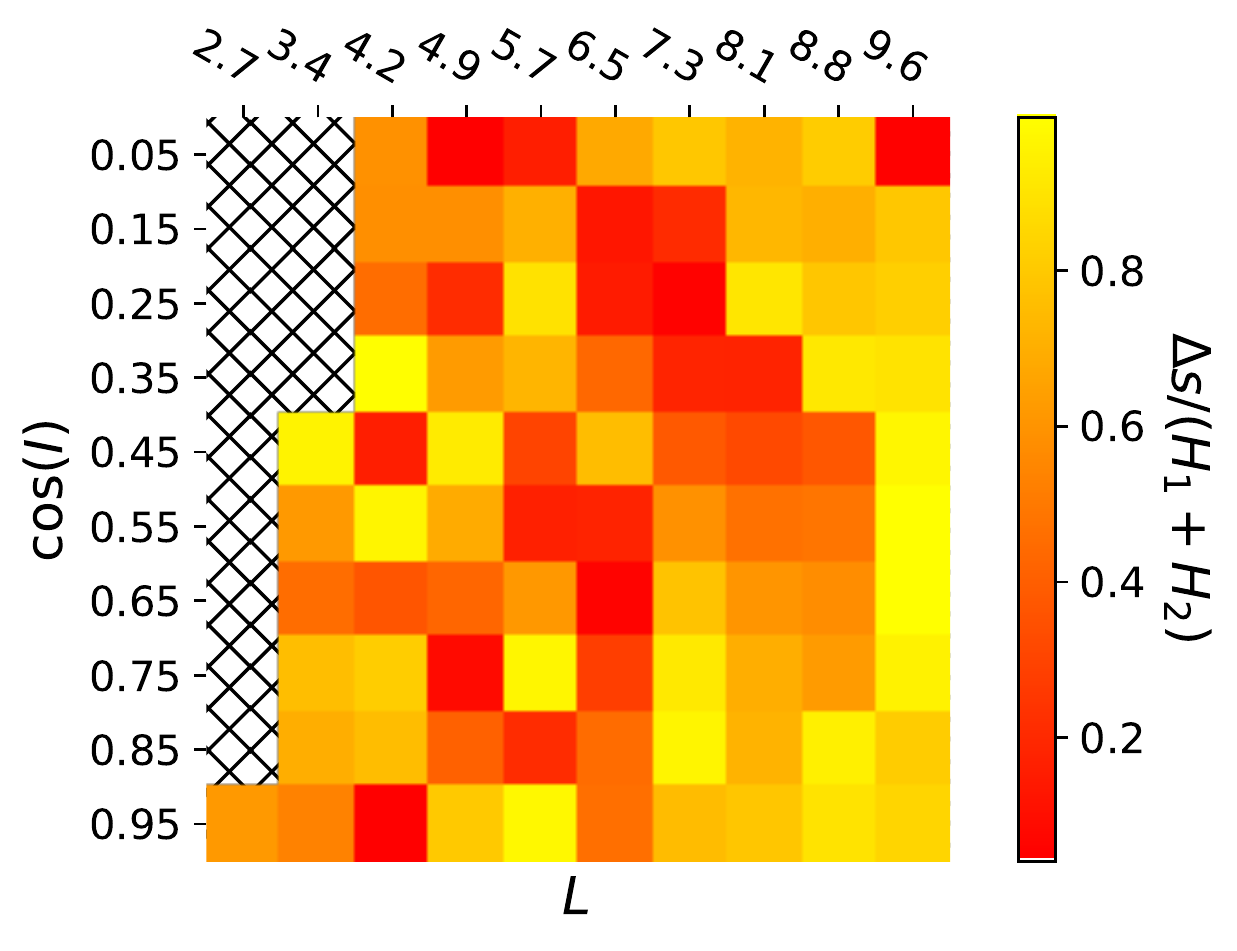}
    \caption{The distribution of (dimensionless) transverse separations $\Delta s/(H_1+H_2)$ for different BH spins: $a=-0.9$ (top panel), $a=0.5$ (middle panel) and $a=0.9$ (bottom panel).}
    \label{fig:dsH}
\end{figure}

In Figs. \ref{fig:H21} and \ref{fig:dsH}, we show $\log_{10}(H_2/H_1)$ and $\Delta s/(H_1+H_2)$, respectively, for cases with different angular momenta $L$ and inclinations $\cos I$. The panels are for different BH spins $a=-0.9$, $a=0.5$ and $a=0.9$. In the case of $a=0$, the orbit is equatorial and hence $\Delta s=0$.

We find that the thickness ratio is close to unity, except for the special cases of nearly plunging angular momenta ($L\lesssim 5$). This is because the stream thickness is specified by the projected distance from \textit{Intersection Line 1}, which is very similar for the colliding ends since they are both located at the point of intersection.





On the other hand, we find that the typical transverse offset $\Delta s$ at the collision point is a small but significant fraction (tens of percent) of the total thickness $H_1 + H_2$, and grazing collisions with $\Delta s/(H_1+H_2)\approx 1$ are very rare. Thus, we expect a large fraction of the gas to be shock-heated in an offset collision, which substantially increases the stream thickness (or directly produces a puffy gaseous envelope). Subsequent secondary shocks will lead to the circularization of the bound gas to form an accretion disk.

We note that a limitation of our study is that the vertical width $H$ does not provide a full description of the transverse density profile of the stream. The consequence is that a small fraction of the gas (the atmosphere of each stream) may still undergo shock interaction even when $\Delta s\gtrsim H_1 + H_2$. It is possible that a weak, oblique shock in a minor collision propagates to the interior of the stream and leads to width expansion. Thus, the criterion for stream collision may actually be somewhat looser than our prescription ($\Delta s/(H_1 + H_2)$ somewhat larger than 1) and the delay time shorter.

\subsection{Relative speed of collision}\label{sec:vrel}
\begin{figure}
    \centering
    \includegraphics[width = 0.4\textwidth]{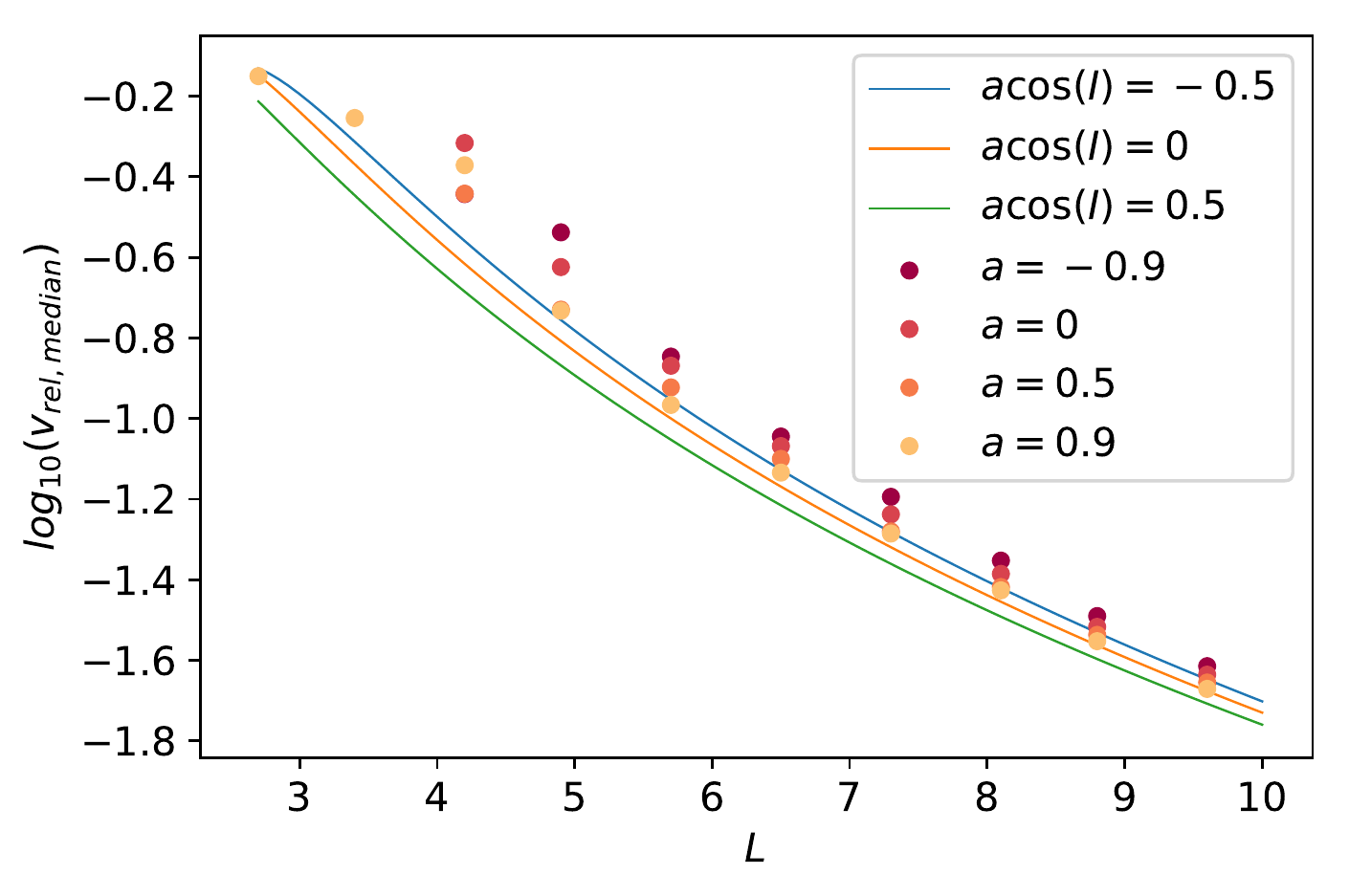}
    \includegraphics[width = 0.4\textwidth]{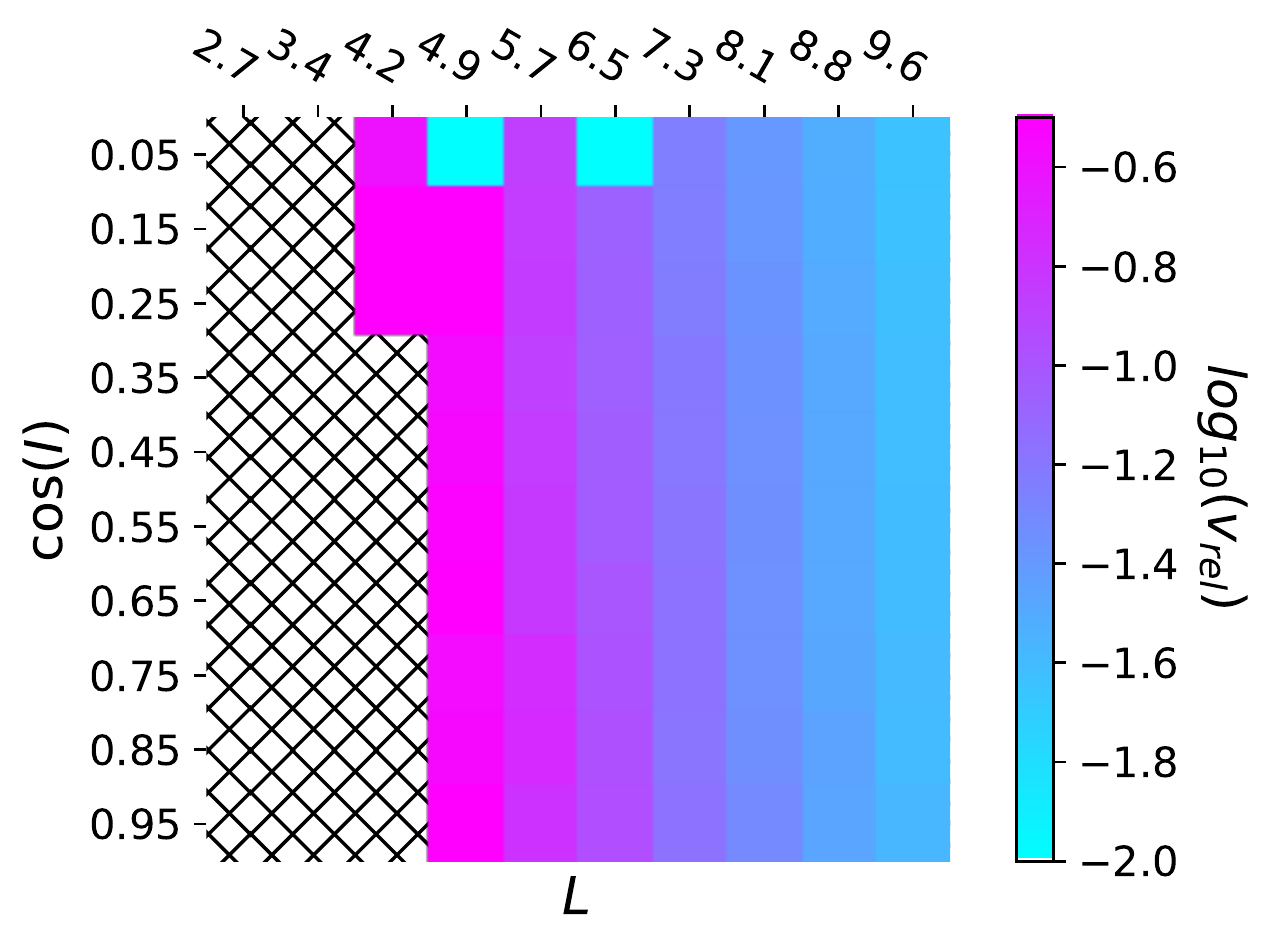}
    \includegraphics[width = 0.4\textwidth]{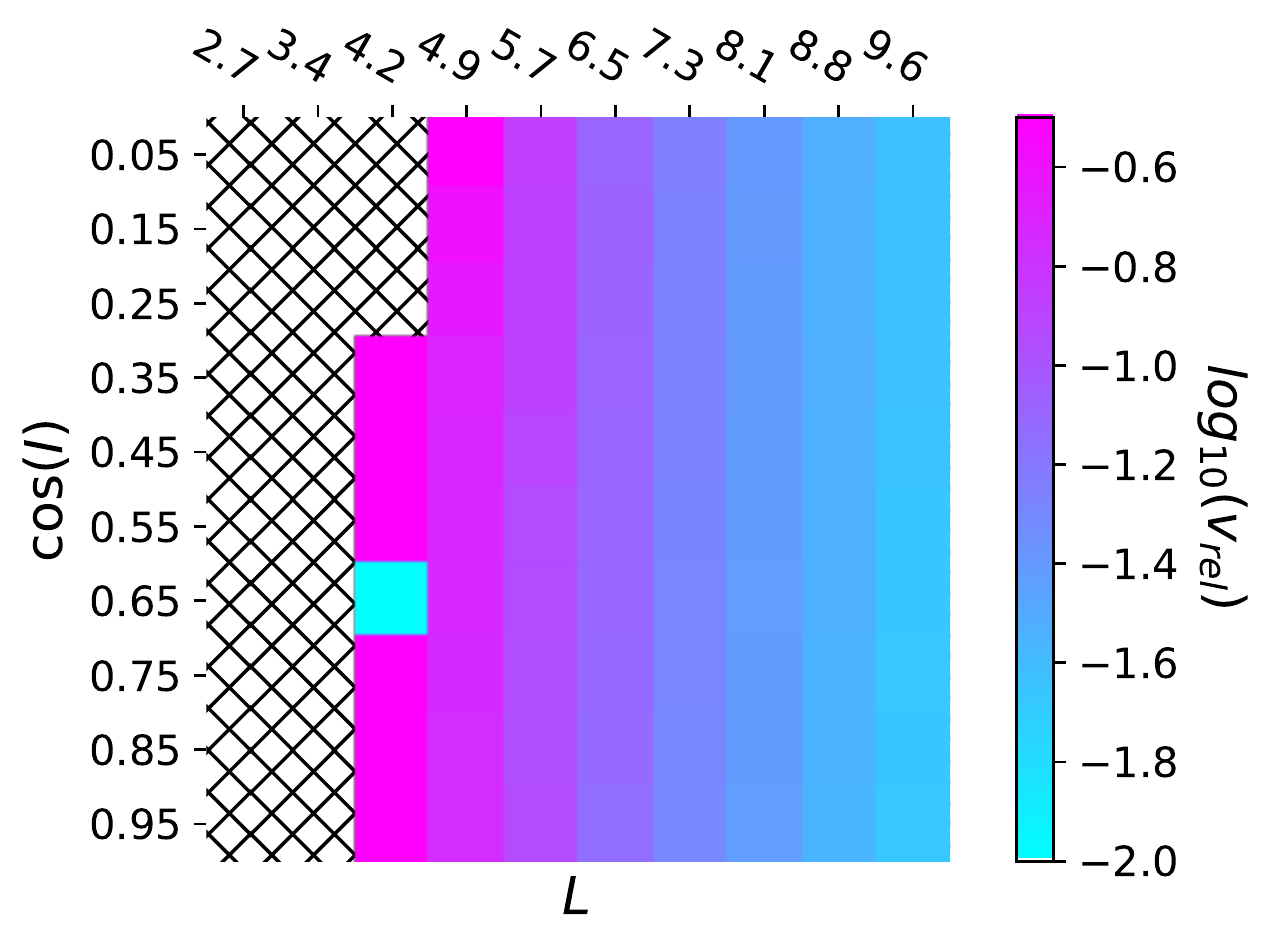}
    \includegraphics[width = 0.4\textwidth]{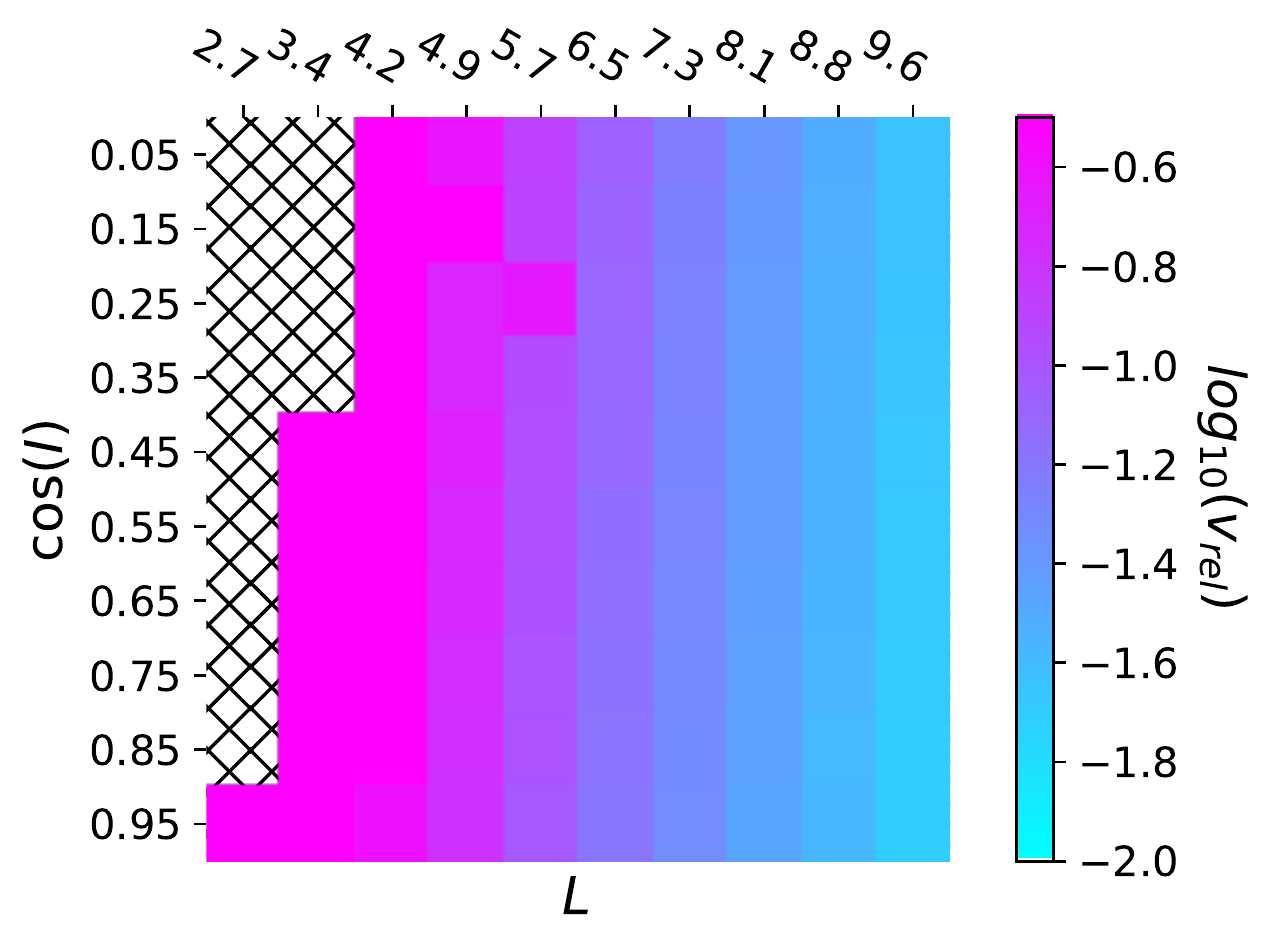}
    \caption{Relative speed of collision in units of $c$. The top plot shows the relative speed for $a=0$, along with the medians of the relative speeds for $a=-0.9$ (second panel), $a=0.5$ (third panel) and $a=0.9$ (bottom panel) over the inclination angles. The solid lines in the top plot illustrate the approximate analytical prediction for the relative speed, as in Eq. (\ref{eq:relative_velocity}).
    }
    \label{fig:rv}
\end{figure}

The (Lorentz invariant) relative speed of the two colliding streams is given by
\begin{eqnarray}
    v_{\rm rel} = {\sqrt{(\vec{p}_1\cdot \vec{p}_2)^2 - 1} \over -\vec{p}_1\cdot \vec{p}_2},
\end{eqnarray}
where $\vec{p}_1$ and $\vec{p}_2$ are the four-velocities of the two streams at the collision point, whose components can be evaluated in any frame as long as the dot product is calculated using the corresponding metric (we compute the dot product in Boyer-Lindquist coordinates).
In Fig. \ref{fig:rv}, we show the relative speeds for all cases considered in this work.  

We find that the relative velocity strongly depends on the total angular momentum of the stream and that the dependence on the inclination and BH spin is weak.
The stream collisions are highly violent in all cases, with relative velocities being a few percent of the speed of light or higher. Note that $v_{\rm rel}/2$ better describes the component of the velocity that is dissipated by shocks, because in the Newtonian limit and when all the gas is shock-heated (for $\Delta s=0$ and $H_1=H_2$), the time-averaged power of shock dissipation is given by\footnote{The stream collision is expected to be intermittent, because after the segment between the two colliding ends is consumed by shocks, the next episode of collision will be delayed by roughly the duration of the previous collision episode.
In the ideal case where the shocked gas does not interact with the cold stream, the shock power alternates between 0 and twice the $\lara{\dot{E}}$ given here. }
$\lara{\dot E} = (1/2)\dot{M} (v_{\rm rel}/2)^{2}=1.9\times10^{43}\mathrm{\, erg\,s^{-1}}\, (\dot{M}/3\,M_\odot\mathrm{\,yr^{-1}}) (v_{\rm rel}/0.03)^2$, where $\dot{M}$ is the mass fallback rate. As pointed out by \citet{jiang16_stream_collision, lu20_self-intersection}, the radiative efficiency of the shocked gas near the collision point is much less than unity (typically a few percent) such that the observed luminosity $L_{\rm sh}\ll \lara{\dot E}$. We see that the shock-powered emission typically falls short of explaining many TDEs with peak optical luminosity of $\sim 10^{44}\rm\, erg\,s^{-1}$ \citep[e.g.,][]{gezari12_ps-10jh} unless the initial orbit of the star is deeply penetrating with pericenter distance much less than the tidal disruption radius.

Remarkably, we find the relative velocity to be nearly independent of whether the collision is in the head-on or rear-end modes. We demonstrate the physical reason in the following
post-Newtonian picture. Let us approximate the orbit right before the stream collision as an ellipse with the BH located at one of the focal points. The stream collision is mainly caused by apsidal precession, and hence the collision points in the head-on and rear-end modes are located at true anomaly  $\pi - \Phi_{\rm ap}/2$ and $\Phi_{\rm ap}/2$, respectively, where the apsidal precession angle $\Phi_{\rm ap}$ is given by Eq. (\ref{eq:apsidal_precession}).


These two collision points (which lie on the same line that goes through the BH) are located at radii
\begin{eqnarray}\label{eq:rint}
    r_{\pm} = {L^2\over 1 \pm e_{\rm orb} \cos(\Phi_{\rm ap}/2)},
    \label{eq:rint}
\end{eqnarray}
where the squared orbital eccentricity is $e_{\rm orb}^2 = 1 - L^2/a_{\rm orb}$ and the semimajor axis is $a_{\rm orb} = [2(1-\tilde E)]^{-1}$. It can be seen that $r_+$ is close to the pericenter radius (corresponding to the rear-end mode) and $r_-$ is typically much larger (the head-on mode). Note that $r_-$ is in agreement with Eq. (7) of \citet{dai15_pn1_precession}.

Conservation of energy and angular momentum along with Eq. (\ref{eq:rint}) imply that the radial components $v_{\mathrm{K},r}$ of the Keplerian velocities at these two collision points are the same, given by
\begin{eqnarray}
    v_{\mathrm{K},r}^2 = -{1\over a_{\rm orb}} + {1 - e_{\rm orb}^2 \cos^2(\Phi_{\rm ap}/2) \over a_{\rm orb}(1-e_{\rm orb}^2)},
\end{eqnarray}
where the dependence on the precession angle is $\cos^2(\Phi_{\rm ap}/2)$ and the sign of $\cos\Phi_{\rm ap}$ does not matter. The velocity component perpendicular to the radial direction is $L/r$, so the collision angle is given by $\tan (\gamma/2) = r\,v_{\mathrm{K},r}/L$. This shows that in the rear-end (or head-on) collision mode, we typically have $\gamma\ll 1\rm\, rad$ (or $\gamma\sim \pi$). 
The relative velocity is twice the radial component of the velocity at the collision points and is given by
\begin{eqnarray}\label{eq:relative_velocity}
\begin{split}
    \left({v_{\rm rel}\over 2}\right)^2
    \approx 2(\tilde E - 1) + {1 - [1+2L^2(\tilde E - 1)] \cos^2(\Phi_{\rm ap}/2) \over L^2}.
\end{split}
\end{eqnarray}
In the limit where $1-e_{\rm orb}\ll 1$ and $\Phi_{\rm ap} \lesssim 1\rm\, rad$, we obtain the approximation
\begin{eqnarray}
    v_{\rm rel}\approx {\Phi_{\rm ap} \over L} \approx {6\pi \over L^3} - {8\pi a\cos I \over L^4}.
    \label{eq:relv}
\end{eqnarray}
This shows that the relative velocity between the two colliding ends is a power-law function of the total angular momentum ($v_{\rm rel}\propto L^{-3}$) or orbital pericenter distance ($\propto r_{\rm p,orb}^{-1.5}$) in the weakly relativistic regime. The relative velocity is slightly smaller for larger, prograde BH spins, as illustrated in Fig. \ref{fig:rv}.


\section{Physical Reason for Intersection}\label{sec:physical_reason}
In this section, we show that the collision is a geometric effect. Throughout this section, we work in the lowest-order post-Newtonian approximation such that $\Phi_{\rm LT}\ll \Phi_{\rm ap}$ (see Eqs. \ref{eq:LT_precession} and \ref{eq:apsidal_precession}) and assume that collision occurs between successive half-orbits $N_1$ and $N_2=N_1+1$.

\begin{figure*}
    \centering
    \includegraphics[width=0.65\textwidth]{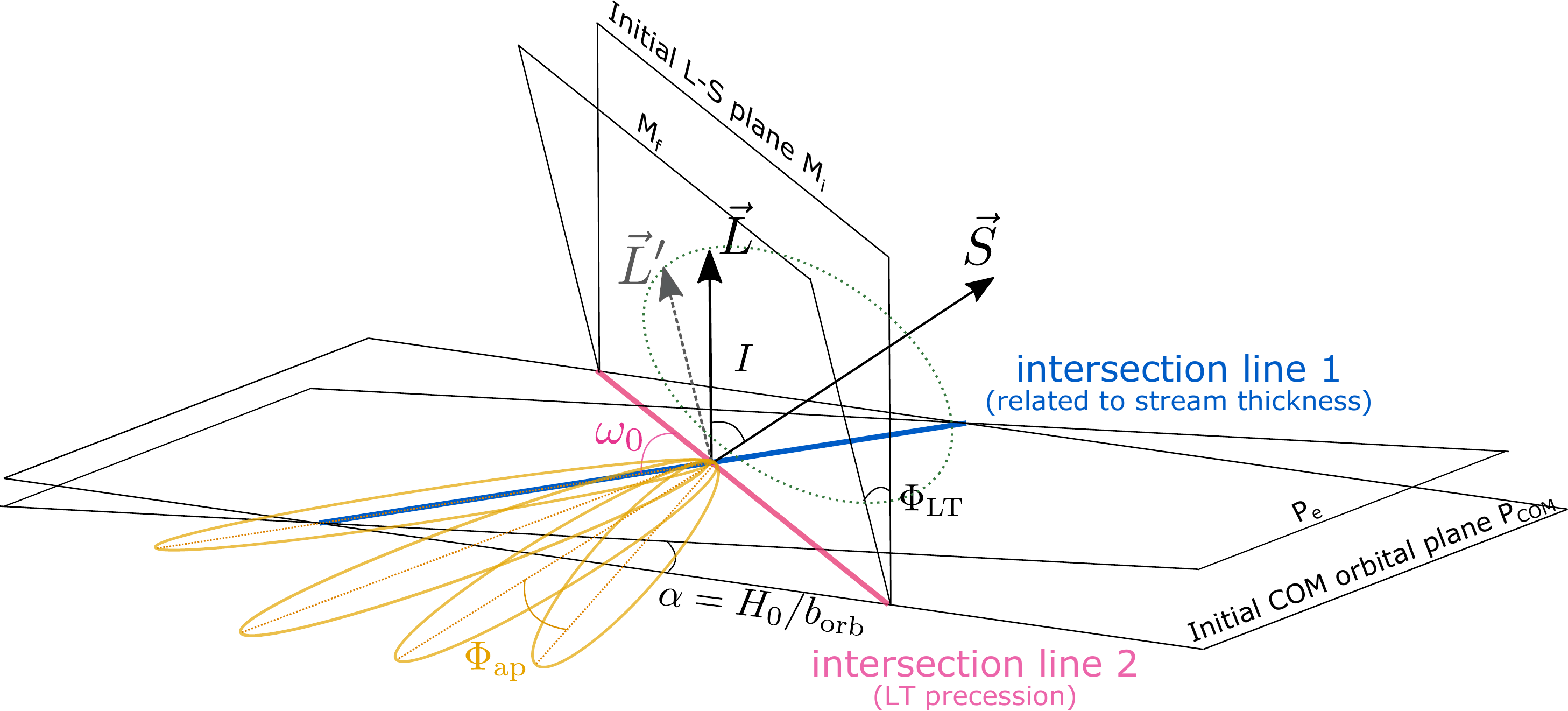}
    \caption{A schematic picture for the geometric viewpoint of stream collision in the post-Newtonian approximation, where the orientation of an elliptical orbit undergoes apsidal and Lense-Thirring (LT) precessions. A collision occurs either due to the increase in thickness $H$ as the major axis of the orbit moves away from Intersection Line 1 (thick blue line), or due to the decrease in vertical separation between adjacent orbital windings as the major axis become more aligned with Intersection Line 2 (pink solid line).}
    \label{fig:schematic}
\end{figure*}

A schematic picture is shown in Fig. \ref{fig:schematic}. The initial ($N=0$) orbital plane at the center of mass (COM) of the fallback stream is denoted as $P_{\rm COM}$, and the particle at the upper edge of the stream lies in the orbital plane labelled as $P_{\rm e}$. These two planes intersect in a line, which is denoted as \textit{Intersection Line 1}. The angle between these two planes ($P_{\rm COM}$ and $P_{\rm e}$) is given by
\begin{eqnarray}
    \alpha = H_0/\borb = {H_0\sqrt{2(1-\tilde{E})}\over L},
\end{eqnarray}
where $H_0$ is the initial stream thickness on the minor axis, $\borb = L\aorb^{1/2}$ is the semi-minor axis, and $\aorb = (2(1-\tilde{E}))^{-1}$ is the semi-major axis as determined by the specific orbital energy $\tilde{E}$. For $H_0=5R_\odot$, $\tilde E=0.9999$, and $L\in (6.5, 9.7)$ (the mildly relativistic cases), we find $\alpha\sim 5\times10^{-3}\rm\, rad$. The stream thickness $H$ at a given position is related to the inclination angle $\alpha$ and the projected distance to Intersection Line 1.
Suppose the collision occurs at radius $r_{\rm int}$ (as given by Eq. \ref{eq:rint}) between the $N_1$-th and $(N_1+1)$-th half-orbits. We project the radial vector $\vec r_{\rm int}$ onto Intersection Line 1 to obtain the stream thickness in the vertical direction (perpendicular to the orbital plane) as
\begin{eqnarray}\label{eq:stream_width}
    H(N_1)=\rint \sin\lrsb{(N_1 + 1)\Phiap/2} \alpha.
\end{eqnarray}

\begin{figure}
\centering
\includegraphics[width = 0.4\textwidth]{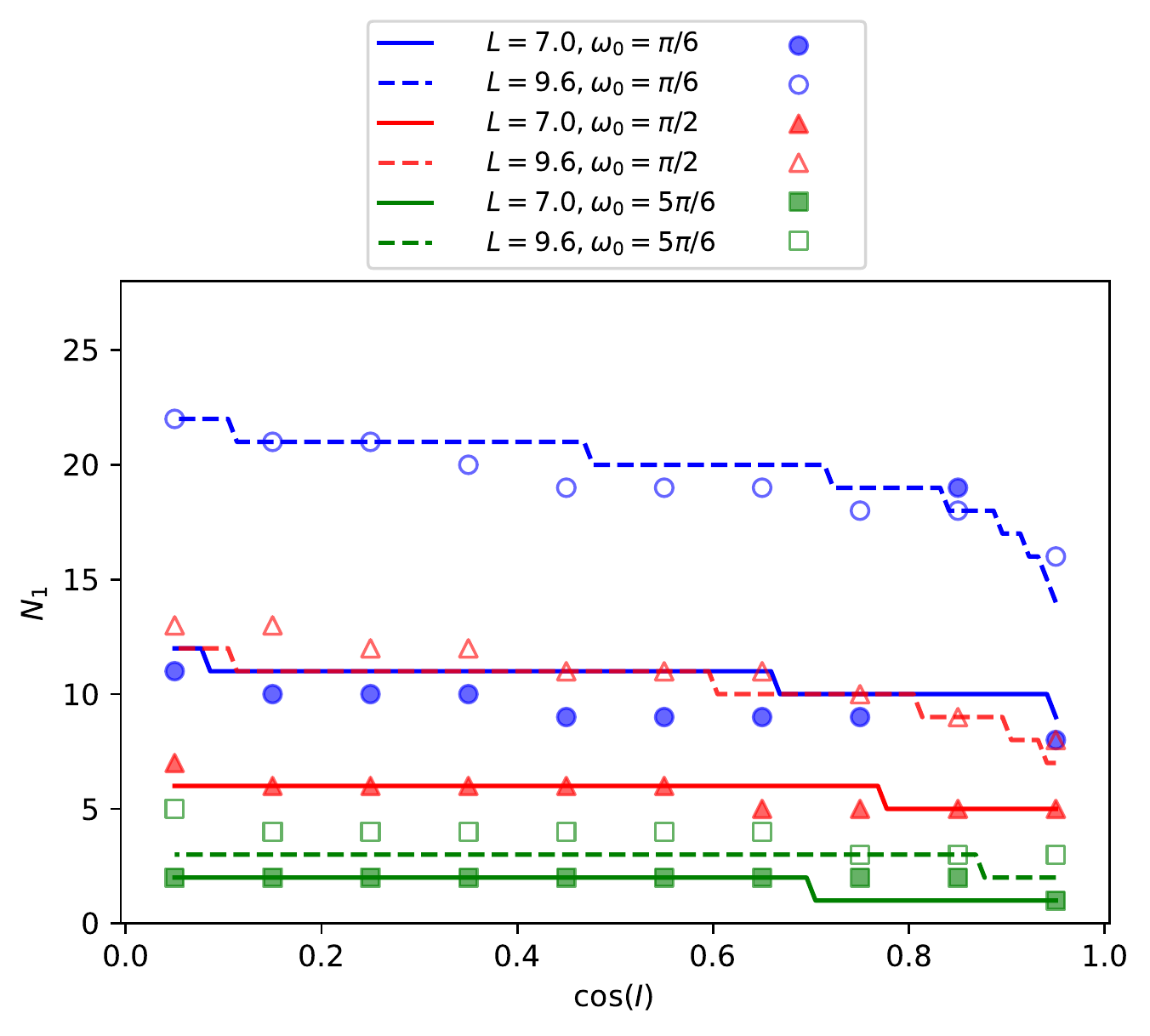}
\includegraphics[width = 0.4\textwidth]{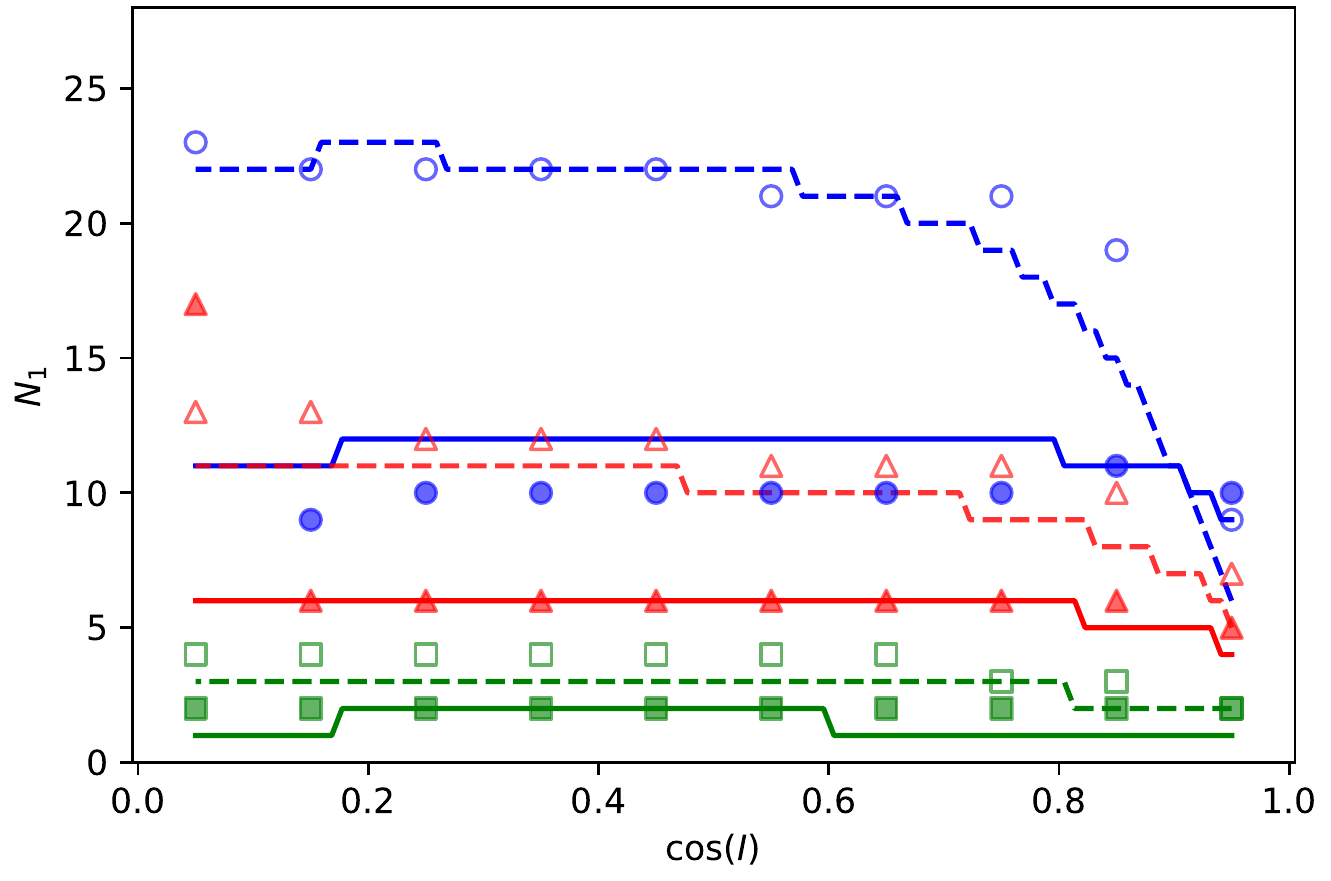}
\includegraphics[width = 0.4\textwidth]{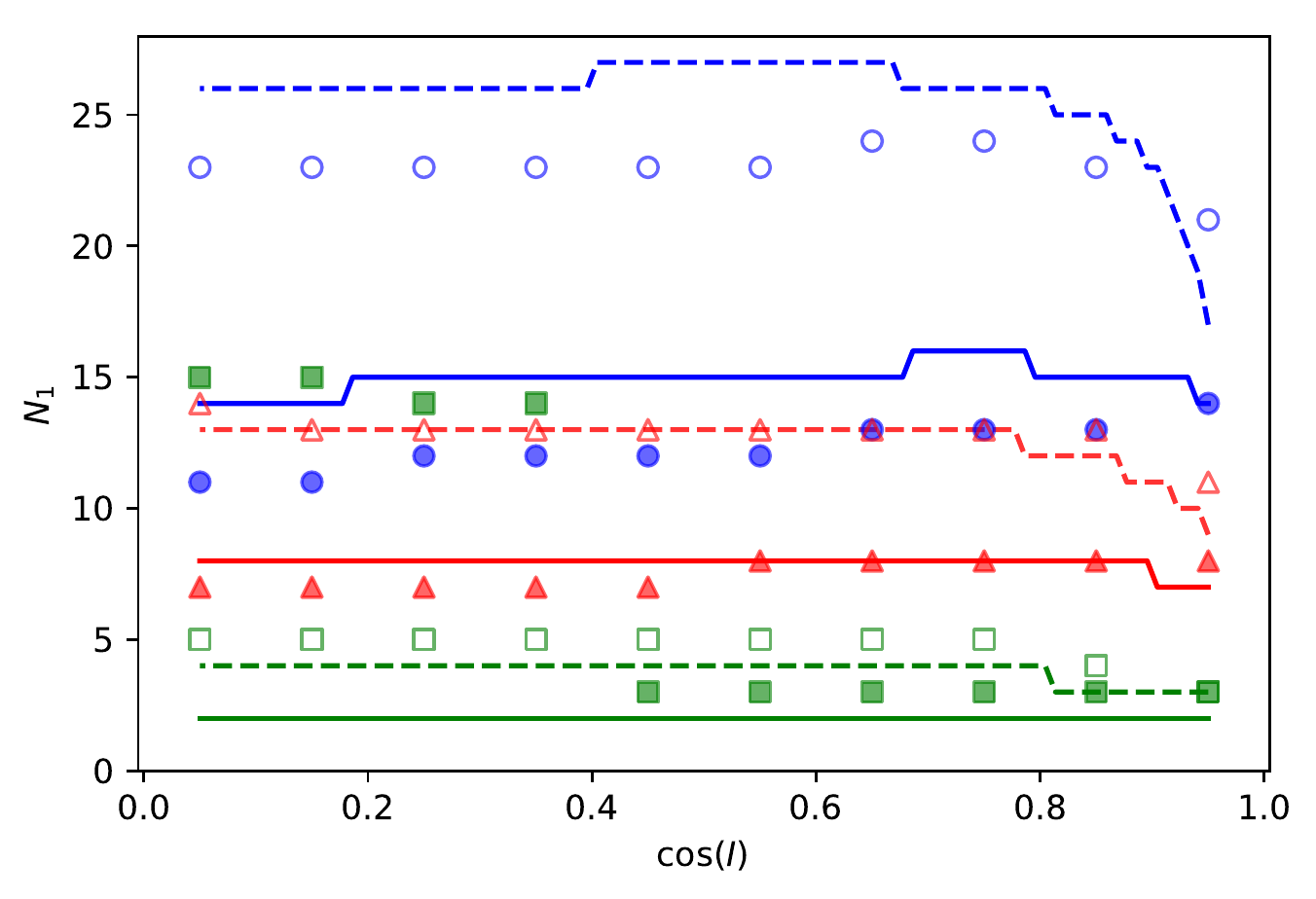}
\caption{Comparison between the analytical prediction for $N_1$ as in Eq. \ref{eq:collision_criterion} (indicated by dashed or dotted lines) and the numerical calculation of $N_1$ (indicated by points, squares or triangles) for various values of $\omega_0$ (in different colors) and BH spin: $a=-0.9$ (top panel), $a=0.5$ (middle panel) and $a=0.9$ (bottom panel). We find good overall agreement between the analytical model and the numerical calculations, except for a few cases with $\cos(I)\approx 0$ and $\omega_0\approx \pi$ under $a=0.9$ and $L=7$. These correspond to the rare cases of nearly polar orbits (i.e., the star's initial major axis is nearly aligned with the BH spin vector), for which the stream thickness evolution can only be captured by using the full tidal tensor and hence both our analytical/numerical results may have large errors.}


\label{fig:n1compare}
\end{figure}

The other quantity that determines the occurrence of stream collision is $\Delta s$, which is the vertical separation between the COMs of the two colliding streams. This can be calculated as follows. Intersection Line $1$ is perpendicular to the initial orbital angular momentum vector $\vec L$ of the orbit. The black hole spin vector $\vec S$ makes an inclination angle $I$ with $\vec L$. The plane defined by $\vec L$ of the initial orbit and $\vec S$ is labelled $M_{\rm i}$. LT precession of $\vec L$ about $\vec S$ causes $\vec L$ to rotate in the direction of $\vec S\times\vec L$, perpendicular to $M_{\rm i}$. The plane defined by the rotated $\vec L'$ and $\vec S$ is labelled by $M_{\rm f}$. For adjacent orbital windings, the angle between $M_{\rm i}$ and $M_{\rm f}$ is $\Phi_{\rm LT}$ (eq. \ref{eq:LT_precession}). $M_{\rm i}$ and $M_{\rm f}$ intersect in \textit{Intersection Line $2$}, which forms an angle $\omega_0$ with \textit{Intersection Line $1$}. The angle $\omega_0\in (0, \pi)$ is entirely set by the initial position and angular momentum of the star when it was fed to the TDE loss cone. It is related to the initial polar angle $\theta_0$ in our Boyer-Lindquist coordinates and the inclination angle $I$ between $\vec L$ and $\vec S$ by
\begin{eqnarray}
    \cos(\theta_0)=\cos(\pi-\omega_0)\sin(I).
\end{eqnarray}
Our numerical simulations as described in \S \ref{sec:methods} and \ref{sec:parameter_exploration} only considered $\theta_0=\pi/2$ (so as to reduce the dimensions of parameter space), which corresponds to $\omega_0=\pi/2$. The probability density distribution of $\d P/\d \omega_0$ is flat in $(0, \pi)$, so it is equally likely that $\omega_0$ is less than or greater than $\pi/2$. According to the post-Newtonian picture described in this section, most of the properties of the stream collision (intersection radius $\rint$, thickness ratio $H_2/H_1$, offset ratio $\Delta s/(H_1+H_2)$, relative velocity $v_{\rm rel}$) do not strongly depend on the choice of $\omega_0$. However, as we now describe, the delay time is strongly affected by $\omega_0$.

In the limit that the cumulative LT precession angle $N_1\PhiLT/2\ll 1\rm\, rad$, the vertical separation between the COMs of the colliding streams is given by
\begin{eqnarray}
    \Delta s'(N_1) = \rint \sin\lrsb{(N_1+1)\Phiap/2 + \omega_0} \PhiLT.
\end{eqnarray}
As we will see later, the number of half orbits before the collision is $N_1 \sim 2/\Phiap$, so cumulative LT precession angle before the stream collision is of the order $\PhiLT/\Phiap\sim a\sin I/L$. Including the cumulative precession of \textit{Intersection Line 2} before the collision leads to a slightly more accurate expression
\begin{eqnarray} \label{eq:ds}
    \Delta s(N_1) = \rint \sin\lrsb{(N_1+1)\Phiap/2 + \omega_0 - N_1\PhiLT/2} \PhiLT,
\end{eqnarray}
which we use (but the two equations above give very similar delay times). Note that as the number of half-orbits $N_1$ increases, the angle between the radial vector $\vec r_{\rm int}$ and Intersection Line 2 (the argument in the sine function above) increases. If $\omega_0 < \pi/2$, $\Delta s$ first increases with $N_1$ and then decreases as the angle between $\vec r_{\rm int}$ and Intersection Line 2 approaches $\pi$. If $\omega_0>\pi/2$, then $\Delta s$ monotonically decreases with $N_1$.

Using Eq. \ref{eq:stream_width} and Eq. \ref{eq:ds}, the criterion for intersection, $\Delta s<H$, is 
\begin{eqnarray}\label{eq:collision_criterion}
\sin\lrsb{{(N_1+1)\Phiap\over 2} + \omega_0 - {N_1\PhiLT\over 2}}{\PhiLT\over \alpha} <\sin\lrsb{{(N_1 + 1)\Phiap\over 2}}.
\end{eqnarray}
Stream crossing occurs at the minimum integer $N_1$ satisfying this equation. Note that above equation includes the cases where $(N_1+1)\Phiap/2 + \omega_0 - N_1\PhiLT/2$ exceeds $\pi$ and when this occurs, $\Delta s$ flips sign meaning that orbital crossing has occurred.
In this geometrical picture, we see that stream collision occurs for different reasons in the following two regimes: $\PhiLT/\alpha < 1$ or $>1$. 

For TDEs with small LT precession angles $\PhiLT$ (due to small BH spin or large orbital angular momentum) or for the parts of the fallback stream that are initially not self-gravitating (the initial thickness is much larger than $5R_\odot$), we may have $\PhiLT/\alpha \ll 1$. In this regime, stream collision occurs mainly due to the increase of the right-hand side of Eq. (\ref{eq:collision_criterion}) (i.e. an increase in the stream thickness), and we can estimate the number of half-orbits before the collision to be $N_1 \simeq 2\Phi_{\rm LT}\sin\omega_0/(\alpha \Phiap) - 1$. On the other hand, for TDEs with large $\PhiLT$ or for the very thin self-gravitating parts of the fallback stream, we may have $\PhiLT/\alpha \gtrsim 1$. In this regime, stream collision occurs mainly due to the decrease of the left-hand side of Eq. (\ref{eq:collision_criterion}) (i.e. a decrease in the vertical separation), and the number of half-orbits before the collision can be estimated by $(N_1+1)\Phiap/2 + \omega_0 \simeq \pi$ or $N_1\simeq 2(\pi-\omega_0)/\Phiap - 1$. Putting these two regimes together, we obtain the estimate
\begin{eqnarray}\label{eq:analyticN1}
    N_1 \simeq \min\lrsb{{\PhiLT\sin\omega_0\over \alpha}, \pi - \omega_0} {2\over \Phiap} - 1.
\end{eqnarray}

Fig. \ref{fig:n1compare} shows a comparison between the analytical prediction for $N_1$ (Eq. \ref{eq:collision_criterion}) and the numerical results of $N_1$ (based on the algorithm in \S \ref{sec:methods}) for various BH spins and three different choices of $\omega_0=\pi/6, \pi/2, 5\pi/6$. We fix the initial stream thickness to be $H_0=5R_\odot$ (the same as in previous sections), which means all the cases are in the $\PhiLT/\alpha \gtrsim 1$ regime. The key trend in this regime is that $N_1$ is a strong function of the orbital angular momentum, $N_1\propto L^2$ (the dependence on the inclination angle is less strong, unless $\cos I \simeq 1$, which corresponds to very vanishing LT precession). Another important trend is that $N_1$ decreases with $\omega_0$, and this is because for larger $\omega_0$ it takes a smaller number of orbital windings to align the major axis of the orbit with Intersection Line 2.

We conclude that the physical reason for stream collisions lies in the geometry of Kerr spacetime: collision occurs when the apsidal precession brings the major axis of the elliptical orbit to be nearly parallel with the plane containing the angular momentum vector and the BH spin vector (minimizing the vertical separation between consecutive orbits) and/or to be nearly perpendicular to the initial major axis (maximizing the stream thickness). An earlier work by \citet{guillochon15_dark_year} suggests that the collision is caused by the increase in stream thickness with the orbital winding number as a result of possible energy dissipation by nozzle shocks. However, in our model, the stream thickness evolves according to the BH's tidal forces and collision still occurs when the stream does not get significantly inflated at the nozzle shocks, as motivated by the recent simulation of \citet{Bonnerot21_stream_width_evolution}.

According to our model, the delay times for TDE stream collisions span a very wide range from a few months up to a decade, which mainly depend on $H_0$ (initial stream thickness), $L$ (orbital angular momentum), and $\omega_0$ (the angle between the initial major axis and the initial $\vec L$-$\vec S$ plane). Short delay times correspond to the cases that satisfy one of the following: $L\lesssim 5$ (highly relativistic TDEs), small $\PhiLT/\alpha$ (thick initial streams or small BH spins), or $\omega_0$ close to $\pi$. On the other hand, long delay times are expected for cases with large $\PhiLT/\alpha$ (thin initial streams) and $\omega_0$ far from $\pi$.

\section{Discussion}\label{sec:discussion}
In this section, we briefly discuss a number of limitations of our current model which may be addressed in future works.


(1) We only considered mono-energetic geodesics, whereas the realistic bound debris has a spread in orbital energy -- the gas falling back at a later time is less bound and has a larger apocenter radius. This causes the two colliding ends of the stream to have different energies and their orbits to have different apocenter radii. Since most of the collisions occur between successive half-orbits ($N_2 = N_1 + 1$), the fallback times of the two colliding ends differ by at most one orbital period. The intersection radius will only be affected by the energy spread if the collision occurs near the apocenter (i.e., for angular momentum $L\sim L_{\rm max}\simeq 9.7$). This is because the apocenter radius of the later colliding end is substantially larger than that of the first one. Accounting for the energy spread requires a more sophisticated scheme that records the positions of different fluid elements with different orbital energies on a grid of global Boyer-Lindquist time. However, for most of the parameter space studied in this work, the intersection radius is sufficiently far from the apocenter and the orbital energies of the two colliding ends are approximately the same.

(2) The stream thickness evolution is approximated by Eq. (\ref{eq:tfe}), which is exact only for the Schwarzschild spacetime in the direction perpendicular to the orbital plane. The stream thickness within the orbital plane is strongly affected by the angular momentum spread across the stream \citep{cheng14_relativistic_tdes}. Its evolution is governed by a different tidal equation with non-inertial terms due to rotation, although the differences from our Eq. (\ref{eq:tfe}) become negligible at sufficiently large radii where the orbits are nearly radial. Additionally, BH spin causes a mixing between the stream thickness perpendicular to the orbital plane and the in-plane thickness, since the tidal tensor is non-diagonal \citep{marck83_tidal_tensor} \citep[see also][]{kesden12_spinning_BH}. Fully capturing the evolution of the cross-sectional shape of the stream requires two dimensional simulations using the full tidal tensor as well as including gas pressure effects, which are beyond the scope of this work.  In Appendix \ref{sec:Newtonian_tidal_equation}, we show the stream thickness evolution under an even simpler, Newtonian tidal equation, and the results are qualitatively similar to those presented in previous sections.

(3) The effects of pressure forces are only crudely accounted for near the nozzle shocks (where $H=0$) by prescribing a perfect bounce of the vertical velocity $\dot{H}$. In reality, pressure forces may have long-lasting effects on the stream thickness evolution. \citet{Bonnerot21_stream_width_evolution} found that the irreversible entropy increase (which is proportional to the adiabatic constant $P/\rho^{5/3}$ for monoatomic gas) at the nozzle shock causes the stream thickness to be affected by pressure forces long after the shock occurrence. Their simulation showed that the long-term effects of pressure forces may only change the stream thickness by a factor of order unity.
Thus, we expect our results to be qualitatively unchanged even when gas pressure is included.


(4) An important caveat is that the delay time distribution (Figs. \ref{fig:N12} and \ref{fig:dpdn1}) depends sensitively on the initial thickness of the fallback stream $H_0$ as well as the angle $\omega_0$ between the major axis and the $\vec{L}$-$\vec{S}$ plane of the initial orbit. As for the initial stream thickness, we took $H_0=5R_\odot$, corresponding to the densest part of the stream that is initially confined by self-gravity. However, the most bound part of the fallback stream is non-self-gravitating and much less dense, and it has a larger thickness by a factor of a few to 10 \citep[see][]{Bonnerot21_stream_width_evolution}. For a larger $H_0$, we expect the delay time before the intersection to be shorter (see eq. \ref{eq:analyticN1}). Even if the most bound part of the stream has a shorter delay time before collision, it is so far unclear how the shocked gas from earlier fallback affects the later, denser parts of the stream. We speculate that, if the orbits of the denser parts are largely unaffected by the surrounding low-density\footnote{The density ratio between the unshocked stream and the surrounding gas is of the order $(r/H)^2 \sim 10^4$.} gas from earlier fallback, then the long delay times (of up to a decade) obtained from our study will lead to delayed mass feeding to the BH. This may potentially explain the late-time X-ray and UV emission \citep{vanVelzen19_late-time_UV, jonker20_late-time_Xray}, without invoking a viscous timescale of a few years or longer (which is difficult to achieve if the gas circularizes near the tidal disruption radius).






\section{Summary}\label{sec:summary}

We have developed an algorithm to find the position of stream collision that results from the tidal disruption of a $1 M_{\odot}$ star by a $10^6 M_{\odot}$ supermassive BH. This algorithm takes as input the BH spin $a$ and the stream's orbital parameters: the specific energy $\tilde E$, the total specific angular momentum $L$, the inclination angle $I$ between the initial orbital angular momentum and the BH spin. The longitudinal motion of the stream follows a geodesic in the Kerr spacetime, and the evolution of the transverse size of the stream is decoupled from the longitudinal evolution. Using an approximate tidal equation (including the effects of pressure forces at the nozzle shocks by assuming a perfect bounce), we calculate the evolution of the vertical thickness of the stream along the geodesic. This allows us to determine the occurrence of collision by comparing the stream thickness and the closest approach separation between different windings.

By performing a parameter space exploration for different $a$, $L$ and $I$ using this algorithm, we study various properties of the stream collision: intersection radius, collision angle, stream thicknesses at collision, transverse separation between the two colliding ends, relative speed at collision, and the delay time between the stellar disruption and stream collision. Our main findings are summarized as follows.


\begin{itemize}
    \item
    Misalignment of BH spin with the orbital angular momentum may lead to delay of the stream collision. The physical reason for the collision is a geometric effect (Fig. \ref{fig:schematic}). Collision occurs when the major axis of the elliptical orbit is brought to be nearly parallel with the initial $\vec L$-$\vec S$ plane (so as to minimize vertical separation between consecutive orbits) or to be nearly perpendicular to the initial major axis right after the disruption (so as to maximize the stream thickness). We calculate the delay time of the collision numerically (fully relativistic) and analytically (in the post-Newtonian limit) and find it to span a wide range from months up to a decade. This means that the thinnest parts of the fallback stream may only join the accretion flow many years after the initial tidal disruption.
    \item Stream self-intersection occurs in two modes: the head-on mode and the rear-end mode. About half of the TDEs are in the head-on mode, where the collision occurs far from the pericenter and the collision angle is close to 180$^{\rm o}$ (or $\cos\gamma\simeq -1$); the other half are in the rear-end mode, where the intersection occurs near the pericenter with a very small collision angle close to $0^{\rm o}$ (or $\cos\gamma\simeq 1$). Very few cases have collision angles close to $90^{\rm o}$.
    \item Intersection typically occurs between consecutive half orbits, i.e. $N_2 = N_1 + 1$, where $N_1 $ and $N_2$ are the half-orbit numbers of the two colliding ends.
    \item The intersection radius $r_{\rm int}$ for the head-on mode generally increases with the total angular momentum but depends less strongly on other parameters (whereas $r_{\rm int}\approx r_{\rm p}$ for the rear-end mode). This is because intersection typically occurs between adjacent orbital windings and the LT precession angle is much smaller than the apsidal precession angle. Larger angular momenta are associated with smaller apsidal precession angles that lead to larger intersection radii. The weak dependence of $r_{\rm int}$ on the BH spin and orbital inclination is such that larger $a\cos (I)$ leads to larger intersection radii.
    \item The ratio of the thicknesses of the two colliding ends $H_2/H_1$ is typically of order unity, and the transverse separation $\Delta s$ is a small but significant fraction of $H_1 + H_2$ (grazing collisions with $\Delta s\approx H_1 + H_2$ are rare). This means that the stream undergoes an offset collision which causes a large fraction of the gas to be shock heated. We expect the collision in the head-on mode to generate a puffy envelope of gas which undergoes secondary shocks and subsequently forms an accretion disk. In the rear-end mode, the stream is expected to be significantly broadened, which leads to further collisions in subsequent orbits and the eventual formation of an accretion disk.
    \item The relative velocity between the two colliding ends is at least a few percent of the speed of light, so the collision is always violent. The relative velocity mainly depends on the total angular momentum of the stream; $v_{\rm rel}\propto L^{-3}$ in the weakly relativistic regime. The weak dependence on the BH spin and inclination is such that larger $a\cos(I)$ leads to smaller relative velocity. The time-averaged energy dissipation rate by the self-crossing shock is of the order $10^{43}\rm\, erg\,s^{-1}$ for angular momentum near the boundary of tidal disruption ($L\sim L_{\rm max}\simeq 9.7$ for a $1M_\odot$ star and a $10^6M_\odot$ BH).
\end{itemize}

This work provides a realistic picture of the aftermath of stellar tidal disruption by a spinning BH. Our results for the intersection radius, collision angle, thickness ratio, and transverse offset can be used as initial conditions for hydrodynamic studies of the stream collision process, the subsequent formation of an accretion disk, and the associated electromagnetic emission \citep[e.g.,][]{bonnerot21_first_light}. Our approach can be improved in the future by including orbital energy spread at different parts of the fallback stream, and employing a more realistic stream width evolution as obtained by hydrodynamic studies.




\vspace{1cm}

\noindent
{\bf Data availability}
The data produced in this study will be shared on reasonable request to the authors.

\vspace{0.5cm}

\noindent
{\it Acknowledgments.} 
GB was supported by the Caltech Summer Undergraduate Research Fellowship. WL was supported by the David and Ellen Lee Fellowship at Caltech and the Lyman Spitzer, Jr. Fellowship at Princeton University. The research of CB was funded by the Gordon and Betty Moore Foundation through Grant GBMF5076. This project has received funding from the European Union’s Horizon 2020 research and innovation programme under the Marie Sklodowska-Curie grant agreement No 836751.

\appendix

\section{Minimum angular momentum}\label{sec:Lmin}

\begin{figure}
    \centering
    \includegraphics[width=3.5in]{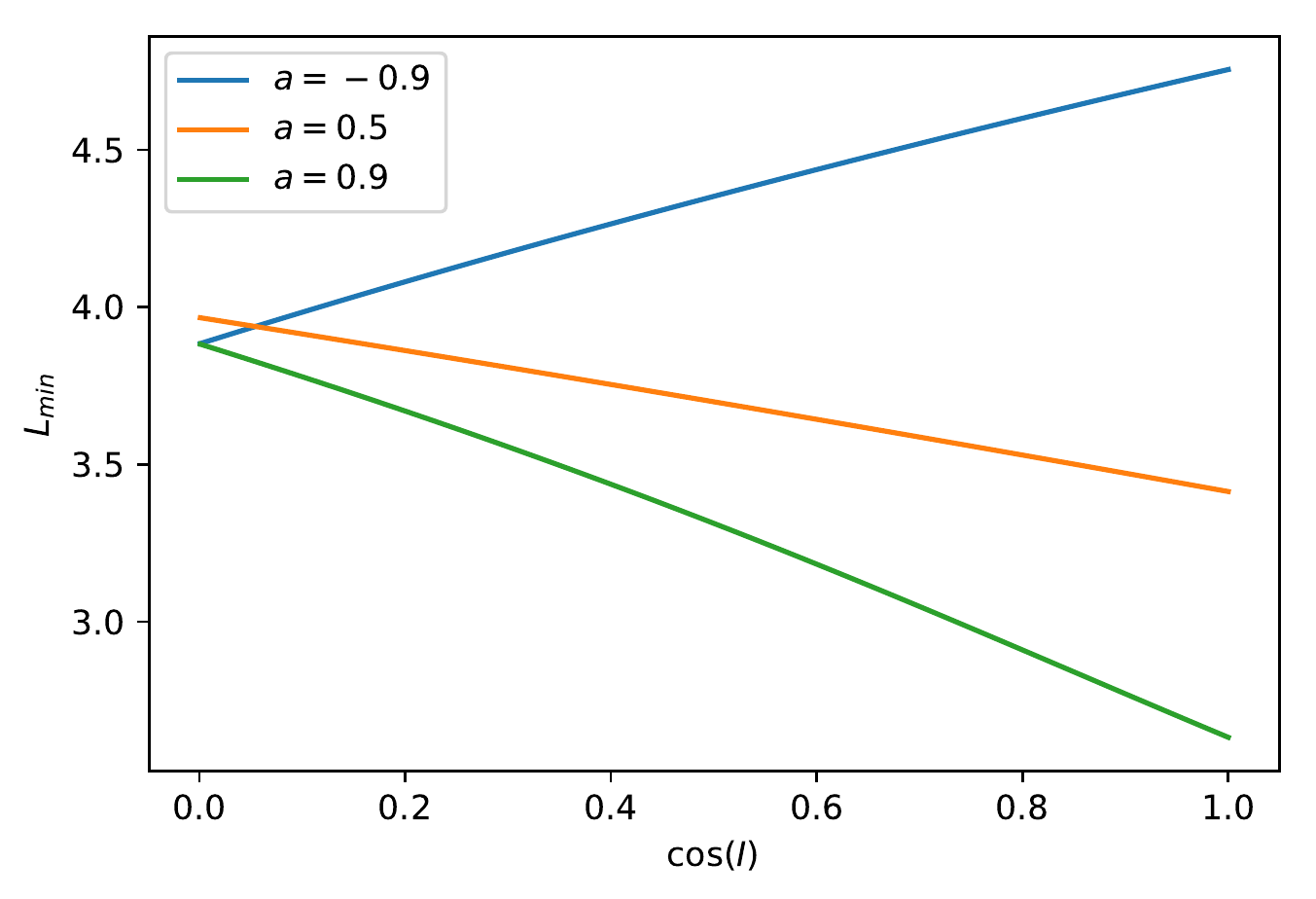}
    \caption{Minimum angular momentum $L_{\rm min}$ as a function of $\cos(I)$ for three different BH spins. }
    \label{fig:lmin}
\end{figure}

Fig. \ref{fig:lmin} shows the minimum angular momentum $L_{\rm min}$ (below which the star is swallowed as a whole) as a function of $\cos(I)$ for the different BH spins $a=-0.9$, $a=0.5$ and $a=0.9$. We numerically solve $V_r=dV_r/dr=0$ for multiple possible solutions ($L_{i}$, $r_{i}$), where $V_r$ is given by Eq. (\ref{eq:Vr}). Then, $L_{\rm min}$ equals the $L_i$ corresponding to the lowest $r_i$ outside the event horizon of the BH.

\section{Newtonian tidal forces}\label{sec:Newtonian_tidal_equation}


For the purpose of understanding the potential errors in adopting the approximate tidal forces in Eq. (\ref{eq:tfe}), it is illustrative to study the thickness evolution using a different model. In this section, we evolve the stream thickness according to the Newtonian tidal equation
\begin{eqnarray}\label{eq:ntfe}
    \frac{d^2H}{d\tau^2}\approx -\frac{H}{r^3},
\end{eqnarray}
where we have retained only the lowest order in the tidal tensor, since the Schwarzschild limit ($r\gg a$) of the vertical tidal tensor component is $C_{22}\approx r^{-3}(1 + 3L^2/r^2)$ and $3L^2/r^2\simeq 6r_{\rm p}/r^2\ll 1$ ($r_{\rm p}$ being the pericenter radius) in the non-relativistic limit. 

\begin{figure}
    \centering
    \includegraphics[width=0.47\textwidth]{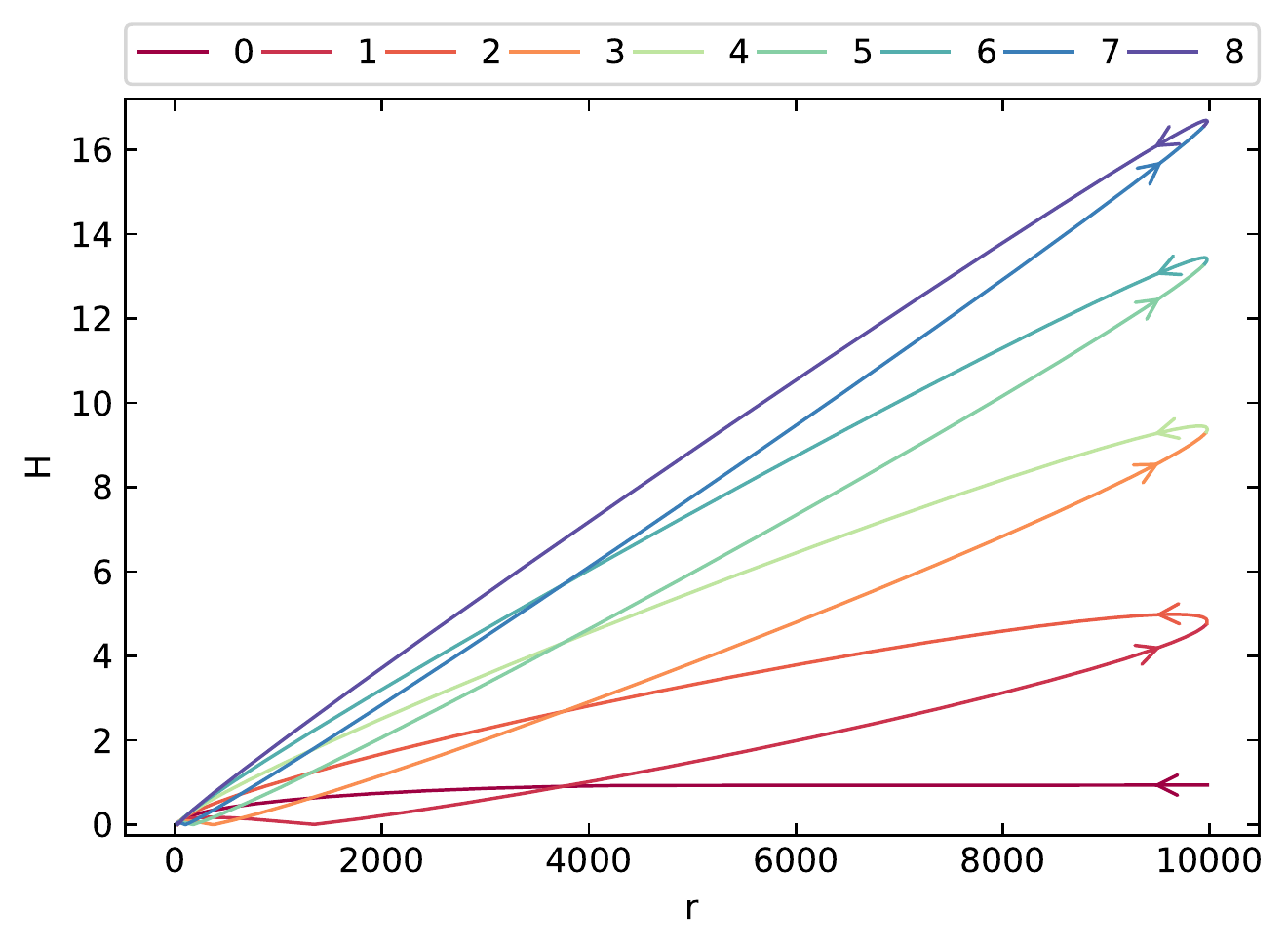}
    \includegraphics[width=0.47\textwidth]{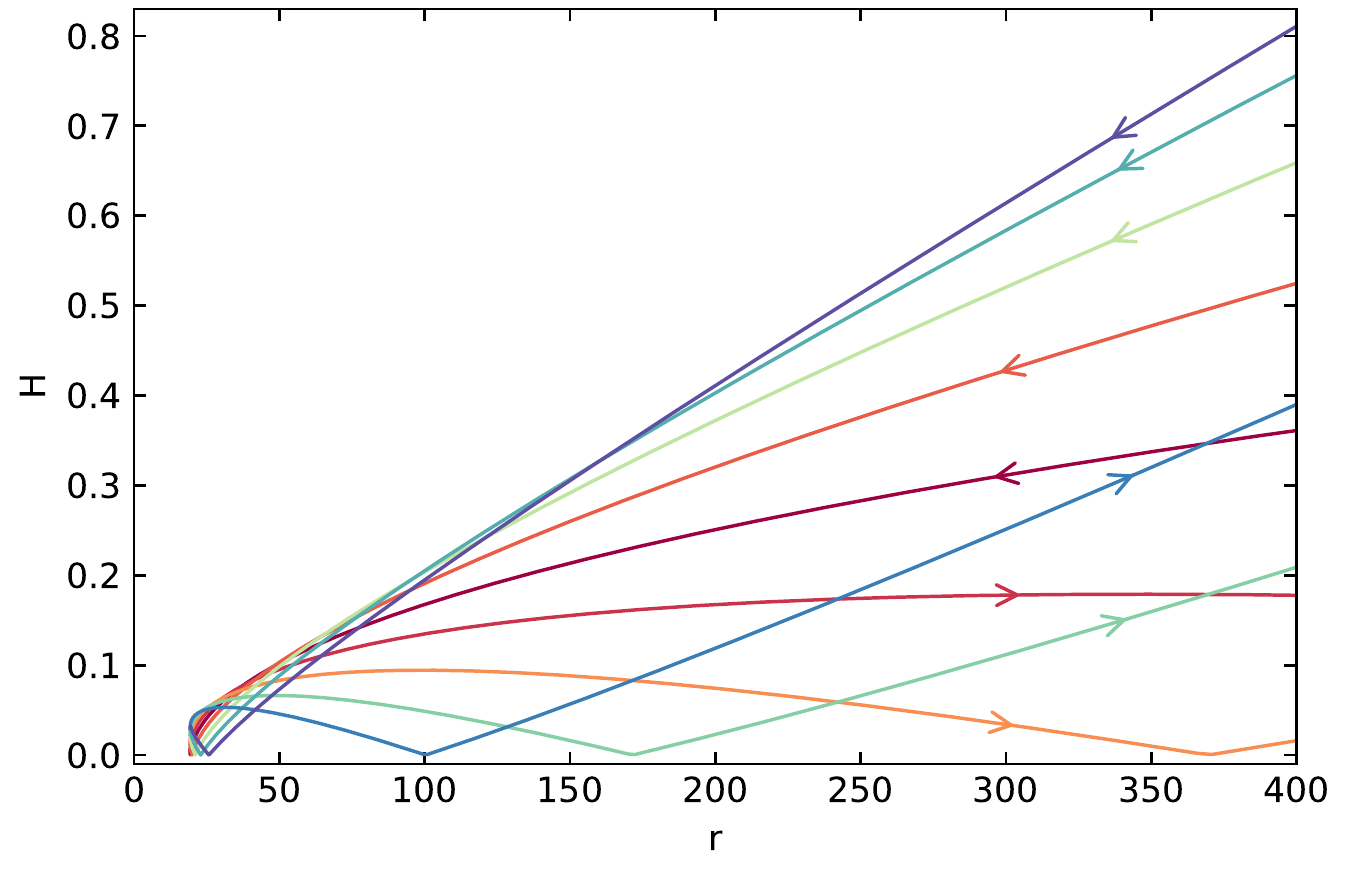}
    \includegraphics[width=0.47\textwidth]{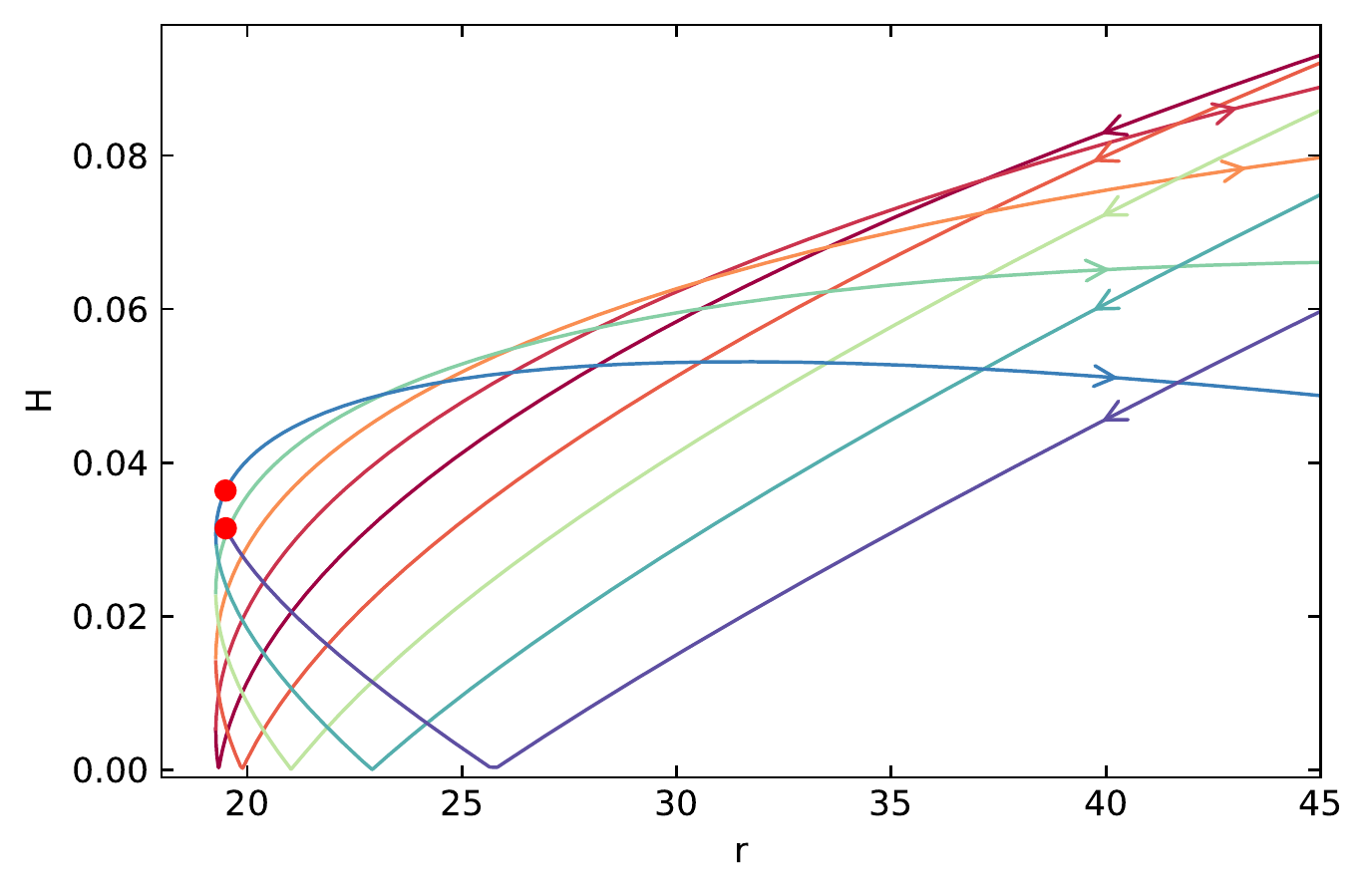}
    \caption{The evolution of stream thickness $H$ as a function of $r$, magnified further to show details, when the Newtonian tidal equation (\ref{eq:ntfe}) is used. On average, the stream becomes thicker with each successive winding, and the thickness ratio between the outward and inward segments of the same winding decreases with time. The geodesic is the same as shown in Fig. \ref{fig:fullorbit}, with $\tilde E=0.9999$, $L=6.5$, $\cos I=0.5$, $a=0.9$. Here we set $\sigma=1$ (see \S \ref{sec:stream_thickness} for its definition). The evolution is terminated when the stream crossing occurs (near the pericenter), as marked by the red dots. Both axes are in units of the gravitational radius $GM/c^2$.}
    \label{fig:nstream}
\end{figure}

We find that the stream thickness evolution under the simplest Newtonian tidal equation is qualitatively similar to the behavior described in \S \ref{sec:stream_thickness}. A representative case is shown in Fig. \ref{fig:nstream}, for the same initial conditions and orbital parameters as in Fig. \ref{fig:H_evolve} in \S \ref{sec:stream_thickness}: total energy $\tilde E=0.9999$, total angular momentum $L=6.5$, inclination angle $\cos I=0.5$, BH spin $a=0.9$, and initial thickness $H_0=5R_\odot$. Under the Newtonian tidal equation, the stream thickness is smaller by about a factor of 2, which causes the self-crossing to occur at a slightly later time between the 7th and 8th half-orbits (as opposed to the 6th and 7th half-orbits under the more sophisticated tidal equation \ref{eq:tfe}). We note that in this example, the collision occurs in the rear-end mode (as opposed to the head-on mode under tidal equation \ref{eq:tfe}). However, the overall result under the Newtonian tidal equation is still that about half of TDEs have rear-end collisions and the other half have head-on collisions.



The details of the collision for this example are provided in Table \ref{tab:approxH}, which should be compared with Table \ref{tab:finalresult}, which was obtained under the more sophisticated tidal equation (\ref{eq:tfe}). The relative speed of collision $v_{\rm rel}$ and the thickness ratio $H_2/H_1$ are very similar in these two choices of tidal equations.

\begin{table}
    \centering
    \begin{tabular}{c c}
    \hline
    $i$ & 8\\
    $j$ & 7\\
    $r$ & 19.63\\
    $\theta\rm\,[rad]$ & 0.53\\
    $\phi\rm\,[rad]$ & 1.89\\
    $\Delta s$ & 0.05\\
    $H_1+H_2$ & 0.07\\
    $\Delta s/(H_1+H_2)$ & 0.79\\
    $H_2/H_1$ & 0.86\\
    $H_1$ & 0.04\\
    $\cos \gamma$ & 0.98\\
    $v_{\rm rel}$ & 0.08\\
    \hline
    \end{tabular}
    \caption{For Newtonian evolution of the stream, properties of the intersection region, including the spatial coordinates ($r,\, \theta,\, \phi$), minimum separation between the two colliding segments $\Delta s$, the stream width $H_1$ and $H_2$, the collision angle $\gamma$, and the relative speed of collision $v_{rel}$, for the parameters $\tilde E=0.9999$, $L=6.5$, $\cos I=0.5$, $a=0.9$, and $\sigma=1$. All results are in units where $G=M=c=1$.
    }
    \label{tab:approxH}
\end{table}

\bibliography{references}
\bibliographystyle{mnras}

\bsp	
\end{document}